\documentclass[aps,twocolumn,superscriptaddress,longbibliography,floatfix,notitlepage]{revtex4-2}
\usepackage{bbm}
\usepackage{physics}
\usepackage{amsmath}
\usepackage{epsfig}
\usepackage{subfigure,mathrsfs}
\usepackage{array}
\usepackage{amssymb}
\usepackage{braket}
\usepackage{lmodern,amssymb}
\usepackage{physics}
\usepackage[dvipsnames]{xcolor}
\usepackage{lipsum}

\newcommand{\beq}[0]{\begin{equation}}
\newcommand{\eeq}[0]{\end{equation}}

\def\be{\begin{equation}}
\def\ee{\end{equation}}
\def\bea{\begin{eqnarray}}
\def\eea{\end{eqnarray}}
\newcommand{\ba}{\begin{eqnarray}}
\newcommand{\ea}{\end{eqnarray}}
\usepackage{hyperref}

\usepackage[bb=boondox]{mathalfa}
\usepackage{graphicx}
\usepackage{booktabs}
\usepackage{tabularx}
\usepackage{array}

\begin{document}

\title{Thermodynamics of Coupled Transport: Quantum Dots as a Model System}

\author{Shuvadip Ghosh}
\thanks{shuvadipg21@iitk.ac.in}
\affiliation{Indian Institute of Technology Kanpur, 
	Kanpur, Uttar Pradesh 208016, India}
\author{Sounak Bandyopadhyay}
\affiliation{Indian Institute of Technology Kanpur, 
	Kanpur, Uttar Pradesh 208016, India}   
\author{Arnab Ghosh}
\thanks{arnab@iitk.ac.in}
\affiliation{Indian Institute of Technology Kanpur, 
	Kanpur, Uttar Pradesh 208016, India}

\begin{abstract}
This review presents a thermodynamic perspective on quantum coupled transport processes in nanoscale systems. Our analysis is formulated within the framework of the entropy production rate, the central thermodynamic quantity that governs non-equilibrium processes and can be expressed in terms of the existing conjugate force–flux pairs. Although thermodynamic laws are valid across classical–quantum boundaries, the present analysis is carried out within a microscopic open quantum system framework, focusing on quantum dots (QDs) coupled to electronic reservoirs. We begin with elementary single transport processes and highlight their strong thermodynamic constraints in the near-equilibrium regime. This motivates the study of coupled transport, where multiple force–flux pairs coexist and interact, giving rise to richer thermodynamic behavior. Employing the entropy production rate as the guiding principle, we analyze coupled energy and particle transport in a minimal two-terminal single–QD setup and show how standard thermoelectric phenomena, such as Seebeck and Peltier effects, as well as thermoelectric heat engines and refrigerators, naturally emerge as a consequence of thermodynamic cross-effects. Next, we extend the framework to a three-terminal coupled quantum dot (CQD) geometry, which provides a versatile platform for exploring coupled transport. Under appropriate constraints, this model reduces to the well-known Sánchez–Büttiker configuration, capturing conventional thermoelectric behavior analogous to that of classical devices. Beyond standard cross-effects, we focus on the conceptually distinct and counter-intuitive phenomenon of inverse currents in coupled transport (ICC), where a current flows against all mutually parallel thermodynamic forces without violating the second law of thermodynamics. We demonstrate that genuine ICC requires breaking the symmetry between two transport channels and identify the necessary and sufficient conditions for its realization within a reduced CQD framework through attractive interdot interactions. Finally, we discuss the implications of the ICC for the design of autonomous quantum thermoelectric engines and refrigerators, highlighting new avenues for unconventional thermodynamic functionality in QD-based systems.
\end{abstract}

\maketitle

\section{Introduction}

\textit{``Thermodynamics is the only physical theory of universal content which I am convinced that, within the framework of applicability of its basic concepts, will never be overthrown.''}
— \textsc{Albert Einstein}

This celebrated remark by \textit{Albert Einstein} eloquently underscores the fundamental and universal character of thermodynamics. Unlike many physical theories whose validity is limited by scale, approximation, or domain of applicability, thermodynamics transcends the classical–quantum boundary \textit{per se}~\cite{callen1985book,kondepudi2015book,gemmer2009book,vinjanampathy2016quantum} and remains relevant across an exceptionally broad spectrum of physical systems. Adopting a thermodynamic perspective to analyze physical processes is, therefore, not only natural but also deeply insightful. Such an approach is often illuminating and reveals model-independent information and governing principles, thereby opening rich avenues for both theoretical exploration and experimental investigation.

{In this review, we \textit{focus} on the \textit{thermodynamic aspects} of a particular class of transport phenomena, namely \textit{coupled quantum transport processes}.} Transport phenomena themselves constitute an extensively studied area of research~\cite{datta1997electronic,imry2002introduction}, with a vast literature that includes both classical ~\cite{degroot1962non} and quantum systems~\cite{nazarov2009quantum}, with fundamentally different microscopic mechanisms and mathematical frameworks~\cite{diventra2008electrical}. {The present review does not aim to address these detailed dynamical descriptions; instead, our emphasis lies entirely on the \textit{thermodynamic characterization of transport processes}.} Although our discussion focuses primarily on quantum transport, the thermodynamic framework~\cite{vinjanampathy2016quantum,seifert2012stochastic} adopted here applies equally to classical and quantum transport, providing a unified framework for coupled transport phenomena.

The primary motivation for studying coupled transport lies in understanding how transport processes are constrained and governed by a clear thermodynamic interpretation. Any transport process is inherently a non-equilibrium phenomenon, involving a departure from an initial equilibrium state, followed by out-of-equilibrium dynamics and the eventual settlement towards a final equilibrium or steady state configuration. The thermodynamic nature of such non-equilibrium processes is encapsulated within a central quantity: the entropy production rate~\cite{degroot1962non,seifert2012stochastic,kondepudi2015book}. This quantity serves as the touchstone of non-equilibrium thermodynamics and determines the thermodynamic permissibility of physical processes. A process is thermodynamically allowed \textit{iff} its entropy production rate is non-negative; violation of this condition would contradict the second law of thermodynamics. It imposes fundamental bounds and limitations on transport properties, deciding both its directionality and efficiency. {While transport phenomena can, in principle, be analyzed across the full non-equilibrium domain, our interest in this review is restricted to the \textit{near-equilibrium regime}, where the analytical characterization of the entropy production rate is feasible.} Beyond near-equilibrium, thermodynamic analysis becomes fairly intricate and lies outside the scope of this present review. 

Any transport process involves a flux or current. From basic thermodynamics, it is clear that a nonzero flux must be driven by a nonzero thermodynamic force. Forces and fluxes, therefore, constitute the basic building blocks of the theory of thermodynamic transport. In the regime of near-equilibrium, these quantities are linked through the entropy production rate and can be written as the sum of the products of the thermodynamic forces and their conjugate fluxes~\cite{callen1985book,degroot1962non,kondepudi2015book,callen1948the-application,prigogine1961book,onsager1931reciprocal-I,onsager1931reciprocal-II}. This formulation provides a transparent and self-consistent framework within which the magnitude, direction, and mutual influence of different currents can be systematically analyzed.

A transport process involving a single force-flux pair is conceptually the simplest; also, its near-equilibrium behavior is comparatively easy to analyze and largely predictable. To access richer and more complex thermodynamic behavior, it is important to consider situations involving multiple force–flux pairs. This leads to the study of coupled transport with at least two distinct transport processes interlinked with one another. In principle, the analysis could be extended to systems involving a larger number of coupled force–flux pairs. However, such frameworks rapidly become intractable and conceptually opaque. {Coupled transport, involving exactly two force–flux pairs, thus represents an optimal compromise: it remains rather simple to permit systematic analysis, yet considerably rich to exhibit a wide variety of nontrivial and physically interesting phenomena. Indeed, it underlies the operational principles of several well-known thermodynamic effects~\cite{monsel2022geometric,josefsson2018a,prete2019thermoelectric,sanchez2011optimal,zhang2015three,thierschmann2015three,thierschmann2016thermoelectrics,erdman2017thermoelectric,jurgens2013thermoelectric}, including Seebeck~\cite{callen1985book,kondepudi2015book,Seebeck1822magnetische,Kelvin1856on,degroot1984non,Mahan1996the,esposito2009thermoelectric,sanchez2011optimal,svensson2012lineshape,whitney2014most} and Peltier effects~\cite{peltier1834nou,callen1985book,degroot1984non,kondepudi2015book,jordan2013powerful,sothmann2012quantum,whitney2014most}, as well as quantum thermoelectric heat engines~\cite{sanchez2011optimal,thierschmann2015three,donsa2014double,dare2017powerful,singha2020realistic} and refrigerators~\cite{zhang2015three,erdman2018absorption,dare2019comparative,mukherjee2020three,barman2021realistic,mayrhofer2021stochastic}. Therefore, a proper thermodynamic understanding of these phenomena becomes an integral part of the theory of coupled quantum transport. In this review, we aim to elucidate these connections in detail. Beyond conventional thermodynamic cross-effects, we will also explore more subtle and counter-intuitive transport phenomena, such as inverse currents in quantum settings~\cite{wang2020inverse,ghosh2026inverse}.}

The scope and purpose of the present review are twofold. The first half of the review (Secs.~\ref{sec2}–\ref{sec4}) provides a clear pedagogical introduction accessible to nonspecialists. Starting from thermodynamics of transport processes, it proceeds to the construction of the model Hamiltonian, survey of various quantum transport methods and the derivation of reduced dynamics in terms of the Lindblad master equation. These sections are intended to provide researchers new to the field of quantum thermal transport with a sufficient background to understand the central theme developed in the second half of the review. Advanced researchers already familiar with open quantum systems and quantum thermodynamics can easily skip these introductory sections and proceed directly to Sec.~\ref{sec5}, which presents the research of topical interest and actual novel contributions of this review: inverse-current in coupled (ICC) quantum transport, possible experimental platforms, their realization, and potential future applications.

The detailed organization of the review is as follows: In Section~\ref{sec2}, we introduce the simplest possible case, namely a single transport process. We then proceed to the central theme of this review --- coupled transport. In Section~\ref{sec3}, we construct a general theoretical framework for investigating quantum coupled transport, including detailed discussions of the system, the environment, and their mutual interactions. In Sec.~\ref{survey}, we present a summary of various quantum transport methods before starting our analysis of coupled transport with a single quantum dot (SQD) in a two-terminal configuration. This section provides readers with a clear understanding of why the Lindblad master equation is particularly well suited to analyzing steady-state thermal currents in strongly coupled quantum systems, compared with other quantum transport methods. Following this, the dynamical evolution and steady-state currents of the SQD system are presented in Sec.~\ref{sec5} using the Lindblad master equation. In Section~\ref{sec5}, we derive a general expression for the entropy production rate, which constitutes the basic framework for the thermodynamic theory presented throughout the review. In Section~\ref{sec6}, we identify the relevant force–flux pairs for the SQD model. We then move on to a detailed analysis of coupled transport in the minimal SQD setup. This analysis is carried out case by case in Section~\ref{sec7}. In Section~\ref{sec8}, we clarify the subtle but important distinction between conventional thermodynamic cross-effects and the phenomenon of ICC. The discussion is then extended to a more general framework in Section~\ref{sec9}, where we analyze coupled transport in a three-terminal coupled QD (CQD) model, which naturally captures the thermodynamics of thermoelectric devices. In Section~\ref{sec10}, we specifically focus on ICC, highlighting their physical origin and thermodynamic implications as one of the most counterintuitive outcomes of transport processes. Finally, we conclude our review in Section~\ref{outlook} with an outlook section, highlighting open questions and future research directions in the field.

\section{Thermodynamics of transport process}\label{sec2}

We start with a general framework of transport phenomena that is equally applicable to macroscopic and microscopic systems. The permissibility of any physical process is strictly governed by the second law of thermodynamics, which dictates that the entropy production rate ($\dot\Sigma$) must be positive for a spontaneous process, while a zero entropy production rate corresponds to a state of thermal equilibrium~\cite{callen1985book,kondepudi2015book}. For transport phenomena, in the near-equilibrium condition, the entropy production rate can be expressed in terms of the force–flux pairs present in the system as
\begin{equation}\label{EPR}
    \dot\Sigma=\sum_i J_i\mathcal{F}_i
\end{equation}
where $J_i$ denotes the $i$-th thermodynamic flux and $\mathcal{F}_i$ is its corresponding conjugate force, interlinked through a cause-and-effect relationship. When a single thermodynamic force is applied, it induces a response in the system in the form of a flux. The resulting flux aligns along the direction of the applied force, which arises from gradients such as temperature or chemical potential. For example, a temperature gradient drives an energy current in the form of heat flow; thus, it acts as a conjugate energy force associated with heat transport.

\subsection{Single Transport Process}

If a single force–flux pair is present, thermodynamic analysis is relatively straightforward. In such a process, the entropy production rate reduces to $\dot\Sigma=J\mathcal{F}$, which implies that the flux should be aligned in the same direction as the conjugate force; otherwise, the entropy production rate would be negative. This can be illustrated using an example of a classical heat transfer between two bodies, say $a$ and $b$. Although the sign of currents is a matter of convention, clearly defining it is beneficial to avoid ambiguity in subsequent sections. Let us define any force or flux aligned from $a$ to $b$ as positive, while the opposite corresponds to a negative sign. Hence, the energy force is defined as 
$\mathcal{F}_a=T_a-T_b$ is positive, indicating that $a$ is the hotter body, and the entropy production rate is $\dot\Sigma=J_a\mathcal{F}_a>0$. This implies that the heat current $J_a$ must be positive, with the flow of heat from the hotter body $a$
to the cooler body $b$, in complete accordance with the second law of thermodynamics. 

The paradigmatic example considered above is the transport of heat (energy), where an energy flux in the form of a heat current arises because of a temperature difference or, more precisely, in the form of an inverse-temperature gradient or thermodynamic force~\cite{degroot1962non,fourier1822the}. Another common example is particle or mass transport, where a particle current is generated by a difference in the chemical potential between two regions. This mechanism underlies diffusion processes in which particles migrate from regions of higher chemical potential to lower chemical potential~\cite{fick1855on,degroot1962non}. Similarly, charge transport along a wire represents a single transport process in which an electric current is driven by a voltage difference (electric potential)~\cite{ohm1827die,ashcroft1976solid}, as observed in electrical conduction through metallic or semiconducting materials~\cite{ashcroft1976solid,ziman1972principles,datta1995electronic}. Single transport behavior also arises in spin transport, where a spin current flows in response to a spin bias or a difference in spin chemical potential (e.g., $\Delta \mu_{\rm S}=\mu_{\uparrow}-\mu_{\downarrow}$)~\cite{zutzic2004spintronics}. Such processes are particularly relevant in magnetic and spintronic systems, where spin degrees of freedom are manipulated independently of charge. In momentum transport, commonly encountered in fluid dynamics, the flux corresponds to momentum flow (expressed through the stress tensor), while the conjugate force is a velocity gradient; this mechanism gives rise to viscous flow in fluids~\cite{landau1987fluid}. Unlike in the case of energy or heat transport, where a gradient is directly related to the corresponding thermodynamic force, this equivalence, however, does not hold for all types of transport. This is because the gradient of a physical quantity and the associated thermodynamic force need not always be identical. For instance, in particle transport, the corresponding thermodynamic force is not simply given by the chemical potential difference itself and may instead depend on the respective temperatures as well. To ensure thermodynamic consistency and correctness, it is therefore essential to identify the appropriate thermodynamic force. To identify the precise form of the conjugate forces, a detailed analysis of the entropy production rate is required, since current alone cannot determine the actual form of the thermodynamic forces. Moreover, in the linear-response regime, a single transport process takes a simple form in terms of \textit{Onsager kinetic coefficient}~\cite{callen1948the-application,onsager1931reciprocal-I,onsager1931reciprocal-II,kondepudi2015book}. For a single force–flux pair, the flux is influenced by its only conjugate force, and it is characterized by the \textit{direct Onsager coefficient} ($L$), via the relation $J=L\mathcal{F}$. Thus, the entropy production rate can be written as $\dot{\Sigma}=L\mathcal{F}^2 \geq 0$, which readily implies that $L$ must be positive.

While the non-negativity of the entropy production rate is universal and remains valid in near equilibrium as well as very far away from equilibrium, the above discussion is primarily valid in the near equilibrium regime. The form of the entropy production rate defined in Eq.~\eqref{EPR} is not applicable for the far-from-equilibrium condition~\cite{kondepudi2015book}. There, a current can flow against its conjugate force in a single transport process, yet it is not forbidden by the laws of thermodynamics. Such an intriguing transport phenomenon is the concept of absolute negative mobility (ANM)~\cite{machura2007absolute,reimann2002brownian,eichhorn2002brownian,nagel2008observation,zhu2004absolute}, in which the response of the system, i.e., the generated current, operates counter to the driving force. However, ANM cannot exist around a thermal equilibrium, as it would create a \textit{perpetual motion machine of the second kind}, capable of performing work with just a single heat bath. Although ANM was initially considered as a consequence of quantum effects ~\cite{hopfel1986negative,keay1995dynamic}, recent studies have explored ANM in various
non-equilibrium classical systems, such as tracer dynamics in steady laminar flows~\cite{sarracino2016nonlinear}, self-propulsion~\cite{ghosh2014giant}, particle separation~\cite{rugueara2012entropic,slapik2019tunable}, as well as various interacting Brownian motions~\cite{slapik2019tunable,cleuren2001ising,cleuren2002random,machura2007absolute,dandigbessi2015absolute,spiechowicz2019coexistence}. Furthermore, there has also been experimental evidence to support ANM in the presence of a single force~\cite{keay1995dynamic,ros2005absolute,nagel2008observation}. Thus, single transport processes can exhibit interesting thermodynamic behavior beyond the near-equilibrium regime, but their detailed analysis lies beyond the scope of this review.

However, if we go beyond single transport, exciting thermodynamic phenomena can occur even within the near-equilibrium situation. The interplay between multiple force–flux pairs enables us to offer a variety of thermodynamic effects. The subsequent sections will focus on coupled transport, which constitutes the main body of this present review.

\subsection{Coupled Transport Process}
Coupled transport refers to a physical situation where two transport processes coexist and are coupled to each other, with pairs of thermodynamic forces and fluxes. Therefore, to investigate coupled transport, one must consider two distinct types of flow --- such as energy and particles, or spin and mass --- and construct a model in which these transport processes are simultaneously present and are intrinsically coupled. Here, without loss of generality, we consider energy and particle currents. Following Eq.~\eqref{EPR}, the general form of the entropy production rate for such a setup takes the simple form: 
\begin{equation}\label{CT1}
\dot{\Sigma}=J_{\rm{E}}\mathcal{F}_{\rm{E}}+J_{\rm{N}}\mathcal{F}_{\rm{N,}}
\end{equation}
where $J_{\rm{E}}(J_{\rm{N}})$ is the energy (particle) current, conjugate to the energy (particle) force, $\mathcal{F}_{\rm{E}}(\mathcal{F}_{\rm{N}})$. It is immediately clear that both currents cannot flow simultaneously against their respective conjugate forces. Such a configuration would render the entropy production rate negative, thus violating the second law of thermodynamics.

However, mere coexistence of two different transport processes does not necessarily imply coupled transport. If the transport processes are independent, each current is driven solely by its own conjugate force; no nontrivial thermodynamic behavior emerges in such a scenario, as the system merely exhibits two independent \textit{single-transport} processes rather than genuine coupled transport. True coupled transport occurs only when each flux responds not only to its conjugate force but also to a non-conjugate force. Therefore, we must construct a model in which two transport processes coexist in an integrated manner. Within the linear-response regime, this mutual interdependence can be expressed through Onsager coefficients. For energy and particle currents, we have
\begin{equation}\label{Onsager}
\begin{pmatrix}
J_{\mathrm{E}}\\
J_{\mathrm{N}}
\end{pmatrix}
=
\begin{pmatrix}
L_{\rm{EE}} & L_{\rm{EN}}\\
L_{\rm{NE}} & L_{\rm{NN}}
\end{pmatrix}
\begin{pmatrix}
\mathcal{F}_{\rm{E}}\\
\mathcal{F}_{\rm{N}}
\end{pmatrix}.
\end{equation}
where, $L_{\mathrm{EE}}$ and $L_{\rm{NN}}$ are the ``\textit{direct Onsager coefficients}", and $L_{\rm{EN}}$ and $L_{\rm{NE}}$ refer to the ``\textit{cross Onsager coefficients}", fulfilling ``\textit{Onsager reciprocity}" condition $L_{\rm{EN}}=L_{\rm{NE}}$, in the absence of magnetic
fields or other time-reversal-odd variables~\cite{onsager1931reciprocal-I,onsager1931reciprocal-II}. 
For a single transport process, from the non-negativity criteria of the entropy production rate, it is clear that the \textit{``direct Onsager coefficients"} must be positive, i.e., $L_{\mathrm{EE}}, \; L_{\rm{NN}} \geq 0$. However, the \textit{cross coefficients} may assume negative values, and it is precisely this property that gives rise to a variety of nontrivial and intriguing transport phenomena~\cite{wang2020inverse}. Moreover, if the time-reversal symmetry is broken, this \textit{Onsager reciprocity} must be replaced by a more general Onsager-Casimir relation which states $L_{ij}(\textbf{B})=\epsilon_i\epsilon_j L_{ji}(\textbf{-B})$; where $\textbf{B}$ is the magnetic field and $\epsilon_i=\pm 1$ denotes the time-reversal parity of the corresponding thermodynamic flux~\cite{casimir1945on}.

In the case of two independent ``\textit{single-transport}" processes, the Onsager cross coefficients vanish, indicating that each flux is unaffected by non-conjugate thermodynamic forces. The existence of a nonzero off-diagonal Onsager coefficient is therefore an absolutely \textit{necessary} criterion. For a genuine coupled transport process, Eq.~\eqref{Onsager} yields a symmetric Onsager matrix which is positive semi-definite in the linear response regime and expressed as $L_{\rm{EE}}L_{\rm{NN}}-L_{\rm{EN}}^2\geq0$. Henceforth, our next objective is to identify a suitable model in which genuine \textit{quantum coupled transport} (QCT) can be thoroughly studied. 

\section{The model for quantum coupled transport}\label{sec3}

\begin{figure}[h!]
    \centering
\includegraphics[width=\columnwidth,height=5.5cm]{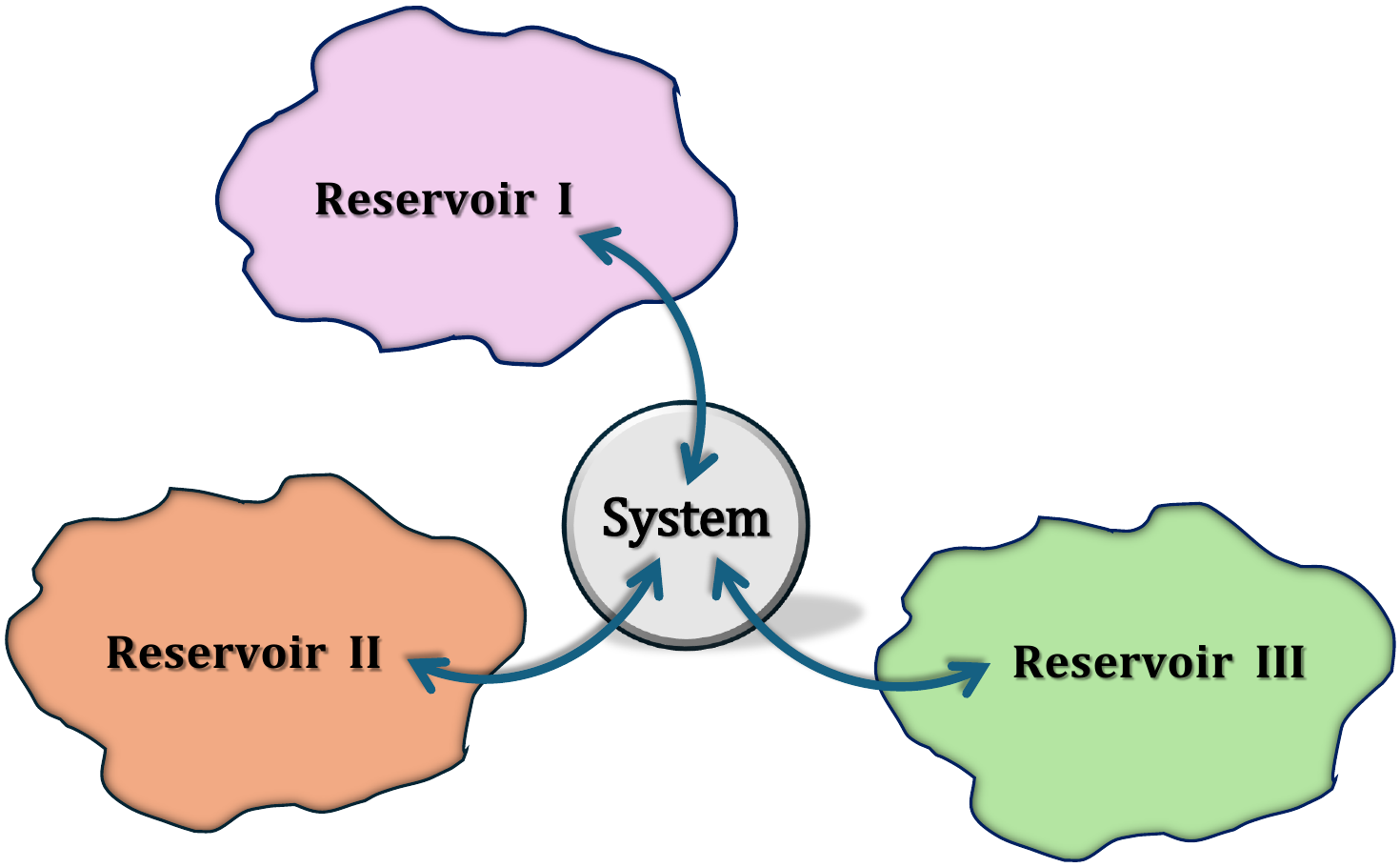}
    \caption{A small quantum system interacts with multiple reservoirs}
    \label{General Model Review}
\end{figure}

The subject of quantum transport has attracted significant attention in recent years because of its rich thermodynamic features and its experimental realizability~\cite{benenti2017fundamental,datta1995electronic,sanchez2011optimal,thierschmann2015three,esposito2009thermoelectric,pekola2021colloquium,sanchez2019thermoelectric,kouwenhoven1997electron,vanDerWiel2002electron,linke2015harnessing}. The theory of quantum transport can be described within the framework of open quantum systems~\cite{breuer2002book,louisell1990book,carmichael1993book,gelbwaser2015thermodynamics,kurizki_kofman_2022,ghosh2017catalysis,gupt2022PRE,gupt2023topranked,gupt2024graph,shuvadip2022univarsal}, where a small quantum system interacts with one or more external reservoirs [Fig.~\ref{General Model Review}]. This general framework consists of three basic constituents: the system, the environment or reservoir, and the interaction between them. Over the years, a variety of standard system–bath models have been proposed in the literature to investigate a wide range of physical processes within this paradigm. Typically, the system is modeled using elementary building blocks such as two-level systems (TLSs) or simple harmonic oscillators, while the reservoirs are composed of bosonic or fermionic degrees of freedom, obeying Bose–Einstein (BE) or Fermi–Dirac (FD) statistics, respectively.

In the present work, our primary focus is on the quantum dot–fermionic reservoir model, which provides a natural platform for studying coupled charge and heat transport. In this setup, quantum dots (QDs)—effectively modeled as TLSs—form the central system, whereas the leads act as fermionic reservoirs composed of electrons. This framework supports both particle (charge) and energy (heat) currents, making it particularly suitable for investigating thermoelectric effects~\cite{seifert2012stochastic,monsel2022geometric,josefsson2018a,prete2019thermoelectric,jaliel2019experimental,nakpathomkun2010thermoelctric,josefsson2019optimal,sanchez2011optimal,zhang2015three,mcConnell2022strong,thierschmann2015three,erdman2017thermoelectric,jurgens2013thermoelectric}, Maxwell’s demon–type feedback mechanisms~\cite{strasberg2013thermodynamic,esposito2012stochastic,annby2022quantum,annby2024maxwell}, and transport phenomena relevant to quantum information processing~\cite{strasberg2013thermodynamic,kutvonen2016thermodynamics,esposito2019thermodynamics}. Consequently, the QD–fermionic reservoir model serves as an ideal candidate to analyze the thermodynamics of coupled transport in nanoscale systems.

Alongside this model, several other well-established theoretical frameworks offer complementary insights into quantum transport phenomena. In particular, the spin–boson~\cite{leggett1987dynamics,weiss2012quantum,agarwalla2017energy,breuer2002book,strasberg2016quantum,gelbwaser2014heat,segal2006heat} and Caldeira–Leggett model~\cite{calderia1981influence,caldeira1983path,calderia1987dynamics} provide canonical descriptions of thermal energy transport and constitute the theoretical foundation for quantum thermal devices such as heat engines~\cite{kosloff2014quantum,nikhil2021statistical,samarth2023introduction}, refrigerators~\cite{kosloff2014quantum,bhardwaj2017in,levy2012quantum,hofer2016autonomous,chen2023optimal,bhattacharyya2025transient}, rectifiers~\cite{li2012colloquium,sanchez2015heat,li2004thermal,scheibner2008quantum,ordonez2017quantum,wang2019thermal,scheibner2008quantum,kalantar2021harmonic,marcos2018thermal,shuvadip2022univarsal,terraneo2002controlling}, and thermal transistors~\cite{neumeier2013single,joulain2016quantum,guo2018quantum,wijesekara2021darlington,ekanayake2023stochastic,gupt2022PRE}. In these models, the reservoirs are composed of bosonic excitations—such as photons, phonons, or magnons—that act as heat carriers, while the system elements, represented by qubits, harmonic oscillators, or interacting spins, govern the energy exchange dynamics. Furthermore, the Landauer–B\"uttiker formalism~\cite{buttiker1986four} is widely used to describe quantum ballistic transport, in which bosonic or fermionic particles can carry coherent current flow. Together, these models form the theoretical basis for understanding nonequilibrium transport in quantum systems.

\subsection{System} 

The key element of our system is the QD. Considering a QD as the building block, one can, however, build more complex systems, like a Double-QD (DQD)~\cite{strasberg2022quantum,vanDerWiel2002electron,dorsch2021heat}, Coupled-QD (CQD)~\cite{gupt2024graph,shuvadip2022univarsal,ghosh2026inverse,joulain2016quantum,sanchez2011optimal}, Series-coupled QDs (SCQD)~\cite{zhang2015thermoelectric,brandes2005coherent,chi2016thermoelectric,zhang2023inverse}, etc. Now, a single QD can be characterized as a TLS by restricting the occupancy of the QD to either zero or one, where the ground state $|g\rangle$ implies an empty QD and the only excited state $|e\rangle$ consists of one particle in the dot. Therefore, $|g\rangle$ and $|e\rangle$ can be rewritten in terms of number states, as $|0\rangle$ and $|1\rangle$, respectively. The system Hamiltonian for a single QD can be defined as
\begin{equation}\label{Hs}
    H_{\mathrm{s}}=\varepsilon \mathcal{N}=\varepsilon d^\dagger d,
\end{equation}
where $\varepsilon$ is the single-particle energy level of the QD and $\mathcal{N}=d^\dagger d$ signifies the number operator of the system, where $d^\dagger(d)$ implies the creation (annihilation) operator of the QD, obeying the anti-commutation relation $\{d,d^\dagger\}=1$. From Eq.~\eqref{Hs}, the energies of the eigenstates $|0\rangle$ and $|1\rangle$ can be evaluated as $0$ and $\varepsilon$, respectively. In a similar way, one can construct Hamiltonians for DQD, CQD, and SCQD systems as well. It is worth pointing out that at this point we are disregarding the spin of the particle to avoid additional complications. One pertinent difference between a TLS and a harmonic oscillator is that a TLS, owing to its bounded energy spectrum, may exhibit a negative absolute temperature due to population inversion~\cite{purcell1951a,ramsey1956thermodynamics,allen1987optical}. This is in sharp contrast to a harmonic oscillator, whose unbounded energy spectrum permits only positive temperatures. However, in this present review, we exclusively focus on positive-temperature and exclude the negative temperature regimes.  

\subsection{Environment}

For our system-bath model of interest, one or more fermionic reservoirs~\cite{ghosh2012canonical,ghosh2011dissipation,ghosh2012fermionic} are considered as the environment. These baths are essentially filled with fermionic particles (e.g., electrons) that follow FD statistics and are characterized by two parameters: temperature and chemical potential. Strictly speaking, the microscopic Hamiltonian of reservoir $\lambda$ is given by
\begin{equation}\label{bath-H}   \mathcal{H}_{\rm{B}}^{\lambda}=\sum_{k}\epsilon^{\lambda}_{k}c_{{\lambda} k}^\dagger c_{{\lambda} k},
\end{equation} 
where, $\epsilon^\lambda_k$ is the bare energy of the non-interacting fermions, $c^\dagger_{\lambda}(c_{\lambda})$ represents the creation (annihilation) operator~\cite{gupt2024graph,shuvadip2022univarsal} and $k$ is the continuous wavenumber. As the baths are also fermionic in nature, bath creation-annihilation operators follow the anti-commutation relation analogous to system operators. Since each reservoir is assumed to remain in grand-canonical equilibrium, its equilibrium density operator is given by
\begin{equation}\label{rho-bath}
\rho_{\lambda}^{\rm{eq}}=\frac{\exp[{-\beta_{\lambda} (\epsilon^{\lambda}_{k}-\mu_{\lambda})c_{{\lambda} k}^\dagger c_{{\lambda} k}}]}{\mathcal{Z_{\lambda}(\beta_{\lambda}})}. 
\end{equation}
Here, $N_{\lambda}=\sum_{k}c_{{\lambda} k}^\dagger c_{{\lambda} k}$ and the grand canonical partition function for the $\lambda$'th reservoir is ${\mathcal{Z_{\lambda}(\beta_{\lambda}})}=\Tr[e^{-\beta_{\lambda}H_{\rm{B}}^{\lambda}}]$, where we absorb the chemical potential into the single-particle energies following Refs.~\cite{bruus2004many,datta2005book}, and re-express the bath Hamiltonian in the following form:  
\begin{equation}\label{H-bath}  H_{\rm{B}}^{\lambda}=\sum_{k} (\epsilon^{\lambda}_{k}-\mu_{\lambda})c_{{\lambda} k}^\dagger c_{{\lambda} k}.
\end{equation} 
If there is more than one bath, the total bath-Hamiltonian can be simply written as $H_{\rm{B}}=\sum_{\lambda}H_{\rm{B}}^{\lambda}$.

\subsection{System-Bath Interaction}

The QD system is `truly' open to the baths, allowing both energy and particle exchange between them. The interaction is implemented through tunneling of the particle. Because this particle tunneling is associated with an energy flow, it generates both particle and energy currents. We are assuming that the system-bath interaction is weak enough to be within the Born-Markov approximation. This, in turn, ensures sequential tunneling~\cite{wang2022cycleflux,gupt2024graph}, which puts a restriction on simultaneous tunneling of more than one particle at a time. Finally, imposing the rotating wave approximation, the tunneling Hamiltonian between any $\alpha$-th QD and a $\lambda$-th bath can be written as
\begin{equation}\label{H-tunn}
H_{{\rm{T}}}^{\alpha\lambda}=\hbar\sum_k[t^{\alpha\lambda}_k c_{\lambda k}^\dagger d_{\alpha}+t^{\alpha\lambda*}_ kd_{\alpha}^{\dagger}c_{\lambda k}],    \end{equation}
where $t_{k}^{\alpha\lambda}$ represents the tunneling amplitude.

\section{Survey of various quantum transport Methods}\label{survey}

To place our discussion in a proper perspective, in this section we briefly outline the common theoretical approaches used to describe particle and charge transport within open quantum systems. This will allow readers to better appreciate how the models discussed in this review, as well as their methods of calculation, fit within the broader field of quantum coupled transport.

\subsection{Non-Equilibrium Green Function (NEGF)}

The NEGF method was pioneered by classic works of Martin, Schwinger~\cite{martin1959theory}, Kadanoff, Baym~\cite{kadanoff1989quantum}, Keldysh~\cite{keldysh1964diagram}, and others from the 1960s. It provides a powerful framework for describing quantum transport based on many-body perturbation theory (MBPT) and diagrammatic techniques~\cite{mahan1981many}. With the advent of mesoscopic physics in the 1980's, it was combined with the seminal work of Landauer~\cite{landauer1957spatial} and has since become an essential tool for nanoelectronics and device modeling~\cite{datta2005book}. Here, we briefly summarize the NEGF method following the formalism developed by Datta \textit{et al.}~\cite{Camsari2023,datta2005book}, which is equivalent to the original Keldysh formulation~\cite{keldysh1964diagram}.

\begin{figure}[t]
    \centering
\includegraphics[width=0.8\columnwidth]{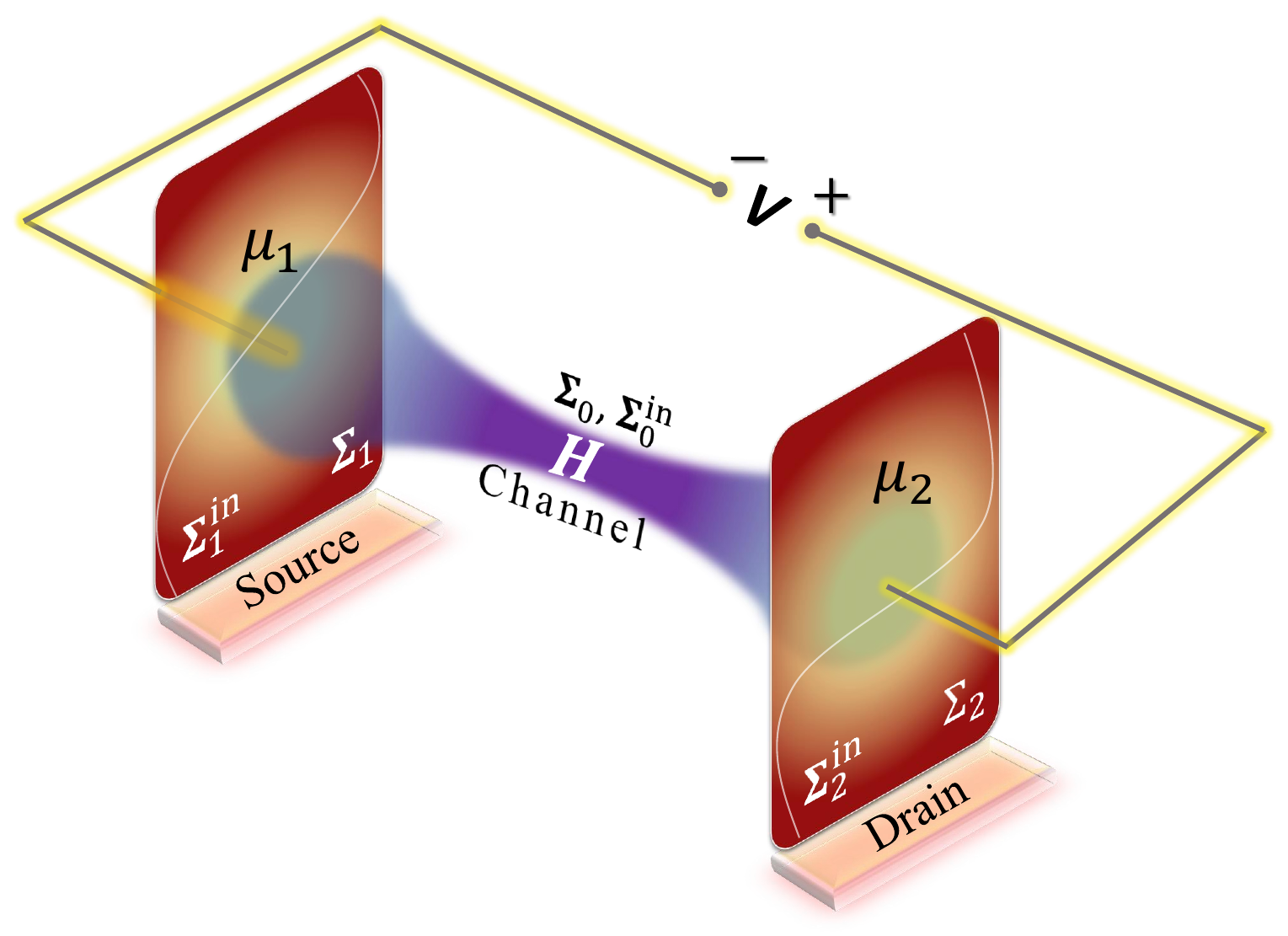}
    \caption{Model of a field-effect transistor comprising a channel with two contacts, labeled by source and drain.}
    \label{FET-model}
\end{figure}

\begin{figure}[b]
    \centering
\includegraphics[width=\columnwidth]{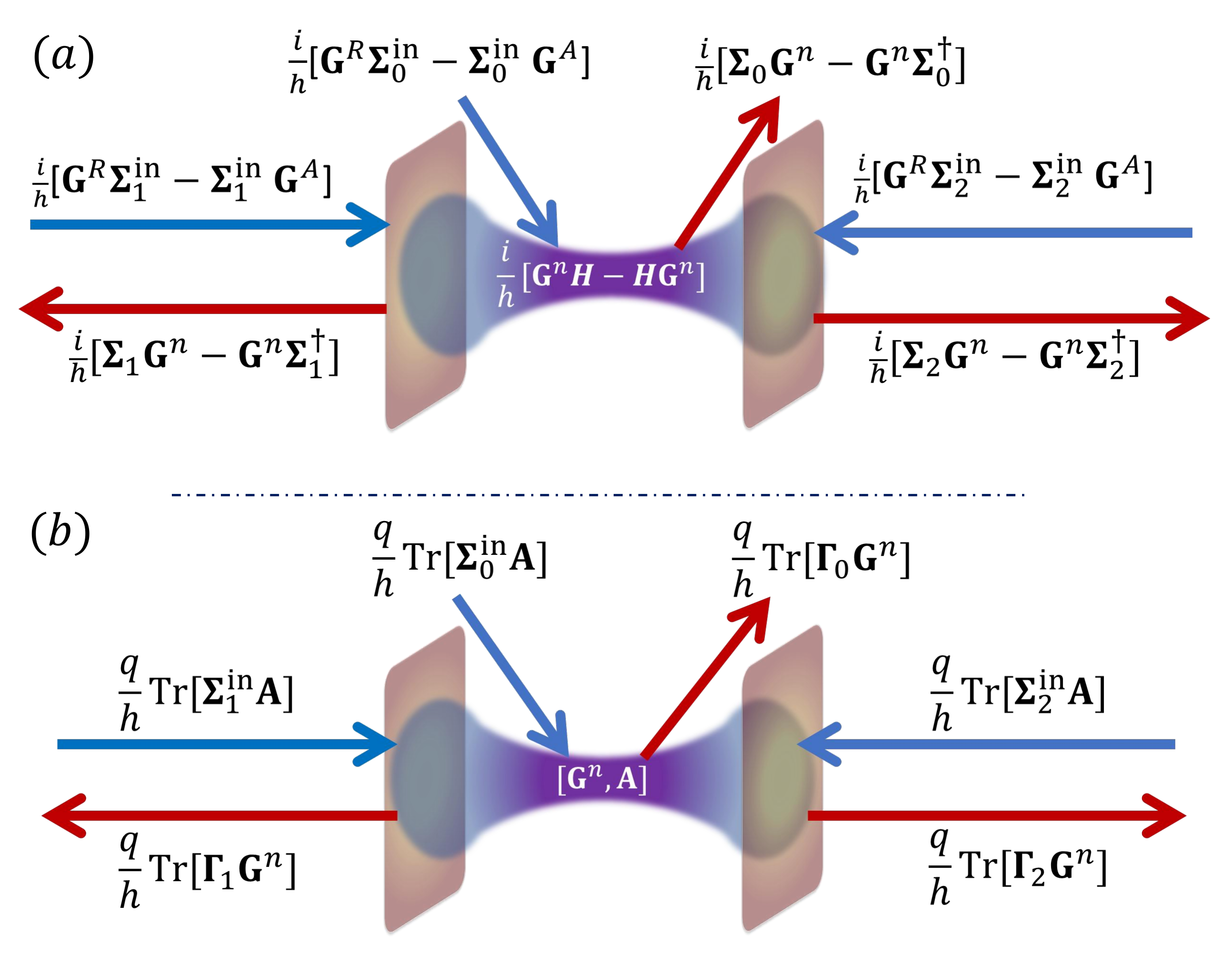}
    \caption{Inflow and outflow from the quantum channel described by the Hamiltonian matrix $[\mathbf{H}]$ and whose coupling to the contacts is described by self-energy functions $\{\pmb{\mathbf{\Sigma}}\}$.}
    \label{NEGF-schematic}
\end{figure}

To illustrate the basic idea, we consider the canonical setup of a field-effect transistor (FET)~\cite{Camsari2023}, where a channel is described by a Hamiltonian $\mathbf {H}$ matrix, while its ``contacts''
are described by self-energy functions $\pmb{\mathbf{\Sigma}}$ and $\pmb{\mathbf{\Sigma}}^{in}$ [Fig.~\ref{FET-model}]; subscripts 1 and 2 represent physical contacts, whereas 0 refers to a dephasing contact. As $\pmb{\mathbf{\Sigma}}=\pmb{\mathbf{\Sigma}}_1+ \pmb{\mathbf{\Sigma}}_2+\pmb{\mathbf{\Sigma}}_0$ is a non-Hermitian matrix, $[E\mathbf{I}-\mathbf{H}-\pmb{\mathbf{\Sigma}}]$ has a well-defined inverse for any real value of $E$. Hence, we can define the retarded and advanced Green functions
\begin{equation}
\mathbf G^R(E)=\left[E\mathbf I-\mathbf H-\pmb{\mathbf{\Sigma}}(E)\right]^{-1},
\qquad \mathbf{G}^A=(\mathbf G^R)^\dagger.
\end{equation}
One may further proceed to define the spectral function and electron density per unit energy as
\begin{equation}
\mathbf A=i(\mathbf G^R-\mathbf G^A);\qquad
\mathbf G^n=\mathbf G^R\pmb{\mathbf{\Sigma}}^{\rm in}\mathbf G^A,
\end{equation}
where $\pmb{\mathbf{\Sigma}}^{\rm in}=\pmb{\mathbf{\Sigma}}_1^{\rm in}+\pmb{\mathbf{\Sigma}}_2^{\rm in}+\pmb{\mathbf{\Sigma}}_0^{\rm in}$. Finally, the current operator $\mathbf{J}_{\text{op}}$ can be expressed as
\begin{eqnarray}
\mathbf{J}_{\text{op}} =-\frac{i}{h} \; \,( [\mathbf{H G}^{n} -\mathbf{G}^{n} \mathbf{H} ]\;+\;  [\mathbf{\Sigma G}^{n} -\mathbf{G}^{n} \pmb{\mathbf{\Sigma}}^{\dagger} ]\; \nonumber\\ +\; \, [\pmb{\mathbf{\Sigma}}^{\text{in}} \mathbf{G}^{A} -\mathbf{G}^{R} \pmb{\mathbf{\Sigma}}^{\text{in}} ]\; ),
\label{eq:currentOp}  \end{eqnarray}
such that a ``current'' of any observable $\mathbf{A}_{\text{op}}$ can be obtained by taking a trace $\text{Tr}(\mathbf{A}_{\text{op}}\mathbf{J}_{\text{op}})$. For charge-current, $\mathbf{A}_{\text{op}}=q$ is a scalar, hence it can be taken outside the trace, and the first term in the above equation (which represents the generation of the quantity $\mathbf{A}_{\text{op}}$ inside the channel [See~Fig.~\ref{NEGF-schematic}(a)]) becomes zero [Fig.~\ref{NEGF-schematic}(b)]. Therefore, the above expression reduces to 
\begin{equation}\label{eq:Jqr-NEGF}
J^r_q= \frac{q}{h} \int dE
\; \mathrm{Tr}\left[
\pmb{\mathbf{\Sigma}}_r^{\rm in}\mathbf A
-\mathbf\Gamma_r\mathbf G^n
\right],
\end{equation}
where $\mathbf\Gamma_r=i(\pmb{\mathbf{\Sigma}}_r-\pmb{\mathbf{\Sigma}}_r^\dagger)$ is the anti-Hermitian part of the self-energy for contact (lead) $r$, which describes the spectral broadening. For the dephasing contact, the condition $\mathrm{Tr}\left[
\pmb{\mathbf{\Sigma}}_0^{\rm in}\mathbf A
\right]=\mathrm{Tr}\left[\mathbf\Gamma_0\mathbf G^n
\right]$ is satisfied, yielding zero current at that contact. If the contacts are in thermal equilibrium, the inscattering functions are given by $\pmb{\mathbf{\Sigma}}^{\text{in}}_r = \mathbf{\Gamma}_rf_r(E)$, where $f_r(E)$ is the Fermi distribution of the reservoir $r$. Cotunneling, involving virtual intermediate states, can naturally be captured through Green's functions, which incorporate the contributions from all relevant states~\cite{datta2005book}. To incorporate dissipation due to inelastic processes, one must include $\left(\pmb{\mathbf{\Sigma}}_0\right)_{\rm diss}$ and $\left(\pmb{\mathbf{\Sigma}}_0^{\rm in}\right)_{\rm diss}$ into the self-energy. In essence, all we need to know is the self-energy $\pmb{\mathbf{\Sigma}}$. Then $\mathbf{\Gamma}$, $\pmb{\mathbf{\Sigma}}^{\text{in}}$ and $\mathbf G^{R}, \mathbf G^{A}$ can be calculated following the prescribed procedures~\cite{Camsari2023}. However, $\pmb{\mathbf{\Sigma}}$ must be approximated and can only be known perturbatively~\cite{majidi2024heat}.  

\subsubsection{Landauer--B\"uttiker formalism}

When the transport is dominated by coherent elastic scattering processes, instead of Eq.~\eqref{eq:Jqr-NEGF}, one can use the Landauer--B\"uttiker expression for the charge current~\cite{buttiker1986four}: 
\begin{equation}\label{LB-current}
J^r_q = \frac{q}{h}\int dE\;
\mathcal T_{lr}(E)[f_l(E)-f_r(E)].
\end{equation}
In going from Eq.~\eqref{eq:Jqr-NEGF} to Eq.~\eqref{LB-current}, we replace $\pmb{\mathbf{\Sigma}}^{\text{in}}_r$ by $\mathbf{\Gamma}_rf_r(E)$, where $\mathcal T_{lr}(E)$ is the \textit{transmission function} between the left and right reservoirs 
\begin{equation}\label{transmission-coeff}
\mathcal T_{lr}(E)=
\Tr[\mathbf{\Gamma}_l(E)\mathbf G^{R}(E)
\mathbf{\Gamma}_r(E)\mathbf G^{A}(E)].
\end{equation}
Equation~\eqref{LB-current} represents the difference between electrons transmitted in opposite directions. For coherent transport (i.e., when phase-breaking processes are negligible), the same transmission function governs both directions, while their occupation factors determine the net current. Equation~\eqref{LB-current} remains valid even for interacting systems as long as inelastic transport is negligible. The scattering theory and MBPT methods provide a general framework for heat, charge, and spin transport in quantum systems, which is capable of incorporating coherence, interactions, transient phenomena, and nonlinear responses~\cite{majidi2024heat}. Non-perturbative methods are required in the Coulomb-blockaded regime where single-electron charging energy $U_0 \gg \mathbf{\Gamma},k_BT$~\cite{datta2005book}.

\begin{table}
\centering 
\caption{Comparison of commonly used approaches for quantum thermodynamics and quantum transport.}
\label{tab:transport_methods}
\begin{tabular}{c c}
\hline \hline 
\textbf{Methods} &  \textbf{Descriptions} \\ 
\hline
\\
\textbf{Scattering theory} & Coherent transport, transmission\\
(Landauer--Büttiker) \quad  & coefficient, and mesoscopic transport\\
\\
\quad &  Versatile quantum-transport framework  \\
\textbf{NEGF (MBPT)} \quad & for coherence, transients dynamics, \\
\quad & cotunneling, and nonlinear response\\
\\
\textbf{QME (LME)} \quad & Weak-coupling, sequential tunneling, \\
\quad & transparent thermodynamic description\\ 
\\
\textbf{Quantum stochastic} & Higher-order cumulants,\\
\textbf{thermodynamics} & trajectory-level information,\\
\textbf{(FCS)} \quad &  and effective stochastic description\\
\hline
\end{tabular}
\end{table}

\subsection{Full Counting Statistics}

Quantum stochastic thermodynamics provides a systematic framework for obtaining the full counting statistics (FCS) of transport events with arbitrary moments and cumulants of currents~\cite{annby2020maxwell,strasberg2022quantum,das2026fluctuations}. The reduced density matrix evolves according to
$\partial_t\rho_t=\mathbb{L}\rho_t:=(\mathbb{L}_0+\mathbb{J}_++\mathbb{J}_-)\rho_t$,
where $\mathbb{L}_0$ describes the coherent system dynamics, and $\mathbb{J}_{\pm}$ denote inward and outward energy-transfer processes. Introducing the $n$-resolved density matrix $p(n,t)$ and the counting field $\chi$,
the master equation for the density matrix $\rho_t(\chi)=\sum_{n}p(n,t)e^{in\chi}$ can be rewritten as
\begin{equation}\label{ME-chi}
\partial_t\rho_t(\chi)=\mathbb{L}(\chi)\rho_t(\chi),
\end{equation}
with the formal solution $\rho_t(\chi)=e^{\mathbb{L}(\chi)t}\rho_0$. With the moment and cumulant generating functions defined as
$\mathcal{M}_t(\chi)=\Tr[\rho_t(\chi)]$, and
$\mathcal{K}_t(\chi)=\ln\mathcal{M}_t(\chi)$, $k$-th moment and cumulant can be obtained as
\begin{equation}
M_n^{(k)}
=\left.\frac{\partial^k\mathcal{M}_t}{\partial(i\chi)^k}\right|_{\chi=0},
\qquad
\kappa_n^{(k)}
=\left.\frac{\partial^k\mathcal{K}_t}{\partial(i\chi)^k}\right|_{\chi=0}.
\end{equation}
In the long-time limit, $\mathcal{K}_t(\chi)\simeq\lambda_0(\chi)t$, where $\lambda_0(\chi)$ is the largest possible eigenvalue of $\mathbb{L}(\chi)$, satisfying $\lambda_0(0)=0$, describes the steady state. Thus, the $k$-th steady-state particle-current cumulant is given by
\begin{equation}\label{k-Cumulant}
\left\langle\left\langle J_N^{(k)}\right\rangle\right\rangle
=\left.\frac{\partial^k\lambda_0(\chi)}
{\partial(i\chi)^k}\right|_{\chi=0},
\end{equation}
where, $\lambda_0(\chi)$ is obtained from the characteristic polynomial equation
$\det[\mathbb{L}(\chi)-\lambda\mathbb{I}]=0$~\cite{das2026fluctuations}.

\subsection{Quantum Master Equation} 

The quantum master equation (QME) provides an alternative framework for quantum transport, particularly in the weak system-reservoir coupling and sequential-tunneling regime~\cite{majidi2024heat}. The reduced dynamics is obtained by tracing out the reservoir degrees of freedom, with the tunneling Hamiltonian treated perturbatively~\cite{breuer2002book}. In the present context, Coulomb interactions between QDs are retained nonperturbatively in the system Hamiltonian, making the approach particularly suitable for Coulomb-coupled QD problems~\cite{shuvadip2022univarsal,gupt2024graph,ghosh2026inverse}.  Once the reduced density matrix is determined, observables can be calculated directly, allowing the study of both transient and steady-state properties in a straightforward manner. In the absence of coherence, the QME reduces to intuitive rate equations or the Pauli master equation that can capture the essential transport mechanisms~\cite{gupt2024graph,gupt2023topranked,ghosh2026inverse}. A key advantage of the completely positive and trace-preserving (CPTP) Lindblad QME employed throughout this review is its thermodynamic transparency~\cite{binder2019thermodynamics}. In this framework, reservoir-induced transitions can be directly associated with exchanges of energy and particles, allowing heat and particle currents, their conjugate thermodynamic forces, and the entropy-production rate to be defined and quantified unambiguously. The same framework naturally connects to the full counting statistics widely used in the quantum stochastic thermodynamics literature~\cite{strasberg2022quantum}. Table~\ref{tab:transport_methods} summarizes these transport methods and provides a brief description of each.

In the following we present the QME framework beginning with the conceptually simple QD–fermionic reservoir model, consisting of a single QD simultaneously tunnel-coupled to two reservoirs, permitting both energy and particle exchange. We then investigate whether this setup provides a suitable framework for studying coupled transport phenomena or not.

\section{Dynamics and steady-state current of a single QD setup}\label{sec4}

\begin{figure}[t]
    \centering
\includegraphics[width=0.86\columnwidth,height=4cm]{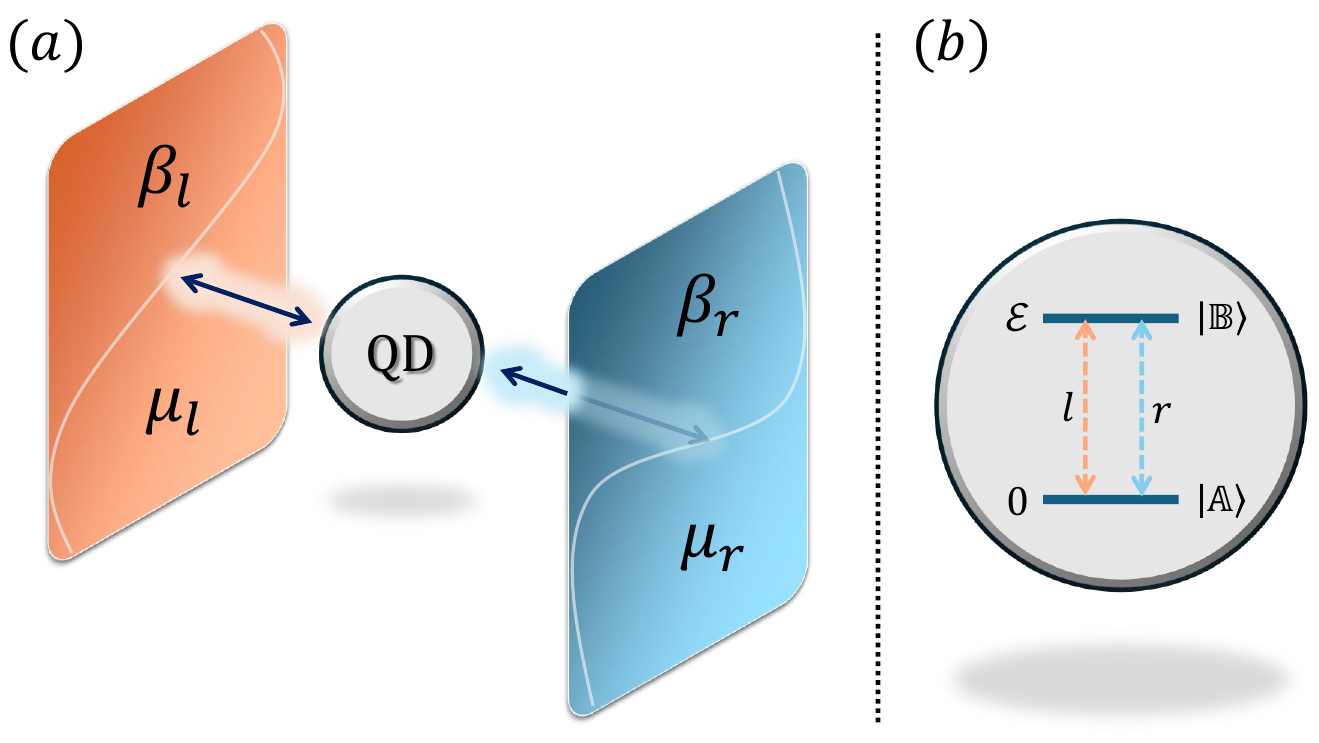}
    \caption{(a) Two-terminal single QD model for coupled transport (b) Two transitions are induced by the reservoirs.}
    \label{Single QD Model}
\end{figure}

The minimal model we consider is a two-terminal setup in which a single QD is tunnel-coupled to two fermionic reservoirs~\cite{pyurbeeva2026quantum}, denoted by $l$ (left) and $r$ (right), as depicted in Fig.~\ref{Single QD Model}(a). The corresponding system Hamiltonian is specified by Eq.~\eqref{Hs}, with two eigenstates, $|\mathbb{A}\rangle = |0\rangle $ and $|\mathbb{B}\rangle = |1\rangle$, with energies 0 and $\varepsilon$, respectively [Fig.~\ref{Single QD Model}(b)]. Consequently, the dynamics allow for only a single excitation process 
$|\mathbb{A}\rangle\rightarrow|\mathbb{B}\rangle$ that occurs through particle exchange from bath to dot and the corresponding de-excitation process 
$|\mathbb{B}\rangle\rightarrow|\mathbb{A}\rangle$, i.e., particle hopping from dot to lead, forming a closed transition cycle $|\mathbb{A}\rangle\rightarrow|\mathbb{B}\rangle\rightarrow|\mathbb{A}\rangle$. Because the QD is simultaneously coupled to both reservoirs, each of these transitions can be mediated by either lead, allowing both energy and particle exchange [Fig.~\ref{Single QD Model}(b)]. Since particle tunneling between a lead and the QD is accompanied by an exchange of energy, the energy transfer is implicitly linked to particle transfer within this model. As a result, the model is suitable for investigating coupled transport between energy and particles. It is worth emphasizing that the simultaneous coupling of the QD to multiple leads is not a requirement for energy transport; however, it is essential to enable net particle transport, a distinction that becomes apparent in the DQD setup [See Sec.~\ref{sec9}], where the particle current arises due to inter-dot particle hopping~\cite{strasberg2022quantum} and not because of simultaneous coupling of one QD to multiple baths.

Because of the simplicity of this model, its qualitative dynamics can be easily understood. If both excitation and de-excitation are mediated by the same reservoir, there is no net particle transfer, and both the particle and energy currents vanish. To generate nonzero particle and energy currents through the QD, the transition cycle must involve different reservoirs for excitation and de-excitation. This condition establishes the basis for studying genuine coupled transport.

To proceed with a quantitative analysis, we must solve the dynamics of the system. Since our system–bath model follows Markovian dynamics, we employ the Lindblad master equation (LME) formalism~\cite{breuer2002book}. Under Born, Markov, and secular (BMS) approximations~\cite{breuer2002book,strasberg2022quantum}, the LME takes the form [See Appendix-~\ref{Appendix-A}]
\begin{equation}\label{LME}
\frac{d}{dt}\rho_{\rm{s}}(t)=\sum_{\lambda}\mathcal{L}_{\lambda}[\rho_{\rm{s}}(t)]\quad ; \quad \lambda=l,r.
\end{equation}
Here, the reduced state of the composite system is represented by the density matrix  $\rho_{\rm{s}}(t)=\Tr_{\rm{B}}\{\rho_{\rm{tot}}(t)\}$, where $\rho_{\rm{tot}}(t)$ is the full density matrix of the system and bath combined, and the form of the Lindblad super-operator $\mathcal{L}_\lambda[\rho_{\rm{s}}(t)]$ is given in the Appendix-~\ref{Appendix-A}. The general expressions for the steady state energy and particle currents for the $\lambda$-th reservoir are given by [See Appendix-~\ref{Appendix-B}]
\begin{equation}\label{S9}
\begin{split}
J^{\lambda}_{\rm{E}}=\Tr_{\rm{s}}[\mathcal{L}_{\lambda}[\rho_{ss}]H_{\rm{s}}];\qquad
J^{\lambda}_{\rm{N}}=&\Tr_{\rm{s}}[\mathcal{L}_{\lambda}[\rho_{ss}]\mathcal{N}],
\end{split}
\end{equation}
where $\rho_{ss}$ signifies the steady-state density matrix. The expression for the steady-state heat current thus simplifies to
\begin{equation}
J^{\lambda}_{\rm{Q}}=J^{\lambda}_{\rm{E}}-\mu_{\lambda}J^{\lambda}_{\rm{N}} =\Tr_{\rm{s}}[\mathcal{L}_{\lambda}[\rho_{ss}]H_{\rm{s}}] - \mu_{\lambda}\Tr_{\rm{s}}[\mathcal{L}_{\lambda}[\rho_{ss}]\mathcal{N}].   
\end{equation}
With the explicit form of the Lindbladians, $\mathcal{L}_{\lambda}[\rho_{ss}]$, defined in Appendix-A [cf.~\eqref{A7}], the general expressions for the energy and particle currents lead to 
\begin{equation}\label{B10}
\begin{split}
J^{\lambda}_{\rm{E}}&=\sum_{\{\omega_{\mathbb{ij}}\}}{\omega_{\mathbb{ij}}}\Gamma_{\mathbb{ij}}^{\lambda+}=\sum_{\{\omega_{\mathbb{ji}}\}}{\omega_{\mathbb{ji}}}\Gamma_{\mathbb{ji}}^{\lambda-};\\
J^{\lambda}_{\rm{N}}&=\sum_{\{\omega_{\mathbb{ij}}\}}\Gamma_{\mathbb{ij}}^{\lambda+}=\sum_{\{\omega_{\mathbb{ji}}\}}\Gamma_{\mathbb{ji}}^{\lambda-},
\end{split}
\end{equation}
where the net transition rate from level $|\mathbb{i}\rangle$ to $|\mathbb{j}\rangle$ driven by tunnelling of the particle from (to) the reservoir $\lambda$ to (from) the $\rm{QD}$ is represented by $\Gamma^{\lambda\pm}_\mathbb{ij}$, which is given by
\begin{equation}\begin{split}\label{C1.3}
\Gamma_\mathbb{ij}^{\lambda\pm}\equiv\Gamma_\mathbb{i\rightarrow j}^{\lambda\pm}=\gamma_{\lambda}f^{\pm}_{\lambda}(\omega_\mathbb{ij})\rho_\mathbb{i}-\gamma_{\lambda}f^{\mp}_{\lambda}(\omega_\mathbb{ji})\rho_\mathbb{j}
=\Gamma_\mathbb{i\rightarrow j}^{\lambda\pm}-\Gamma_\mathbb{j\rightarrow i}^{\lambda\mp}.
\end{split}
\end{equation}

The first term denotes the transition from the eigenstate $|\mathbb{i}\rangle$ to $|\mathbb{j}\rangle$ driven by particle excitation (or de-excitation), while the second term represents the reverse transition $|\mathbb{j}\rangle \rightarrow |\mathbb{i}\rangle$, driven by particle de-excitation (or excitation) w.r.t. the system. From Eq.~\eqref{C1.3}, it follows $\Gamma_\mathbb{ij}^{\lambda+}=-\Gamma_\mathbb{ji}^{\lambda-}$, where $f^{\pm}_{\lambda}(\omega_{\mathbb{ij}})$ denotes the Fermi–Dirac distribution (FDF) function. It is associated with the transition $|\mathbb{i}\rangle \rightarrow |\mathbb{j}\rangle$ induced by the reservoir $\lambda$, with the transition energy $\omega_{\mathbb{ij}}$. Throughout this article~\cite{sanchez2011optimal,zhang2021inverse,ghosh2026inverse}, the notation “$\pm$” is used where the “$+$” sign corresponds to particle excitation (i.e., tunnelling into the system), whereas the “$-$” sign refers to particle de-excitation (i.e., tunnelling out of the system). If the transition $|\mathbb{i}\rangle \rightarrow |\mathbb{j}\rangle$ is governed by particle excitation w.r.t. the system, the reverse transition corresponds to particle tunnelling from the system into the lead. This observation allows us to establish a relation between the FDFs for forward and reverse processes~\cite{shuvadip2022univarsal,tesser2022heat}:

For the single–QD system connected to two terminals, the dot possesses only two eigenstates, 
$|\mathbb{A}\rangle$ and $|\mathbb{B}\rangle$, and is tunnel-coupled to the left $l$ and right $r$ leads. Under these conditions, the general form of the energy current [cf.~\eqref{B10}] simplifies to
\begin{equation}
J^{l}_{\rm{E}}={\omega_{\mathbb{AB}}}\Gamma_{\mathbb{AB}}^{l+}=\varepsilon\Gamma_{\mathbb{AB}}^{l+}\quad;\quad J^{r}_{\rm{E}}={\omega_{\mathbb{AB}}}\Gamma_{\mathbb{AB}}^{r+}=\varepsilon\Gamma_{\mathbb{AB}}^{r+}.  
\end{equation}
The non-zero energy and particle transport between the reservoirs occurs exclusively through two transition paths --- excitation by lead $l$ and de-excitation by lead $r$, or vise versa. Thus, in the steady state, the associated transition rates obey
\begin{equation}
\Gamma_{\mathbb{AB}}^{l+}=\Gamma_{\mathbb{BA}}^{r-}\equiv\Gamma_{+}  \quad;\quad \Gamma_{\mathbb{AB}}^{r+}=\Gamma_{\mathbb{BA}}^{l-}\equiv\Gamma_{-},
\end{equation}
implying $\Gamma_{+}=-\Gamma_{-}$. Finally, we obtain the energy current as
\begin{equation}\label{JE}
J^{l}_{\rm{E}}=-J^{r}_{\rm{E}}=\varepsilon \Gamma_{+}  
\end{equation}
Similarly, the particle current defined in Eq.~\eqref{B10} yields
\begin{equation}\label{JN}
J^{l}_{\rm{N}}=-J^{r}_{\rm{N}}=\Gamma_{+}  
\end{equation}
From Eqs.~\eqref{JE}~and~\eqref{JN}, it is evident that the energy and particle currents are \textit{linearly} dependent via a relation $J_{\mathrm{E}}^{l(r)}=\varepsilon J_{\mathrm{N}}^{l(r)}$. Having derived analytical expressions for the currents in this minimal QD–reservoir model, we determine the corresponding thermodynamic forces by analyzing the entropy production rate. We first carry out this analysis in a general open quantum system framework with multiple reservoirs and then specialize the result to the present two-terminal single–QD model.

\section{Deriving entropy production rate for general QD-fermionic reservoir model}\label{sec5}

In this section, we derive the fundamental thermodynamic relations for the entropy production rate. We consider a general setup, where a QD system (it may be a CQD or DQD as well) is coupled with multiple baths through particle and energy exchange.
Starting with the definition of the entropy change of the system, one can write down the following relation~\cite{landi2021irreversible}
\begin{equation}\label{EP2}
\varDelta\mathcal{S}_{\rm{s}}(t)=\Sigma(t)+\Phi(t),
\end{equation}
where the \textit{entropy production} term $\Sigma(t)$ signifies the irreversible contribution of the system entropy change and the $\Phi(t)$ term refers to the \textit{entropy flux}. At close to equilibrium, we equate the von Neumann entropy (times $k_B$) with thermodynamic entropy as
\begin{equation}\label{S1}
\begin{split}
 \mathcal{S}_{\rm{s}}(t)=-k_B \Tr_{\rm{s}}[\rho_{\rm{s}}(t)\ln{\rho_{\rm{s}}(t)}].
\end{split}   
\end{equation}
From Eq.~\eqref{EP2}, we can evaluate
\begin{equation}\label{C1}
\begin{split}
 \varDelta\mathcal{S}_{\rm{s}}(t)=&\mathcal{S}_{\rm{s}}(t)-\mathcal{S}_{\rm{s}}(0)\\
 =&-k_B \Tr[\rho_{\rm{tot}}(t)\ln{\rho_{\rm{s}}(t)}]+k_B \Tr[\rho_{\rm{tot}}(0)\ln{\rho_{\rm{s}}(0)}].
\end{split}
\end{equation}
In the absence of initial system--bath correlations or entanglement, the total density operator is assumed to be factorized as
\begin{equation}\label{C2}
\rho_{\rm{tot}}(0)=\rho_{\rm{s}}(0)\prod_{\lambda}\rho_{\lambda}^{\rm{eq}}.
\end{equation}
This factorized initial state is a standard assumption in the derivation of Markovian QME. It is physically well justified because the reservoirs are macroscopically large compared to the system size. Moreover, the system-bath coupling is weak, and the bath relaxation time is much faster than the system relaxation time, consistent with the Born--Markov approximation~\cite{thierschmann2015three,sanchez2011optimal,gupt2024graph,ghosh2026inverse}. Consequently, this assumption is not only mathematically convenient but also physically and experimentally well motivated (See also Section.~\ref{survey}C). Substituting Eq.~\eqref{C2} in Eq.~\eqref{C1},
the final expression of $\varDelta\mathcal{S}_{\rm{s}}(t)$ can be evaluated as
\begin{equation}\label{C5}
\begin{split}
\varDelta\mathcal{S}_{\rm{s}}(t)=&-k_B \Tr[\rho_{\rm{tot}}(t)\ln\left\{{\rho_{\rm{s}}(t)}\prod_{\lambda}\rho_{\lambda}^{\rm{eq}}\right\}]\\
    &+k_B \Tr[\rho_{\rm{tot}}(t)\ln{\rho_{\rm{tot}}(t)}]\\
    &+k_B \sum_{\lambda}\Tr_{\lambda}[\{\rho_{\lambda}(t)-\rho_{\lambda}^{\rm{eq}}\}\ln{\rho_{\lambda}^{\rm{eq}}}].    
\end{split}
\end{equation}
In deriving the above equation, we have used the relation $\rho_{\rm{tot}}(t)=\mathcal{U}\rho_{\rm{tot}}(0)\mathcal{U}^\dagger$, i.e., the states $\rho_{\rm{tot}}(t)$ and $\rho_{\rm{tot}}(0)$ are unitarily related to each other. 
The last term in Eq.~\eqref{C5} can be identified as the \textit{entropy flow} ($\Phi$), representing the reversible contribution to the system entropy change due to heat exchanges. A comparison of the above equation with Eq.~\eqref{EP2}, help us to define the \textit{entropy production} and the \textit{entropy flow} as,
\begin{equation}\label{C6}
\begin{split}
\Sigma(t)=&\Tr[\rho_{\rm{tot}}(t)\ln\{\rho_{\rm{tot}}(t)\}]\\
&-\Tr[\rho_{\rm{tot}}(t)\ln\left\{\rho_{\rm{s}}(t)\prod_{\lambda}\rho_{\lambda}^{\rm{eq}}\right\}],\\
\Phi(t)=&\sum_{\lambda}\Tr_{\rm{\lambda}}\left[\left\{\rho_{\lambda}(t)-\rho_{\lambda}^{\rm{eq}}\right\}\ln{\rho_{\lambda}^{\rm{eq}}}\right],
\end{split}
\end{equation}
where, $\rho_{\lambda}^{\rm{eq}}$ and $\rho_{\lambda}(t)$ represent the density operators of the $\lambda$'th reservoir under equilibrium and \textit{near equilibrium} conditions, respectively. Moreover, $\rho_{\lambda}^{\rm{eq}}$ can be written as
\begin{equation}\label{EP4}
\rho_{\lambda}^{\rm{eq}}=\frac{e^{-\beta_{\lambda}H_{\rm{B}}^{\lambda}}}{\mathcal{Z_{\lambda}(\beta_{\lambda}})}, 
\end{equation}
where ${\mathcal{Z_{\lambda}(\beta_{\lambda}})}=\Tr[e^{-\beta_{\lambda}H_{\rm{B}}^{\lambda}}]$ is the the grand canonical partition function for the $\lambda$'th reservoir. As a result, $\Phi(t)$ is identified as the heat exchange with the reservoir 
\begin{equation}\label{EP5}
\Phi(t)
=k_B\sum_{\lambda}\beta_{\lambda}\varDelta Q_{\lambda}(t),
\end{equation}
between the final and initial time: $\varDelta Q_{\lambda}(t)={\langle H_{\rm{B}}^{\lambda}\rangle}_{0}-{\langle H_{\rm{B}}^{\lambda}\rangle}_{t}$, with ${\langle H_{\rm{B}}^{\lambda}\rangle}_{t}=\Tr_{\rm{\lambda}}[\rho_{\lambda}(t)H_{\rm{B}}^{\lambda}]$. Thus, we can rewrite Eq.~\eqref{EP2} as
\begin{equation}\label{EP6}
\varDelta\mathcal{S}_{\rm{s}}(t)=\Sigma(t)+k_B\sum_{\lambda}\beta_{\lambda}\varDelta Q_{\lambda}(t),
\end{equation}
where we replace the second term by Eq.~\eqref{EP5}. The entropy production rate can then be deduced from the above equation as follows,
\begin{equation}\label{EP7}
\begin{split}
\dot{\Sigma}(t)=\frac{d}{dt}\left\{\varDelta\mathcal{S}_{\rm{s}}(t)\right\}-k_B\sum_{\lambda}\beta_{\lambda}J_{\rm{Q}}^\lambda(t).
\end{split}
\end{equation}
where, we define $J_{\rm{Q}}^\lambda(t)=\frac{d}{dt}\left\{\varDelta Q_{\lambda}(t)\right\}$. At the \textit{steady state}, $\dot{\Sigma}$ becomes minimum~\cite{esposito2010entropy,landi2021irreversible,strasberg2022quantum} and $\frac{d}{dt}\left\{\varDelta\mathcal{S}_{\rm{s}}(t)\right\}=0$. Therefore, Eq.~\eqref{EP7} yields
\begin{equation}\label{EP8}
\dot{\Sigma}=-k_B\sum_{\lambda}\beta_{\lambda}J_{\rm{Q}}^\lambda.
\end{equation}
Again, if steady-state heat currents are defined in terms of energy and particle currents as $J_{\rm{Q}}^\lambda=J_{\rm{E}}^\lambda-\mu_{\lambda}J_{\rm{N}}^\lambda$, the above equation can be expressed as 
\begin{equation}\label{EP9}
\dot{\Sigma}=-k_B\sum_{\lambda}\beta_{\lambda}J_{\rm{E}}^\lambda+k_B\sum_{\lambda}\bigg(\frac{\beta_{\lambda}}{\mu_{\lambda}}\bigg)J_{\rm{N}}^\lambda.
\end{equation}
To analyze the preceding expression, we require explicit forms of energy and particle currents, namely~$J_{\mathrm{E}}^{\lambda}$~and~$J_{\mathrm{N}}^{\lambda}$. The corresponding currents have already been evaluated [Cf.~Eqs.\eqref{JE} and~\eqref{JN}] for a minimal single--QD two-terminal configuration. Therefore, examining the resulting $\dot{\Sigma}$ allows us to identify the thermodynamic forces corresponding to energy and particle transport in this setup. This, in turn, enables a formulation of coupled transport within the force--flux framework. Accordingly, in the next section, we derive the explicit force--flux pairs for our model and proceed to the central goal of this work: a detailed thermodynamic analysis of coupled transport phenomena.

\section{Identifying the thermodynamic force-flux pairs for two-terminal single QD setup}\label{sec6}

Based on the general expression of the entropy production rate defined in Eq.~\eqref{EP9}, we identify the force-flux pairs within the two-terminal single QD model as 
\begin{equation}\label{R2}
\dot{\Sigma}=-k_B[\beta_{l}J_{\rm{E}}^{l}+\beta_{r}J_{\rm{E}}^{r}]+k_B[\beta_{l}\mu_{l}J_{\rm{N}}^{l}+\beta_{r}\mu_{r}J_{\rm{N}}^{r}] .
\end{equation}
As there is no external source or sink, the net energy and particle currents are conserved in the steady state, i.e., $\sum_{\lambda}J_{\rm{E}}^\lambda=\sum_{\lambda}J_{\rm{N}}^\lambda=0$. This leads to four equivalent expressions for the entropy production rate:
\begin{equation}
	\dot{\Sigma} =
	\left\{
	\begin{aligned}
		J_{\mathrm{E}}^{l}\mathcal{F}_{\mathrm{E}}^{l} + J_N^{l}\mathcal{F}_{\mathrm{N}}^{l} \\
		J_{\mathrm{E}}^{l}\mathcal{F}_{\mathrm{E}}^{l} + J_N^{r}\mathcal{F}_{\mathrm{N}}^{r} \\
		J_{\mathrm{E}}^{r}\mathcal{F}_{\mathrm{E}}^{r} + J_{\mathrm{N}}^{l}\mathcal{F}_{\mathrm{N}}^{l} \\
		J_{\mathrm{E}}^{r}\mathcal{F}_{\mathrm{E}}^{r} + J_{\mathrm{N}}^{r}\mathcal{F}_{\mathrm{N}}^{r}
	\end{aligned}
	\right.
\end{equation}
where the thermodynamic forces are defined as:
\begin{equation}\label{R5}
\begin{split}
    \mathcal{F}_{\rm{E}}^{l}&=-\mathcal{F}_{\rm{E}}^{r}=k_B(\beta_r-\beta_{l}),\\
\mathcal{F}_{\rm{N}}^{l}&=-\mathcal{F}_{\rm{N}}^{r}=k_B(\beta_l\mu_l-\beta_r\mu_r).
\end{split}
\end{equation}
Once more, let us reiterate the convention used for forces and fluxes: When a current ($J_{\rm{E}}^\lambda, J_{\rm{N}}^\lambda$ or $J_{\rm{Q}}^\lambda$) enters the system from reservoir $\lambda$, the sign of the current is considered positive, i.e., $J^{\lambda}>0$. In the case of the forces, it follows from Eq.~\eqref{R5} that a positive energy force $\mathcal{F}_{\rm E}^{l} > 0$ implies $\beta_l < \beta_r$, whereas $\mathcal{F}_{\rm E}^{r} > 0$ implies $\beta_r < \beta_l$. Although these two forces describe the same physical situation, they are related by $\mathcal{F}_{\rm E}^{l} = -\mathcal{F}_{\rm E}^{r}$. Therefore, to determine the direction of the forces solely from their signs, one must adopt a form of the entropy production rate that uses consistent labels (either $l$ or $r$) for both energy and particle fluxes and forces. This leaves us with two equivalent choices:
\begin{equation}\label{R6}
\begin{split}
\dot{\Sigma}=J_{\rm{E}}^{l}\mathcal{F}_{\rm{E}}^{l}+J_{\rm{N}}^{l}\mathcal{F}_{\rm{N}}^{l}=J_{\rm{E}}^{r}\mathcal{F}_{\rm{E}}^{r}+J_{\rm{N}}^{r}\mathcal{F}_{\rm{N}}^{r}.
\end{split}
\end{equation}
Without loss of any generality (w.l.o.g), we choose $\dot{\Sigma}=J_{\rm{E}}^{l}\mathcal{F}_{\rm{E}}^{l}+J_{\rm{N}}^{l}\mathcal{F}_{\rm{N}}^{l}$.

To proceed, we clarify a subtle but important distinction between gradients and thermodynamic forces. Although related, they are \textit{distinct thermodynamic quantities}. For energy transport, the thermodynamic force coincides with the temperature gradient in the linear-response (near-equilibrium) regime with $\mathcal{F}_{\rm E}^l = k_B(\beta_r - \beta_l) = \frac{\Delta T}{T_l T_r}$. In contrast, the particle force $\mathcal{F}_{\rm N}^l = k_B(\beta_l \mu_l - \beta_r \mu_r)$ reduces to the chemical potential gradient only when $T_l = T_r = T$. For $\Delta T \neq 0$, we have $\mathcal{F}_{\rm N}^l = \frac{\Delta \mu}{T_l} - \mu_r \mathcal{F}_{\rm E}^l.$ Thus, $\mathcal{F}_{\rm N}^l$ approximates $\Delta \mu$ only for small $\Delta T$. Its sign can even differ from that of $\Delta \mu$ when $\frac{\mu_r}{T_r} |\Delta T|  > |\Delta \mu|$. This raises a natural question: why are the chemical potential gradient and the particle force so often used interchangeably~\cite{callen1985book,kondepudi2015book,sanchez2011optimal,walldorf2017thermoelectrics,pyurbeeva2026quantum} in the literature?  The reason is practical. In macroscopic and most mesoscopic systems, this condition is rarely met, so the two are used interchangeably. However, on the nanoscale, with $\mu_r$ and $\Delta \mu$~$\sim10-80$ $\mu$eV and $T_r$ and $\Delta T$~$\sim10-100$ mK, the above inequality can easily be satisfied, which is precisely the regime relevant for the quantum transport models considered here. Thus, interpreting transport in this regime solely based on $\Delta \mu$ can be misleading: a particle current may appear to flow against the gradient while actually flowing along its true thermodynamic force. Since entropy production is defined in terms of thermodynamic forces rather than gradients, transport --- especially at the nanoscale --- must be characterized using the correct thermodynamic forces.

\section{Thermodynamics of coupled transport in two terminal single QD setup}\label{sec7}

We now explore the thermodynamics of coupled transport in a two-terminal single–QD model, involving one energy and one particle force–flux pair. For a systematic analysis, we examine the behavior case by case.

\subsection{Both $\mathcal{F}_{\rm E}^l$ and $\mathcal{F}_{\rm N}^l$ are zero}

As expected, when both forces are zero, no driving mechanism exists in the system, and consequently both the energy and the particle currents are zero, implying a state of thermal equilibrium, with $\dot{\Sigma}=0$. According to Eq.~\eqref{R5}, $\mathcal{F}_{\rm E}^l=\mathcal{F}_{\rm N}^l=0$ corresponds to both leads being at the same temperature and chemical potential. Microscopically, under these conditions, the transition cycle $|\mathbb{A}\rangle\rightarrow|\mathbb{B}\rangle\rightarrow|\mathbb{A}\rangle$ is equally likely to be initiated by the left or right reservoir. A cycle in which excitation is governed by the left lead and de-excitation by the right lead is equally likely to its reverse process. There is no net particle transfer between the system and either reservoir, and the average particle current vanishes. Consequently, the energy current also vanishes, as the two currents are related through the proportionality factor $\varepsilon$. In summary, when both forces are zero, we obtain the trivial result, marked as the origin in the 2D diagram in Fig.~\ref{Chart}.

\subsection{Seebeck effect in presence of only energy force}

Now, in the presence of only the energy force $\mathcal{F}_{\rm{E}}^{l}$ (i.e., on the x-axis of Fig.~\ref{Chart}), the energy current $J_{\rm{E}}^{l}$ must be in the same direction to make the entropy production rate non-negative. Without loss of generality, if we consider 
$\mathcal{F}_{\rm{E}}^{l}>0$ by choosing 
$\beta_r>\beta_l$ [cf.~\eqref{R5}], $J_{\rm{E}}^{l}$ must also be positive. This is the expected thermodynamic behavior under an applied temperature gradient.
However, there is no thermodynamic restriction on the direction of $J_{\rm{N}}^{l}$, since $\mathcal{F}_{\rm N}^l=0$. Hence, $J_{\rm{N}}^{l}$ may assume either sign, implying that it can flow from left to right or from right to left. Thus, the model supports a nonzero particle current even in the absence of a particle force. In this case, the particle current is driven by the nonzero energy force. This intriguing phenomenon is known as a thermodynamic cross effect, wherein a nonzero thermodynamic current is driven by a non-conjugate force, in the absence of its conjugate force. 

Within the linear response regime, this situation can be explained via Onsager coefficients [cf.~Eq.~\eqref{Onsager}]. Since $J_{\rm{N}}^{l}=L_{\mathrm{NE}}\mathcal{F}_{\rm{E}}^{l}+L_{\mathrm{NN}}\mathcal{F}_{\rm{N}}^{l}$ and $\mathcal{F}_{\rm{N}}^{l}=0$, it reduces to $J_{\rm{N}}^{l}=L_{\mathrm{NE}}\mathcal{F}_{\rm{E}}^{l}$. This situation corresponds to the well-known Seebeck effect, wherein a temperature gradient induces a particle (or charge) current. As the Onsager cross coefficient can assume positive or negative values, depending on the sign of the cross coefficient $L_{\mathrm{NE}}$, two distinct regimes can be identified [Fig.~\ref{Chart}]. For 
$L_{\mathrm{NE}}>0$,
$J_{\rm{N}}^{l}$ flows along the direction of $\mathcal{F}_{\rm{E}}^{l}$. This is the conventional Seebeck effect~\cite{callen1985book,kondepudi2015book,Seebeck1822magnetische,Kelvin1856on,degroot1984non,esposito2009thermoelectric,sanchez2011optimal,whitney2014most,gupt2024graph}. In contrast, for 
$L_{\mathrm{NE}}<0$, the particle current flows opposite to the applied energy force, a situation that can be identified as the unconventional Seebeck effect.

The Seebeck effect forms the basis for the fundamental thermodynamic principle underlying a thermoelectric heat engine. In the present two-terminal single–QD model, however, the direction of the particle current $J_{\rm N}^l$ is constrained by the relation $J_{\rm E}^l = \varepsilon J_{\rm N}^l$. Since $\varepsilon > 0$, both the energy and particle currents are necessarily positive and aligned in the same direction. Thus, we obtain $J_{\rm N}^l>0$ even when $\mathcal{F}_{\rm N}^l=0$. The model therefore supports the normal Seebeck effect. An unconventional Seebeck effect cannot occur within this model, since the particle and energy currents are constrained to flow in the same direction.

\subsection{Peltier effect in presence of only particle force}

We now consider the case where $\mathcal{F}_{\rm E}^l = 0$ but $\mathcal{F}_{\rm N}^l$ is nonzero, i.e, on the y-axis of Fig.~\ref{Chart}. To analyze this case, we take $\mathcal{F}_{\rm N}^l > 0$, which corresponds to $\beta_l \mu_l > \beta_r \mu_r$ [cf. Eq.~\eqref{R5}]. As anticipated, the resulting particle current $J_{\rm N}^l$ is positive, indicating particle flow from the left to the right reservoir. This follows directly from Eq.~\eqref{Onsager}: $J_{\rm N}^l= L_{\rm NN}\mathcal{F}_{\rm N}^l$. With $\mathcal{F}_{\rm N}^l, \; L_{\rm NN} > 0$, the, $J_{\rm N}^l$ shares the sign of its conjugate force, $\mathcal{F}_{\rm N}^l$.

In contrast, the energy current $J_{\rm E}^l=L_{\rm EN}\mathcal{F}_{\rm N}^l$ is governed by the Onsager cross coefficient $L_{\rm EN}$, which may be positive or negative [Fig.~\ref{Chart}]. Thus, even for $\mathcal{F}_{\rm N}^l > 0$, the energy current $J_{\rm E}^l$ can flow in either direction [cf. Eq.~\eqref{Onsager}]. If $L_{\rm EN} > 0$, then $J_{\rm E}^l > 0$, corresponding to energy flow from left to right --- the conventional Peltier effect~\cite{peltier1834nou,callen1985book,degroot1984non,kondepudi2015book,jordan2013powerful,sothmann2012quantum,whitney2014most,gupt2024graph,whitney2018quantum}. If $L_{\rm EN} < 0$, the energy current is driven opposite to the particle force ($J_{\rm E}^l < 0$), corresponding to a unconventional Peltier effect. 

The Peltier effect constitutes the thermodynamic principle that underlies the operation of a refrigerator. However, in the present two-terminal single–QD model, the currents are constrained by $J_{\rm E}^l = \varepsilon J_{\rm N}^l$. As a result, the energy and particle currents are necessarily aligned. Therefore, only a positive Peltier effect can occur; an unconventional Peltier effect is excluded by this proportionality constraint. In the following, we consider the regime in which both thermodynamic forces are nonzero and investigate whether this setup can function as a thermodynamic heat engine or a refrigerator, operating on the basis of the Seebeck and Peltier effects.

\subsection{In the presence of both energy and particle force}

\begin{figure}[h!]
    \centering   
    \includegraphics[width=\columnwidth,height=9cm]{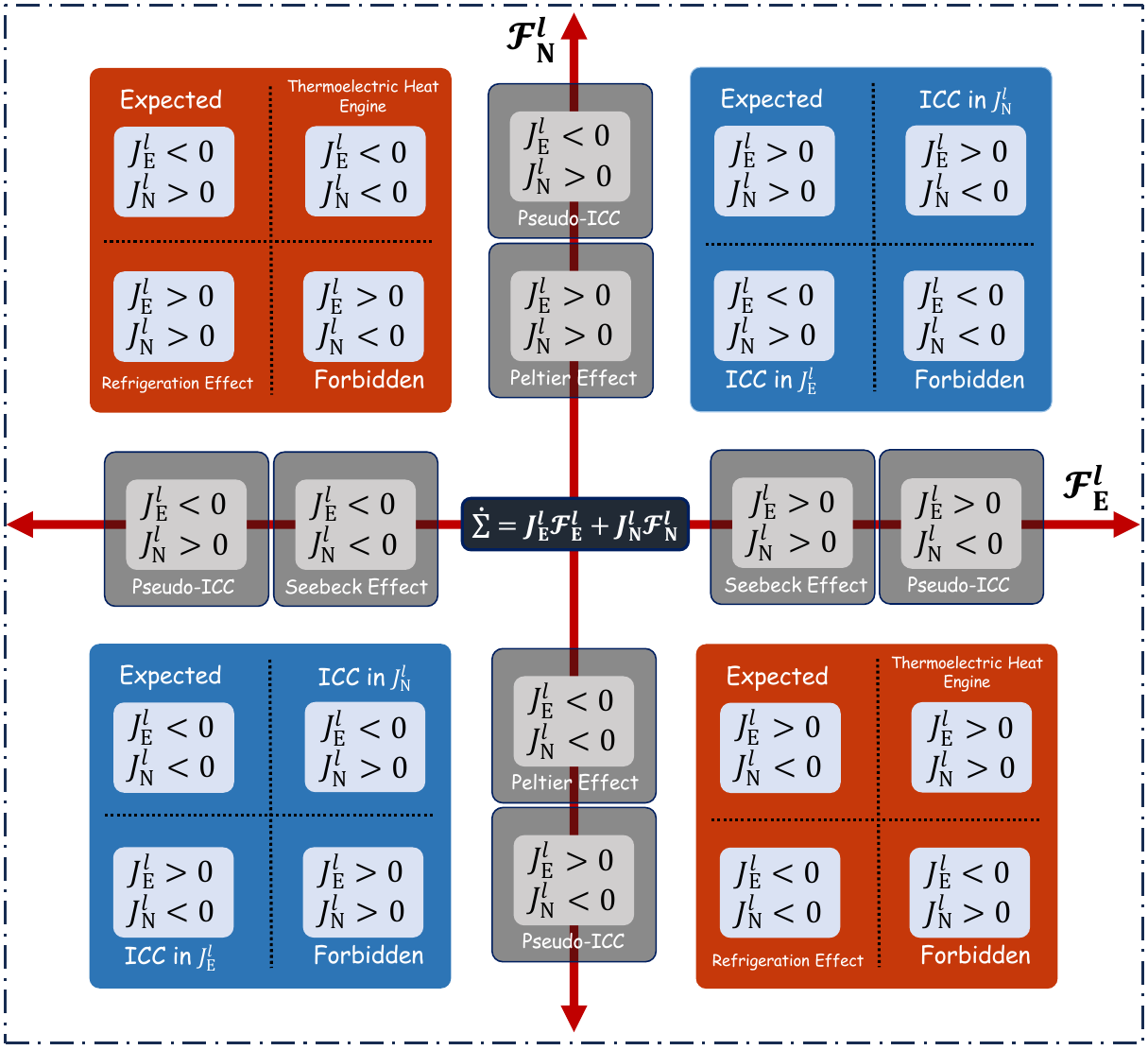}
    \caption{Pictorial representation of various thermodynamic phenomena governed by the entropy production rate $\dot{\Sigma} =J_{\rm{E}}^{l}\mathcal{F}_{\rm{E}}^{l}+J_{\rm{N}}^{l}\mathcal{F}_{\rm{N}}^{l}$, i.e., normal cross-effect (red shaded), ICC (blue shaded), pseudo-ICC, Seebeck and Peltier effects that can be obtained.}
    \label{Chart}
\end{figure}

When both thermodynamic forces are nonzero, the coupled transport phenomena become particularly rich, and the parameter space naturally divides into four quadrants [Fig.~\ref{Chart}]. The first and third quadrants are equivalent, with mutually parallel forces. As the overall sign is merely a matter of convention,  the first and third quadrants describe the same physical behavior. Similarly, the second and fourth quadrants are isomorphic in nature, with one force acting against the other. We first restrict our attention to the first quadrant, where both forces are positive, and then discuss the difference with the other quadrants.

\begin{enumerate}
\item In the first quadrant, when both $\mathcal{F}_{\rm E}^{l}$ and $\mathcal{F}_{\rm N}^{l}$ are positive, two possibilities can arise:
(a) Both energy and particle currents are positive, which is the most natural and expected thermodynamic response.
(b) One of the currents is negative. The latter case is highly nontrivial and counterintuitive, as it corresponds to a situation in which a particular flux flows against both applied forces. Nevertheless, this behavior is fully consistent with thermodynamics, provided $\dot{\Sigma} \geq 0$. This result is a direct manifestation of coupled transport and is known as the inverse current in coupled (ICC) transport, a phenomenon that was recently identified in a classical Hamiltonian system by Wang \textit{et. al.}~\cite{wang2020inverse}. Later, this phenomenon has also been investigated in quantum transport processes~\cite{zhang2021inverse,zhang2023inverse}. A general definition of ICC, which remains valid for multi-terminal models as well, can be stated as follows: \textit{when two thermodynamic forces are mutually parallel, one of the induced currents flows against both forces}. Although the sign convention of a current is arbitrary, the defining characteristic of ICC is that \textit{a current must simultaneously oppose both mutually parallel thermodynamic forces}. It is therefore evident that negative cross effects act as a necessary precursor for the emergence of a genuine ICC. Specifically, if a current can flow against its non-conjugate force in the absence of the conjugate force, then there exists the possibility for that current to also flow against its conjugate force when both forces are present. Thus, an \textit{unconventional Seebeck} or \textit{unconventional Peltier effect} inherits the possibility of realizing an ICC in the energy or particle current, respectively. Such behavior may be classified as pseudo-ICC. However, it does not constitute a true or genuine ICC, due to the absence of the conjugate force itself.
For this reason, regimes described by pseudo-ICC serve as a necessary prerequisite for the emergence of genuine ICC behavior. Realizing a genuine ICC is considerably more challenging than achieving a pseudo-ICC. Thus, it represents a highly counterintuitive and nontrivial manifestation of coupled transport.

In the present two-terminal single–QD setup, we have already ruled out the possibility of realizing either an unconventional Seebeck or an unconventional Peltier effect. Consequently, this minimal model cannot support an ICC phenomenon. In subsequent discussions, we will construct a model capable of exhibiting the ICC phenomenon. Although achieving ICC is highly nontrivial, the realization of ICC in the particle current or the energy current naturally corresponds to the operation of an autonomous thermoelectric engine or a refrigerator, respectively~\cite{ghosh2026inverse}. Therefore, it carries profound thermodynamic significance in the study of coupled transport processes.

\item We now consider the fourth quadrant of Fig.~\ref{Chart}, where the energy force is positive, acting from the left reservoir to the right, while the particle force is negative, acting in the opposite direction. As noted earlier, a positive energy force corresponds to $\beta_r > \beta_l$ [cf.~Eq.~\eqref{R5}], while $\beta_r \mu_r > \beta_l \mu_l$ implies a negative particle force [cf.~Eq.~\eqref{R5}]. The thermodynamically favored response is that each current flows along its conjugate force: the energy current is positive, corresponding to heat flow from the hot left reservoir to the cold right reservoir, and the particle current is negative, indicating transport from right to left. This configuration guarantees a strictly positive entropy production rate and is therefore consistent with the second law. However, alternative current configurations are also allowed, provided that the entropy production remains non-negative. The energy current may remain aligned with its force while driving the particle current in the same direction, so that the particle current flows against its own force. In contrast, the particle current may align with its force and drag the energy current in the same direction, causing the energy current to flow against its conjugate force. These two regimes correspond, respectively, to the operation of a thermoelectric heat engine and a refrigerator. In both cases, the behavior arises from thermodynamic cross effects, in which a non-conjugate force drives a current against its own conjugate force.

Therefore, in the two-terminal single–QD model, the fourth quadrant allows either engine or refrigeration behavior. An analogous situation arises in the second quadrant (Fig.~\ref{Chart}), where the energy force is negative, and the particle force is positive. In this regime also, two thermodynamically allowed outcomes are possible: both currents may be positive, corresponding to refrigeration, or both may be negative, indicating thermoelectric heat-engine operation.

\end{enumerate}

From the above discussion, we identify two fundamentally distinct outcomes of coupled transport: thermodynamic cross effects and inverse current phenomena. Although closely related, these effects differ in a subtle but crucial way. While physical consequences of cross effects have been extensively studied over several decades~\cite{benenti2017fundamental,onsager1931reciprocal-I,onsager1931reciprocal-II,callen1948the-application,kondepudi2015book,disalvo1999thermoelectric}, with thermoelectric heat engines and refrigerators serving as canonical examples, inverse current phenomena, namely, ICC --- remain largely unexplored. In the following sections, we first clarify the distinction between these two phenomena and therefore seek a more realistic model capable of exhibiting richer thermodynamic behavior, including both pseudo-ICC and genuine ICC. 

\section{Cross-effect Vs ICC}\label{sec8}

At first glance, the distinction between ICC and thermodynamic cross effects may appear subtle, since in both cases a current flows against its conjugate force. For example, when a particle current flows against its conjugate particle force, the system functions as a thermoelectric heat engine, whereas when an energy current flows against a temperature gradient, refrigeration occurs. These are the familiar, “traditional” manifestations of thermodynamic cross effects, where an induced current (generated by its non-conjugate force) is driven against its conjugate force. By contrast, the defining feature of ICC is that a current flows against both driving forces --- more precisely, against two mutually parallel thermodynamic forces. In this regime, the current is not assisted by any non-conjugate force; instead, it opposes both forces simultaneously. This requirement makes ICC much more counterintuitive and challenging. Although ICC can realize \textit{autonomous} engine or refrigeration behavior, such devices are fundamentally different from traditional ones: both driving forces oppose a flux, and the system operates autonomously, without any force aiding the current.

\section{Three terminal CQD-model for coupled transport}\label{sec9}

In this section, we focus on a capacitively coupled three-terminal CQD setup~\cite{gupt2024graph,shuvadip2022univarsal,ghosh2026inverse,joulain2016quantum,sanchez2011optimal}
rather than a conventional DQD system~\cite{strasberg2022quantum,vanDerWiel2002electron,dorsch2021heat}. While coupled transport in a DQD configuration can arise from weak inter-dot hopping, such systems consist of two nearly independent single-QD Hamiltonians connected by a small tunneling term. In contrast, strong capacitive coupling in a CQD fundamentally modifies the system’s Hilbert space, rendering it a genuinely composite quantum system whose Hamiltonian is diagonal in the joint eigenbasis of the two QDs ' number operators. We further extend this two terminal single-QD model to a three-terminal CQD configuration, where an auxiliary QD is tunnel-coupled to an additional electronic (fermionic) reservoir without directly participating in charge transport between the primary leads. This model resembles to the well-established three-terminal Sánchez–Büttiker model~\cite{sanchez2011optimal}, which has been widely studied \cite{timm2007gauge,walschares2013optimally,strasberg2013thermodynamic,kutvonen2016thermodynamics,esposito2019thermodynamics,thierschmann2015three,whitney2016thermoelectricity,erdman2017thermoelectric,werlang2014optimal,zhang2017three,aligia2020heat,tesser2022heat,esposito2012stochastically,dutta2017thermal,dutta2020single} and experimentally realized as a three-terminal energy-harvesting device by Thierschmann et al.~\cite{thierschmann2015three}. In the following, we first discuss thermoelectric effects in a three-terminal setup, then briefly review the novel ICC effect in a similar setup.

\subsection{Thermoelectric effects in three terminal CQD setup}

In this section, we study coupled transport in a three-terminal CQD model. We first discuss the model and then develop its thermodynamic force–flux formulation to analyze coupled energy and particle transport within this setup. In a suitable limit, the setup is reduced to the two-terminal single-QD case, retaining all the corresponding thermoelectric effects.

\subsubsection{Model}

\begin{figure}[t]
    \centering
\includegraphics[width=\columnwidth,height=4.8
cm]{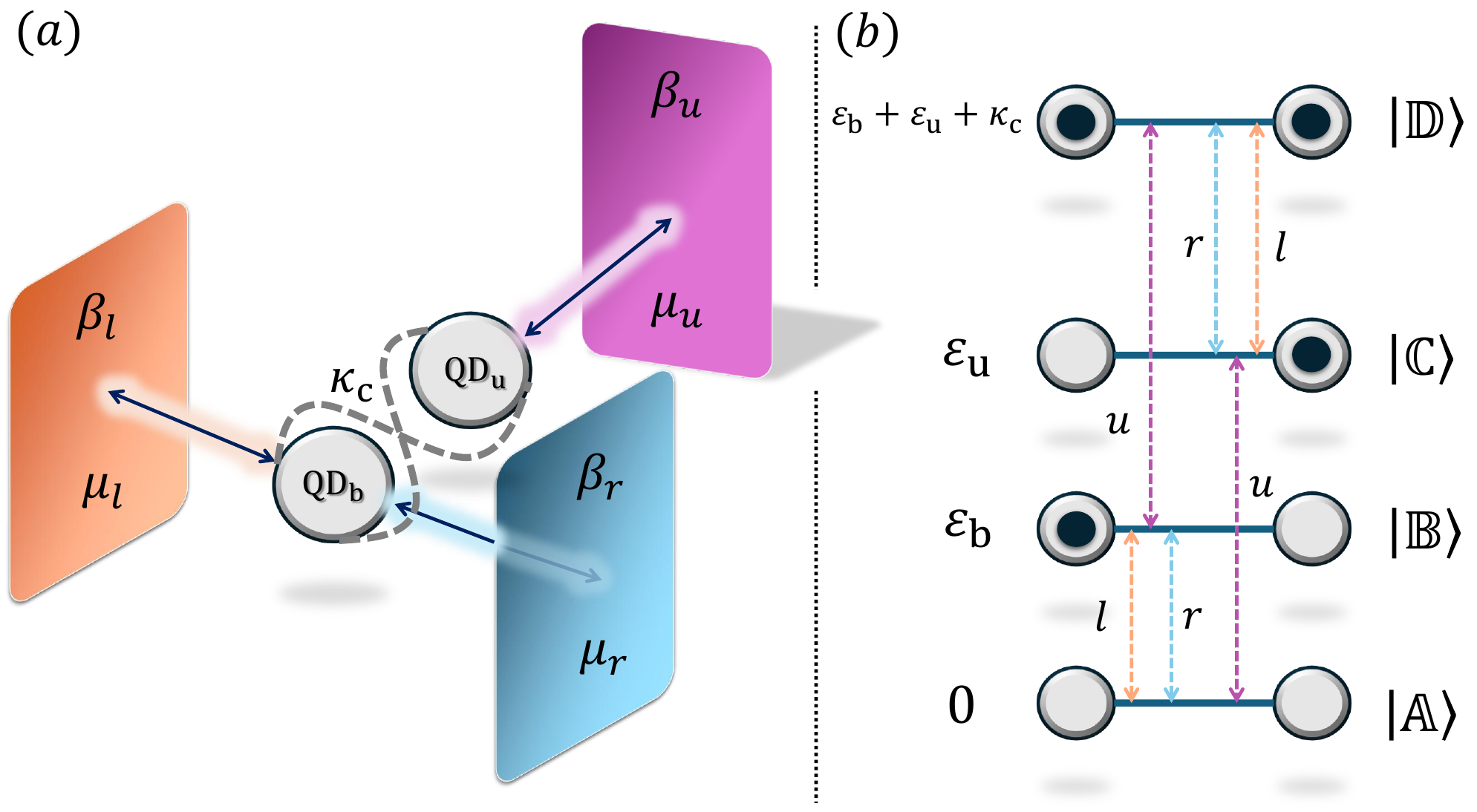}
    \caption{(a) Schematic of the three-terminal CQD model. (b) Energy level diagram of CQD where transitions are mediated by the baths within the eigen-states of the composite system.}
    \label{CQD Model}
\end{figure}

The model consists of two strongly and capacitively coupled QDs, denoted as the upper dot (${ \rm QD_u }$) and the bottom dot (${ \rm QD_b }$), as schematically illustrated in Fig.~\ref{CQD Model}(a). The QDs interact exclusively through long-range Coulomb interaction, suppressing direct particle exchange due to Coulomb blockade, while enabling energy exchange between the dots via a capacitive coupling of strength $\kappa_{\rm c}$~\cite{ruokola2011single,zhang2017three,shuvadip2022univarsal}. Although the QDs are strongly and capacitively coupled, we still assume that the system-reservoir interaction is weak enough so that the Markovian dynamics can be followed for the reduced system dynamics. The ${\rm QD_b}$ is simultaneously tunnel-coupled to two fermionic reservoirs, labeled $\textit{l}$ (left) and $\textit{r}$ (right), thereby allowing particle transport between these two leads~\cite{sanchez2011optimal,ghosh2012fermionic,zhang2018coulomb,shuvadip2022univarsal}. In contrast, ${\rm QD_u}$ is tunnel-coupled to a single electronic reservoir, allowing particle and energy exchange between the dot and the reservoir. If we focus on the lower subsystem, namely ${\rm QD}_b$, which is tunnel-coupled to both reservoirs $l$ and $r$, it effectively reproduces the two-terminal single-QD model discussed in Sec.~\ref{sec4}. For spinless electrons, $\kappa_{\rm c}$ is necessarily positive. Therefore, the Hamiltonian describing the CQD system is given by~\cite{werlang2014optimal,shuvadip2022univarsal,gupt2023topranked,gupt2024graph}
\begin{equation}\label{HS}
\begin{split}
 H_{\rm{S}}=\varepsilon_{\rm{b}} \mathcal{N}_{\rm{b}}+\varepsilon_{\rm{u}} \mathcal{N}_{\rm{u}}+\kappa_{c}\mathcal{N}_{\rm{b}}\mathcal{N}_{\rm{u}}.
\end{split}
\end{equation}
In Eq.~\eqref{HS}, $\varepsilon_\alpha>0$ $(\alpha= \rm{u},\rm {b})$ denotes the single-particle energy level of the $\alpha$'th ${\rm QD}$, which is taken to be positive without loss of generality. Since the electron density in the dots is low, the occupancy is restricted to either zero or one~\cite{whitney2018quantum,gupt2023topranked,gupt2024graph,shuvadip2022univarsal}. Consequently, the eigenstates of ${\rm QD_\alpha}$ are $|0\rangle$ and $|1\rangle$, with eigenenergies $0$ and $\varepsilon_{\alpha}$, respectively. The corresponding number operators for ${\rm QD_\alpha}$ are 
$\mathcal{N}_{\rm{b}}= d^\dagger_{\rm{b}} d_{\rm{b}}$ and $\mathcal{N}_{\rm{u}}=d^\dagger_{\rm{u}} d_{\rm{u}}$, where $d^\dagger_{\rm{b}}$ ($d_{\rm{u}}$) and $d^\dagger_{{\rm{b}}}$ ($d_{{\rm{u}}}$) denote the electron creation (annihilation) operators for the respective ${\rm QD_\alpha}$, obeying to the anti-commutation relations $\{d_{\rm{b}},d^\dagger_{\rm{b}}\}=1= \{d_{{\rm{u}}},d^\dagger_{{\rm{u}}}\}$. Note that the interaction term ($d^\dagger_{\rm b} d_{\rm b} d^\dagger_{\rm u} d_{\rm u}$) is quartic in system operators and contributes only when both QDs are occupied. This nonlinear quartic term describes a strong capacitive (Coulomb) coupling between the dots while prohibiting particle exchange between them. In contrast, a conventional bilinear coupling would allow coherent tunneling of electrons between the two QDs, leading to the well-studied DQD model~\cite{strasberg2022quantum,vanDerWiel2002electron,dorsch2021heat}. Since our primary interest is the transport of heat in the absence of particle transport between the dots, the quartic interaction is essential for realizing the desired effects. As the two QDs are strongly and capacitively coupled, the overall system Hamiltonian is diagonal in the eigenbasis of the individual ${\rm QD}$. This can be represented by the tensor product of the number operator's eigenbasis of the coupled ${\rm QDs}$. For convenience, the four eigenstates $\{|0\rangle, \ket{1}\}\otimes\{|0\rangle, \ket{1}\}$,  are labeled by $|\mathbb{A}\rangle=\ket{00}$, $|\mathbb{B}\rangle=\ket{1 0}$, $|\mathbb{C}\rangle=\ket{01}$, $|\mathbb{D}\rangle=\ket{11}$  and their corresponding eigenenergies ($\varepsilon_\mathbb{i}$, $\mathbb{i=A,B,C,D}$) are $\varepsilon_\mathbb{A}=0$, $\varepsilon_\mathbb{B}=\varepsilon_{\rm{b}}$, $\varepsilon_\mathbb{C}=\varepsilon_{\rm{u}}$ and $\varepsilon_\mathbb{D}=\varepsilon_{\rm{b}}+\varepsilon_{\rm{u}}+\kappa_{\rm c}$ respectively [FIG.~\ref{CQD Model}(b)]. Finally, the energy of the most excited state of the composite system is $\varepsilon_{\rm{u}}+\varepsilon_{\rm{b}}+\kappa_{\rm c}$, where we assume w.l.o.g., $\varepsilon_{\rm{b}} < \varepsilon_{\rm{u}}$.

The total Hamiltonian of all three reservoirs is given by $H_{\rm{B}}=H_{\rm{B}}^{u}+H_{\rm{B}}^{l}+H_{\rm{B}}^{r}$. The QDs are weakly coupled to the reservoirs, allowing sequential tunnelling ~\cite{shuvadip2022univarsal,gupt2023topranked,gupt2024graph,dutta2017thermal,dutta2020single,thierschmann2015three}. The tunnel-coupled Hamiltonians characterised by the coupling constants $t_{k}^{\alpha\lambda}$ are given by Eq.~\eqref{H-tunn}, which describes the tunnelling of a particle between the $\alpha$'th ${\rm QD}$ and the reservoir $\lambda$. Furthermore, the transitions between $|\mathbb{B}\rangle\leftrightarrow|\mathbb{C}\rangle$ and $|\mathbb{A}\rangle\leftrightarrow|\mathbb{D}\rangle$ are prohibited due to the Coulomb blockade and the sequential tunnelling approximation~\cite{zhang2017three,zhang2018coulomb,shuvadip2022univarsal}. So, there are in total four authorised transitions between the eigenstates of the composite system, and each of the three reservoirs $l$, $r$, and $u$ controls two transitions, i.e., $|\mathbb{A}\rangle\leftrightarrow|\mathbb{B}\rangle$ and $|\mathbb{C}\rangle\leftrightarrow|\mathbb{D}\rangle$ are triggered by leads $l$ and/or $r$, while reservoir $u$ governs the transitions between $|\mathbb{A}\rangle\leftrightarrow|\mathbb{C}\rangle$ and $|\mathbb{B}\rangle\leftrightarrow|\mathbb{D}\rangle$, respectively. The transition energy between the eigenstates $|\mathbb{i}\rangle \rightarrow |\mathbb{j}\rangle$ is defined as $\omega_\mathbb{ij}=\varepsilon_\mathbb{j}-\varepsilon_\mathbb{i}$, where $\varepsilon_\mathbb{i}$ is the eigenenergy of $H_{\rm{s}}$ with the eigenstate $\{|\mathbb{i}\rangle\}$ and the four transition energies are, respectively $\omega_{\mathbb{AB}}=\varepsilon_{\rm{L}}$, $\omega_{\mathbb{AC}}=\varepsilon_{\rm{R}}$, $\omega_{\mathbb{CD}}=\varepsilon_{\rm{L}}+\kappa_{\rm c}$, and $\omega_{\mathbb{BD}}=\varepsilon_{\rm{R}}+\kappa_{\rm c}$.

The dynamics can be solved using QME, similar to our previous two-terminal single-QD model.
As usual, the state of the composite system is described by the reduced density matrix  $\rho_{\rm{s}}(t)=\Tr_{\rm{B}}\{\rho_{\rm{tot}}(t)\}$, where $\rho_{\rm{tot}}(t)$ is the total system and bath density matrix. The time evolution of $\rho_{\rm{s}}(t)$ is given by~\cite{breuer2002book,shuvadip2022univarsal,strasberg2022quantum,gupt2023topranked} (See Appendix~\ref{Appendix-A}):
\begin{equation}\label{LME}
\frac{d}{dt}\rho_{\rm{s}}(t)=\sum_{\lambda}\mathcal{L}_{\lambda}[\rho_{\rm{s}}(t)]; \quad \lambda=l,r,u.
\end{equation}
where we utilise the strong coupling between the QDs, while maintaining the weak coupling between the system and its surrounding environment. Such a procedure, which is obtained by tracing out the environmental degrees of freedom, naturally gives rise to both dissipation and decoherence~\cite{breuer2002book}. However, since the system Hamiltonian [cf.~Eq.~\eqref{HS}] is diagonal in the eigenbasis of the individual QDs, the BMS approximation can be implemented, allowing the above LME [cf.~Eq.~\eqref{LME}] to be derived solely based on $H_{\rm S}$. Consequently, the reduced density matrix of the system contains no off-diagonal (coherence) terms, and our analysis is exclusively concerned with the dissipative dynamics of each QD. These dynamics are influenced not only by the coupling of each QD to its own bath but also by the interactions between QDs themselves. Importantly, it is crucial for accurately characterizing heat flow across a wide range of system parameters~\cite{gupt2024graph} where decoherence effects are not a primary requirement for the phenomena discussed below. The Lindblad super-operator $\mathcal{L}_\lambda[\rho_{\rm{s}}(t)]$ in Eq.~\eqref{LME} is defined in the Appendix~\ref{Appendix-A}, and following the same treatment done in the single QD model, we can calculate the steady-state currents. The general analytical expressions of the steady-state energy and particle are already defined in Eq.~\eqref{B10}, which yields the general heat current expression as
\begin{equation}\label{B10.1}
\begin{split}
J^{\lambda}_{\rm{Q}}=\sum_{\{\omega_{\mathbb{ij}}\}}({\omega_{\mathbb{ij}}}-\mu_{\lambda})\Gamma_{\mathbb{ij}}^{\lambda+}=\sum_{\{\omega_{\mathbb{ji}}\}}({\omega_{\mathbb{ji}}}-\mu_{\lambda})\Gamma_{\mathbb{ji}}^{\lambda-},  
\end{split}
\end{equation}
where, $\Gamma_{\mathbb{ij}}^{\lambda+}=-\Gamma_{\mathbb{ji}}^{\lambda-}$. For this CQD-three terminal setup, at steady state ($\dot{\rho}_\mathbb{i}=0$), the transition rates satisfy:
\begin{equation}\label{gamma}
\begin{split}
\Gamma^{lr+}_\mathbb{AB}&=\Gamma^{u+}_\mathbb{BD}=\Gamma^{lr-}_\mathbb{DC}=\Gamma^{u-}_\mathbb{CA}=\Gamma_{\mathbb{ABCDA}}\equiv\Gamma_{+};\\
\Gamma^{u+}_\mathbb{AC}&=\Gamma^{lr+}_\mathbb{CD}=\Gamma^{u-}_\mathbb{DB}=\Gamma^{lr-}_\mathbb{BA}=\Gamma_{\mathbb{ACDBA}}\equiv\Gamma_{-},
\end{split}
\end{equation}
implying $\Gamma_{+}=-\Gamma_{-}$. With the help of Eq.~\eqref{B10}, one can verify
\begin{equation}\label{Jer-Jnr}
\begin{split}
J_{\rm{E}}^u=&\varepsilon_{\rm{u}}\Gamma_{+}+(\varepsilon_{\rm{u}}+\kappa_{\rm c})\Gamma_{-}=\kappa_{\rm c}\Gamma_{+};\\
\quad\quad J_{\rm{N}}^u=&\Gamma_\mathbb{AC}^{u+}+\Gamma_\mathbb{BD}^{u+}=\Gamma_{+}+\Gamma_{-}=0,
\end{split}
\end{equation}
as $\rm QD_u$ is coupled only to the lead $u$, resulting in $J_{\rm{N}}^u = 0$. This immediately gives $J_{\rm{Q}}^u=J_{\rm{E}}^u=\kappa_{\rm c}\Gamma_{+}$. 
As $\rm QD_u$ is only coupled with lead $u$, there will be no net particle current driven by $u$ in the steady state, i.e., $J_{\rm{N}}^u = 0$. However, $\rm QD_b$ is simultaneously coupled with leads $l$ and $r$, allowing particle flow across $l$ and $r$ in the steady state. Since there is no external source or sink, the total energy and particle currents must be conserved in the steady state. Hence, we summarise our first set of results:
\begin{equation}\label{conservation-current}
\begin{split}
\sum_{\lambda=l,r,u}J_{\rm{E}}^\lambda=0, \quad \text{and} \quad \sum_{\lambda=l,r,u} J_{\rm{N}}^\lambda=0,
\end{split}
\end{equation}
which yields
\begin{equation}
  J_{\rm{E}}^u=-J_{\rm{E}}^{lr}=\kappa_{\rm c}\Gamma_{+}; \quad \text{and} \quad  J_{\rm{N}}^l=-J_{\rm{N}}^r.
\end{equation}
The heat current is not conserved as $J_{\rm{Q}}^{l}+J_{\rm{Q}}^{r}\equiv J_{\rm{Q}}^{lr}\neq J_{\rm{E}}^{lr}=-\kappa_{\rm c}\Gamma_{+}$. 

\subsubsection{Identifying thermodynamic forces}

For the present model, the general form of the entropy production rate defined in Eq.~\eqref{EP9} reads
\begin{equation}\label{CQD1}
\begin{split}
\dot{\Sigma}=&-k_B[\beta_{l}J_{\rm{E}}^{l}+\beta_{r}J_{\rm{E}}^{r}+\beta_{u}J_{\rm{E}}^{u}]\\
&+k_B[\beta_{l}\mu_{l}J_{\rm{N}}^{l}+\beta_{r}\mu_{r}J_{\rm{N}}^{r}+\beta_{u}\mu_{u}J_{\rm{N}}^{u}].
\end{split}
\end{equation}
Since $J_{\rm{N}}^{u}=0$, Eq.~\eqref{CQD1} reduces to
\begin{equation}\label{CQD2}
\dot{\Sigma}=-k_B[\beta_{l}J_{\rm{E}}^{l}+\beta_{r}J_{\rm{E}}^{r}+\beta_{u}J_{\rm{E}}^{u}]+k_B[\beta_{l}\mu_{l}J_{\rm{N}}^{l}+\beta_{r}\mu_{r}J_{\rm{N}}^{r}].
\end{equation}
Furthermore, both energy and particle currents are conserved in the steady state. This conservation follows directly from Eq.~\eqref{conservation-current}. These conditions lead to six equivalent representations of the entropy production rate. However, unlike the single–QD model, a subtlety arises in the present three-terminal setup. While the particle force can be identified uniquely, the energy biases are not unique and depend on the specific choice of independent energy currents used to express the entropy production rate. This non-uniqueness originates from the presence of two independent energy biases and one particle force in the three-terminal geometry. For the particle sector, one may choose $J_{\rm N}^{l}$ or $J_{\rm N}^{r}$ as the independent current; since they are equal in magnitude and opposite in direction, i.e., $J_{\rm N}^{l}=-J_{\rm N}^{r}$. So, particle current can be identified unambiguously, and substituting this relation into Eq.~\eqref{CQD2} yields the following
\begin{equation}\label{CQD3}
\dot{\Sigma}
= -k_B\!\left[\beta_l J_{\rm E}^l + \beta_r J_{\rm E}^r + \beta_u J_{\rm E}^u\right]
+
\begin{cases}
k_B(\beta_l\mu_l - \beta_r\mu_r)\,J_{\rm N}^l,\\[4pt]\quad\quad\qquad
\text{or}\\[4pt]
k_B(\beta_r\mu_r - \beta_l\mu_l)\,J_{\rm N}^r.
\end{cases}
\end{equation}
A comparison with the phenomenological form of the entropy production rate [cf.~Eq.~\eqref{EPR}] identifies the particle force uniquely as
\begin{equation}\label{CQD4}
\mathcal{F}_{\rm{N}}^{l}=-\mathcal{F}_{\rm{N}}^{r}=k_B(\beta_{l}\mu_{l}-\beta_{r}\mu_{r}).
\end{equation}
Since both definitions are thermodynamically equivalent, either choice of particle force may be employed. For the energy sector, the situation is a little more nuanced. Due to energy conservation $J_{\rm{E}}^{l}+J_{\rm{E}}^{r}+J_{\rm{E}}^{u}=0$, only two of the three energy currents are independent. Consequently, the entropy production rate can be expressed using different pairs of energy currents, leading to three distinct --- yet physically equivalent --- representations of the 
energy biases.
To proceed, we fix the particle force as $\mathcal{F}_{\rm N}^{l}$ without loss of generality. Equation~\eqref{CQD3} then yields
\begin{equation}\label{CQD5}
\begin{split}
\dot{\Sigma}=-k_B[\beta_{l}J_{\rm{E}}^{l}+\beta_{r}J_{\rm{E}}^{r}+\beta_{u}J_{\rm{E}}^{u}]+J_{\rm{N}}^{l}\mathcal{F}_{\rm{N}}^{l}.
\end{split}
\end{equation}
If we eliminate $J_{\rm{E}}^{l}$, it gives the following results
\begin{equation}\label{CQD6}
\begin{split}
\dot{\Sigma}&=k_B(\beta_{l}-\beta_{r})J_{\rm{E}}^{r}+k_B(\beta_{l}-\beta_{u})J_{\rm{E}}^{u}+J_{\rm{N}}^{l}\mathcal{F}_{\rm{N}}^{l}\\
&\equiv J_{\rm{E}}^{r}\mathcal{F}_{\rm{E}}^{r}+J_{\rm{E}}^{u}\mathcal{F}_{\rm{E}}^{u}+J_{\rm{N}}^{l}\mathcal{F}_{\rm{N}}^{l}
\end{split}
\end{equation}
which identifies the energy biases as
\begin{equation}
\mathcal{F}_{\rm{E}}^{r}=k_B(\beta_{l}-\beta_{r})\quad;\quad \mathcal{F}_{\rm{E}}^{u}=k_B(\beta_{l}-\beta_{u}).  
\end{equation}
Alternatively, eliminating $J_{\rm{E}}^{r}$ yields
\begin{equation}\label{CQD8}
\begin{split}
\dot{\Sigma}&=k_B(\beta_{r}-\beta_{l})J_{\rm{E}}^{l}+k_B(\beta_{r}-\beta_{u})J_{\rm{E}}^{u}+J_{\rm{N}}^{l}\mathcal{F}_{\rm{N}}^{l}\\
&\equiv J_{\rm{E}}^{l}\mathcal{F}_{\rm{E}}^{l}+J_{\rm{E}}^{u}\mathcal{F}_{\rm{E}}^{u}+J_{\rm{N}}^{l}\mathcal{F}_{\rm{N}}^{l}
\end{split}
\end{equation}
leading to
\begin{equation}\label{CQD9}
\mathcal{F}_{\rm{E}}^{l}=k_B(\beta_{r}-\beta_{l})\quad;\quad \mathcal{F}_{\rm{E}}^{u}=k_B(\beta_{r}-\beta_{u}).  
\end{equation}
Finally, eliminating $J_{\rm{E}}^{u}$ yields
\begin{equation}
\mathcal{F}_{\rm{E}}^{l}=k_B(\beta_{u}-\beta_{l})\quad;\quad \mathcal{F}_{\rm{E}}^{r}=k_B(\beta_{u}-\beta_{r})  
\end{equation}
These results show that the form of the energy biases depends on the chosen representation. Nevertheless, the thermodynamic behavior remains unaltered, since only two independent energy biases are sufficient to fully characterize the three-terminal setup. This non-uniqueness of thermodynamic forces and currents is not a mathematical inconsistency but an intrinsic feature of general three-terminal transport. Although explicit forms of forces depend on the choice of force–flux pairs used to express the entropy production rate, all such choices are thermodynamically equivalent and lead to the same physical behavior. Accordingly, w.l.o.g, we adopt the representation in Eq.~\eqref{CQD8}, where the conjugate forces associated with the flux set $\{J_{\rm E}^{l}, J_{\rm E}^{u}, J_{\rm N}^{l}\}$ are given by Eq.~\eqref{CQD9}. Alternative but equivalent representations may be constructed by choosing any two independent energy fluxes (e.g., $\{J_{\rm E}^{u}, J_{\rm E}^{r}\}$, $\{J_{\rm E}^{u}, J_{\rm E}^{l}\}$, or $\{J_{\rm E}^{r}, J_{\rm E}^{l}\}$) together with one particle flux (either $J_{\rm N}^{l}$ or $J_{\rm N}^{r}$), and their corresponding conjugate forces. Importantly, the quantities $\{\mathcal{F}_{\rm E}^{u}, \mathcal{F}_{\rm E}^{r}, \mathcal{F}_{\rm N}^{r}\}$ should not be interpreted as physical thermal or particle forces. Rather, they are \textit{entropic energy} and \textit{particle biases} --- defined within the entropy representation --- and are conjugate to the full set of fluxes $\{J_{\rm E}^{u}, J_{\rm E}^{r}, J_{\rm N}^{r}\}$~\cite{ghosh2026inverse}. In systems with a single force-flux pair, this distinction vanishes, and the entropic biases coincide uniquely with the physical thermodynamic forces.

However, the presence of three independent force–flux pairs complicates the thermodynamic analysis and can obscure the essential features of coupled transport. We therefore seek to reduce the three-terminal setup to an effective framework that captures core principles with minimal complexity. In the following, this reduction can be carried out systematically, making the model an ideal platform to study coupled energy–particle transport.

\subsubsection{Reduction to an ideal model for coupled transport}

\begin{figure}[t]
    \centering
\includegraphics[width=\columnwidth,height=4.8cm]{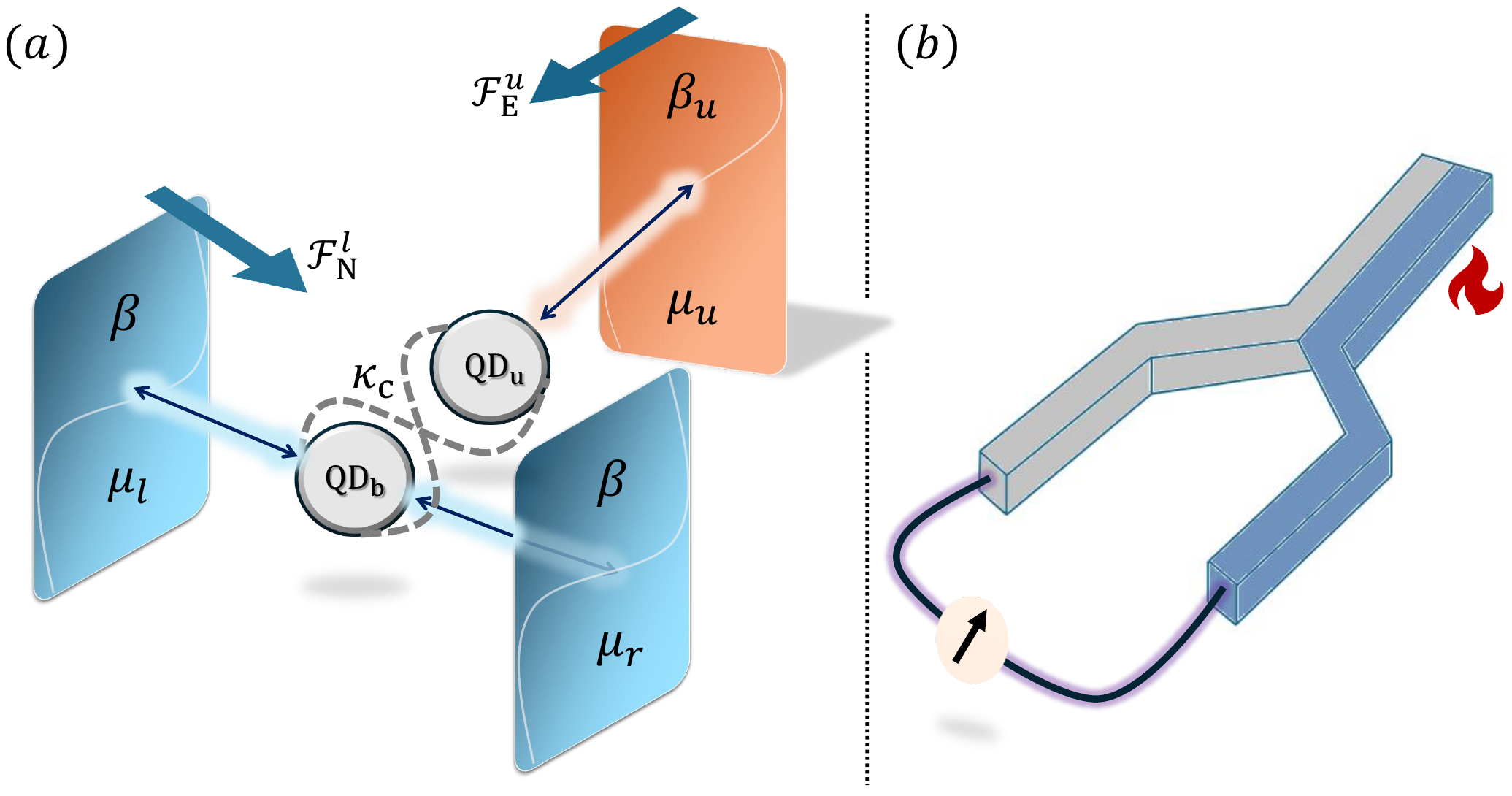}
    \caption{(a) Schematic of the reduced CQD-three terminal setup as a model for a quantum thermoelectric device, (b) Classical thermoelectric device that motivates the reduced model.}
    \label{CQD_thermo Model}
\end{figure}

Our objective is to effectively reduce the complexity of the model by imposing suitable constraints such that only two independent force–flux pairs remain. Since our primary interest lies in coupled transport involving two distinct transport processes, i.e., energy and particle transport, we must eliminate one of the energy force–flux pairs. This requires imposing an appropriate restriction on one of the energy biases, thereby reducing the three-terminal setup to an effective two-force framework suitable for analysing coupled transport [Fig.~\ref{CQD_thermo Model}(a)]. This is analogous to the classical thermoelectric device [Fig.~\ref{CQD_thermo Model}(b)]. Since we are interested in thermoelectric effects, we assume that the temperatures of the leads in the lower subsystem are equal. This implies $\beta_l=\beta_r\equiv\beta$. Under this condition, one of the entropic energy biases $\mathcal{F}_{\rm E}^{l}$ becomes zero, while the remaining two entropic biases are unambiguously reduced to the energy force
$\mathcal{F}_{\rm E}^{u} = k_B(\beta - \beta_u)$
and a particle force
$\mathcal{F}_{\rm N}^{l} = k_B \beta (\mu_l - \mu_r):=k_B\beta\Delta\mu$
as depicted in Fig.~\ref{CQD_thermo Model}(a). The above choice of the parameters renders the present three-terminal CQD setup exactly equivalent to the well-known Sánchez–Büttiker model~\cite{sanchez2011optimal}. In this limit, the general expression for the entropy production rate in Eq.~\eqref{CQD8} simplifies to
\begin{equation}\label{sigma-SB-model}
\dot{\Sigma} = J_{\rm N}^{l} \mathcal{F}_{\rm N}^{l}+J_{\rm E}^{u} \mathcal{F}_{\rm E}^{u} .  \end{equation}

It is important to emphasise that although Eq.~\eqref{CQD8} does not provide a unique representation of the entropy production rate, once one of the entropic biases vanishes (here $\mathcal{F}_{\rm E}^{l}=0$), the remaining force–flux pairs acquire a direct physical interpretation. In this regime, Eq.~\eqref{sigma-SB-model} describes genuine coupled transport governed by well-defined thermodynamic energy and particle forces. Therefore, the above situation can be analysed in close analogy with the single–QD model. Consequently, all thermodynamic outcomes identified in the single-QD setup can also be recovered within the present reduced model. In this sense, the reduced CQD three-terminal configuration serves as an appropriate framework for investigating coupled transport phenomena. In addition, similar to the minimal single–QD model, this reduced setup can operate as both a thermoelectric heat engine and a refrigerator. Due to these capabilities, the model is widely regarded as an efficient and experimentally relevant platform to study thermoelectric transport and energy conversion at the nanoscale~\cite{sanchez2011optimal,thierschmann2015three,walldorf2017thermoelectrics,benenti2017ffundamental,whitney2016thermoelectricity}.

Although this reduced model bears a close resemblance to the single–QD setup, an important distinction must be emphasised. In the earlier minimal models, the thermodynamic forces acted mutually parallel to each other. In contrast, as indicated by Eq.~\eqref{sigma-SB-model}, the forces in the present reduced configuration are non-parallel, and the associated energy and particle currents are also non-parallel. This feature is precisely what makes the model analogous to classical thermoelectric devices, where electrical and thermal driving forces act along different directions. However, this characteristic also implies that the reduced model is not suitable for ICC, which, by definition, requires mutually parallel thermodynamic forces. Nevertheless, the reduced model remains highly effective for studying thermoelectric phenomena. It can capture all conventional thermoelectric effects, including Seebeck and Peltier effects, as well as their unconventional counterparts. However, these negative thermoelectric responses should not be interpreted as pseudo-ICC effects, since the driving forces are inherently non-parallel in this configuration. However, because of its clear physical interpretation and experimental relevance, this reduced model is one of the most suitable platforms for investigating thermoelectric transport.

Finally, we are left with a single remaining objective: to identify a model in which genuine ICC phenomena can be explored. Although the present reduced model does not permit this, the full CQD three-terminal framework offers additional flexibility and admits alternative reductions capable of exhibiting coupled transport under mutually parallel forces. Accordingly, in the next section, we focus on constructing and analysing such an ideal reduced model, specifically designed to investigate the ICC effect.

\section{Inverse Current in Coupled Transport}\label{sec10}

\begin{figure}[t]
    \centering
\includegraphics[width=\columnwidth,height=4cm]{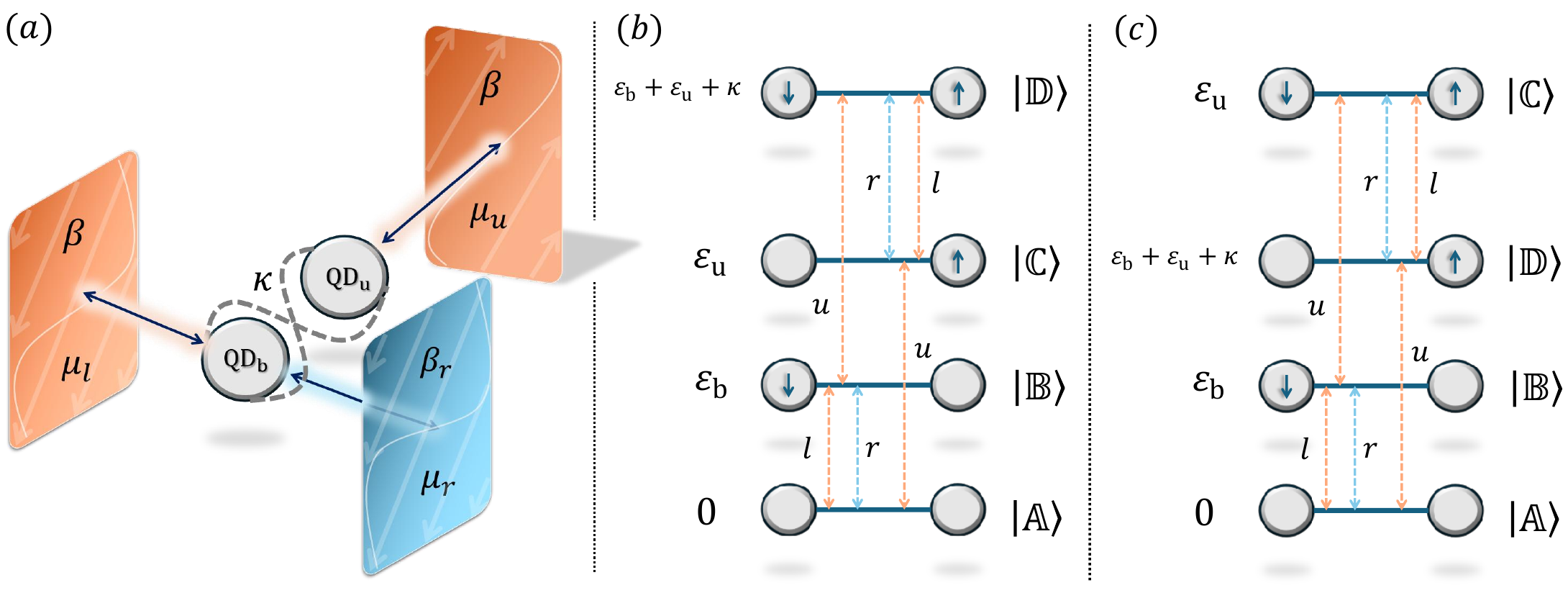}
    \caption{(a) Ideal schematic of three-terminal CQD setup for realising ICC effect. The energy level diagram of the composite system eigenstates for (b) $\kappa>0$ and (c) $\kappa<0; \; \text{but} \;|\kappa|>\varepsilon_{\rm{b}}$.}
    \label{CQD_ICC Model}
\end{figure}

To investigate ICC as a genuinely distinct effect, it is essential to impose the constraint such that it yields a reduced model with two mutually parallel forces. This requirement can be fulfilled naturally if both the energy and particle forces are associated with the same reservoir. In the present formulation, following Eq.~\eqref{CQD8}, the single particle force $\mathcal{F}_{\rm N}^{l}$ is already associated with the left lead. Consequently, to obtain mutually parallel forces, the remaining energy force must also be associated with the left lead. This can be achieved by imposing a thermodynamic constraint in which the other energy force, $\mathcal{F}_{\rm E}^{u}$, vanishes. From Eq.~\eqref{CQD9}, this condition is satisfied by $\beta_r=\beta_u=\beta$, which means that the right and upper reservoirs are held at the same temperatures. Under this condition, the general CQD three-terminal setup reduces to an effective two-force coupled-transport system, in which the entropy production rate can be expressed in terms of mutually parallel energy and particle forces, as 
\begin{equation}\label{ICC}
\dot{\Sigma} = J_{\rm E}^{l} \mathcal{F}_{\rm E}^{l} + J_{\rm N}^{l} \mathcal{F}_{\rm N}^{l},  \end{equation}
where the two real \textit{thermodynamic forces} are defined as
\(\mathcal{F}_{\rm E}^{l} = k_B(\beta_l - \beta)\) and
\(\mathcal{F}_{\rm N}^{l} = k_B(\beta_l \mu_l - \beta \mu_r)\), respectively
[Fig.~\ref{CQD_ICC Model}(a)]. Without loss of generality, we consider that both forces are mutually parallel and satisfy
\(\mathcal{F}_{\rm E}^{l} > 0\) and \(\mathcal{F}_{\rm N}^{l} > 0\). In this condition, either \(J_{\rm E}^{l}\) or \(J_{\rm N}^{l}\) may flow against both forces. The resulting negative contribution to the entropy production rate can be compensated by the remaining current flowing along the forces, such that the total entropy production remains non-negative \(\dot{\Sigma} \geq 0\). Accordingly, Fig.~\ref{CQD_ICC Model}(a) illustrates the ideal configuration for ICC, where one current is driven simultaneously against two mutually parallel thermodynamic forces, yet the second law of thermodynamics is fully satisfied.

Interestingly, the resulting expression for the $\dot{\Sigma}$ is formally identical to that of the minimal two-terminal single-QD model. Thus, one expects the standard thermodynamic cross-effects of coupled transport---such as the Seebeck and Peltier effects---to be observed within this reduced framework. Moreover, in the minimal single-QD model and the S\'anchez--B\"uttiker configuration, the system can operate either as a thermoelectric heat engine or as a refrigerator. However, these conventional cross-effects have already been extensively studied ~\cite{monsel2022geometric,josefsson2018a,prete2019thermoelectric,jaliel2019experimental,nakpathomkun2010thermoelctric,josefsson2019optimal,sanchez2011optimal,zhang2015three,thierschmann2015three,erdman2017thermoelectric,jurgens2013thermoelectric}. We are interested here in whether the present reduced model can support the emergence of ICC, a genuinely distinct and counterintuitive thermodynamic phenomenon, and under what conditions such an effect can be realized. In the single-QD model, although the thermodynamic forces are mutually parallel, the strict constraint linking the energy and particle currents precludes the emergence of either pseudo-ICC or genuine ICC behavior, rendering inverse currents impossible. Since the present reduced model yields an entropy production rate formally identical to that of the two-terminal single-QD setup, a natural question arises: \textit{does this model necessarily inherit the same limitations, or can it support ICC behaviour despite this apparent equivalence?}

Despite their formal equivalence in force–flux space, the two models differ fundamentally in structure and dynamics. The previous model consists of a single QD, acting as a TLS, whereas the reduced model considered here is based on a CQD architecture, comprising two strongly and capacitively coupled QDs. This structural complexity significantly alters the internal dynamics of the system. Consequently, the solution of the LME [cf. {Eq.~\eqref{LME}}] for the CQD system yields a dynamical behaviour that is qualitatively distinct from that of the single-QD model. A detailed and comprehensive analysis of this CQD setup has already been carried out in Ref.~\cite{ghosh2026inverse}. The steady-state energy and particle currents in this case are given by
\begin{equation}\label{cqd-e-n}
\begin{split}
J_{\rm{E}}^{l}&=\varepsilon_{\rm{b}}\Gamma_\mathbb{AB}^{l+}+(\varepsilon_{\rm{b}}+\kappa)\Gamma_\mathbb{CD}^{l+}=\varepsilon_{\rm{b}}J_{\rm{N}}^l+\kappa\Gamma_\mathbb{CD}^{l+};\\
J_{\rm{N}}^l&=(\Gamma^{l+}_\mathbb{AB}+\Gamma^{l+}_\mathbb{CD})
\end{split}
\end{equation}
where the $\Gamma_{\mathbb{ij}}^{\lambda+}$ stands for the net excitation rates from the state $|\mathbb{i}\rangle$ to $|\mathbb{j}\rangle$ induced by reservoir $\lambda$ [Cf.~Eq.~\eqref{C1.3}]. It is evident from Eq.~\eqref{cqd-e-n} that the energy and particle currents in this model are no longer proportional to each other, i.e., $J_{\rm E}^{l} \neq \varepsilon J_{\rm N}^{l}$. In this case, an additional term appears that depends on the strength of the inter-dot interaction $\kappa$. The presence of this new term may lead to the possibility of realizing a genuine ICC effect within the present model. A detailed analysis in Ref.~\cite{ghosh2026inverse} shows that for positive $\kappa$, the transport remains qualitatively similar to the single-QD case and no inverse-current behavior is observed. In contrast, negative inter-dot interactions can open a qualitatively new transport regime. However, a negative $\kappa$ alone is not sufficient to guarantee ICC. Rather, ICC becomes feasible \textit{iff} the symmetry between energy and particle exchange is broken. This occurs when the ordering of the system eigenstates $|\mathbb{C}\rangle$ and $|\mathbb{D}\rangle$ is inverted, which requires
$\varepsilon_{\rm b}+\kappa<\varepsilon_{\rm{b}}$.

The remaining question is how to break the symmetry between energy and particle exchange. As discussed above, this requires a negative interaction parameter $\kappa$, corresponding to an effective attraction between the QDs. However, in the standard three-terminal CQD model, the interdot Coulomb interaction is inherently repulsive, yielding only $\kappa >0$. To overcome this limitation, we modified the model to allow for effective attractive interactions. Specifically, we consider spin-polarised electronic reservoirs instead of spinless ones. While spinless electrons permit only repulsive interdot interactions, spin-polarised reservoirs~\cite{gupt2024graph,wang2022cycleflux} enable more general interaction scenarios and allow for $\kappa <0$. In this setup, ${\rm QD_b}$ is coupled to spin-down ($\downarrow$) electronic reservoirs, whereas ${\rm QD_u}$ is coupled to a spin-up ($\uparrow$) reservoir [FIG.~\ref{CQD_ICC Model}(a)]. Under these assumptions, the Hamiltonian of the coupled QD system takes the form
\begin{equation}\label{HS-ICC}
\begin{split}
H_{\rm{S}}=\varepsilon_{\rm{b}} \mathcal{N}_{\rm{b}\downarrow}+\varepsilon_{\rm{u}} \mathcal{N}_{\rm{u}\uparrow}+\mathcal{N}_{\rm{b}\downarrow}\mathcal{N}_{\rm{u}\uparrow} \left[\kappa_{c} \mathbb{1} + \kappa_{s}\sigma^{z}_{\rm{b}\downarrow}\sigma^{z}_{\rm{u}\uparrow} \right].
\end{split}
\end{equation}
Here, $\varepsilon_\alpha>0$ $(\alpha=\rm{b},\rm{u})$ is the single-particle energy level of the $\alpha$'th QD. Due to the low electron density~\cite{whitney2018quantum,gupt2023topranked,gupt2024graph,shuvadip2022univarsal}, each dot can host at most one electron. So, the eigenstates of ${\rm QD_{b(u)}}$ are $|0\rangle$ and $\ket{\downarrow} (\ket{\uparrow})$, with energies $0$ and $\varepsilon_{\rm{b}} (\varepsilon_{\rm{u}})$, respectively. The corresponding number operators are $\mathcal{N}_{\rm{b}\downarrow}= d^\dagger_{\rm{b}\downarrow} d_{\rm{b}\downarrow}$ and $\mathcal{N}_{\rm{u}\uparrow}=d^\dagger_{\rm{u}\uparrow} d_{\rm{u}\uparrow}$, where $d^\dagger_{\alpha\sigma}$ ($d_{\alpha\sigma}$) creates (annihilates) an electron with spin $\sigma={\uparrow,\downarrow}$ in the dot $\alpha$. These operators satisfy the canonical anti-commutation relations ${d_{\alpha\sigma},d^\dagger_{\alpha'\sigma'}}=\delta_{\alpha\alpha'}\delta_{\sigma\sigma'}$. The spin–spin interaction term~\cite{daroca2025role} is expressed by $\sigma^{z}_{\rm{b}\downarrow}=1-2\mathcal{N}_{\rm{b}\downarrow}$ and $\sigma^{z}_{\rm{u}\uparrow}=2\mathcal{N}_{\rm{u}\uparrow}-1$, with the standard relations $\sigma^{z}_{\rm{b}\downarrow}|\downarrow\rangle=-|\downarrow\rangle$ and $\sigma^{z}_
{\rm{u}\uparrow}|\uparrow\rangle=+|\uparrow\rangle$. The additional factor $\mathcal{N}_{\rm b\downarrow}\mathcal{N}_{\rm u\uparrow}$ explicitly projects the spin-spin interaction onto the doubly occupied sector. The underlying physical picture is that the spin-spin interaction should only be operative when both QDs are occupied. An empty QD does not carry an electron spin and, therefore, should not contribute a spin-dependent interaction energy. Hence, the spin-spin term vanishes identically for $\ket{00}$, $\ket{\downarrow0}$, and $\ket{0\uparrow}$ states
because one of the two number operators has eigenvalue zero. For the state \(|\downarrow\uparrow\rangle\), both number operators have eigenvalue one, and therefore the spin-spin interaction term contributes an amount $\kappa_s
\left(-1\right)\left(+1\right)
=-\kappa_s.$ Due to strong capacitive coupling~\cite{wang2022cycleflux,gupt2024graph,shuvadip2022univarsal}, the system Hamiltonian $H_{\rm s}$ is diagonal in the tensor-product eigenbasis of the individual QD number operators. The four occupation states ${|0\rangle,\ket{\downarrow}}\otimes{|0\rangle,\ket{\uparrow}}$ are labelled as $|\mathbb{A}\rangle=\ket{00}$, $|\mathbb{B}\rangle=\ket{\downarrow 0}$, $|\mathbb{C}\rangle=\ket{0\uparrow}$, $|\mathbb{D}\rangle=\ket{\downarrow\uparrow}$ 
with corresponding eigenenergies $\varepsilon_\mathbb{A}=0$, $\varepsilon_\mathbb{B}=\varepsilon_{\rm b}$, $\varepsilon_\mathbb{C}=\varepsilon_{\rm u}$, and $\varepsilon_\mathbb{D}=\varepsilon_{\rm b}+\varepsilon_{\rm u}+\kappa$ [FIG.~\ref{CQD_ICC Model}(b),~\ref{CQD_ICC Model}(c)] respectively, where $\kappa=\kappa_{\rm c}-\kappa_{\rm s}$; both $\kappa_{\rm c}$ and $\kappa_{\rm s}$ are positive and w.l.o.g, we assume $\varepsilon_{\rm b}<\varepsilon_{\rm u}$. As a result, $\kappa$ can be repulsive ($\kappa>0$) or attractive ($\kappa<0$). When $\kappa<0$ and $|\kappa|>\varepsilon_{\rm b}$, the order of the energy of the states $|\mathbb{C}\rangle$ and $|\mathbb{D}\rangle$ is inverted [FIG.~\ref{CQD_ICC Model}(c)], signaling a breakdown of the symmetry between energy and particle exchange. The central result is that the condition
$-\kappa>\varepsilon_{\rm{b}}>0$ is required for the appearance of ICC in both energy and particle currents. This inequality corresponds to an interchange of the energy levels associated with the states $\ket{0\uparrow}$ ($|\mathbb{C}\rangle$) and $\ket{\downarrow\uparrow}$ ($|\mathbb{D}\rangle$). For repulsive interdot interactions ($\kappa>0$), the eigenstates are ordered such that particle excitation (de-excitation) necessarily implies energy excitation (de-excitation), as shown in Fig.~\ref{CQD_ICC Model}(b). In contrast, when the attractive interaction satisfies $-\kappa > \varepsilon_{\rm b} > 0$, the eigenstate ordering is inverted [Fig.~\ref{CQD_ICC Model}(c)]. A particle excitation process, $\ket{0\uparrow}\rightarrow\ket{\downarrow\uparrow}$, is then accompanied by energetic de-excitation, and vice versa. This breaking of symmetry between energy and particle exchange is the key mechanism that allows one current to flow against its conjugate thermodynamic force while remaining fully consistent with the second law. 

As a result, a genuine ICC can be achieved in both energy-and spin-polarized particle currents when the two mutually parallel thermodynamic forces. This opens up promising avenues for unconventional autonomous nano-thermoelectric engines and refrigerators based on ICC effects in coupled QD systems. In particular, our analysis reveals that attractive interactions between QDs constitute a necessary requirement for realizing such autonomous quantum thermal machines~\cite{hamo2016electron,tabatabaei2018charge,little1964possibility,prawiroatmodjo2017negativeU}.

In summary, this review presents a comprehensive account of the thermodynamics of quantum-coupled transport (QCT). We begin with single transport processes and progress to coupled transport, highlighting the richer thermodynamic behaviour that emerges. We present a detailed thermodynamic analysis of how relevant force–flux pairs can be systematically identified by examining the form of the entropy production rate in the near-equilibrium regime. This methodology is general and can be straightforwardly extended to more complex systems, where a force–flux description may reveal richer thermodynamic behaviour associated with multidimensional transport processes. A subtle yet pertinent point addressed in this review is the distinction between thermodynamic forces and gradients, and which of these quantities truly characterize the thermodynamics of current flow. Although these terms are often used interchangeably, they are not, in general, thermodynamically equivalent. As demonstrated throughout this review, the correct thermodynamic description is rooted in the entropy production rate, where each flux is paired with its conjugate thermodynamic force. This establishes the thermodynamic force and not the mere gradient as the fundamental driver of transport. While gradients may coincide with thermodynamic forces in special cases and thus serve as useful proxies, a rigorous and consistent analysis --- particularly in multi-terminal or complex systems --- must be formulated in terms of properly defined thermodynamic forces. A central result of this review is the detailed analysis of the ICC phenomena. We establish ICC as a rare and counterintuitive thermodynamic effect --- fully consistent with the second law, yet fundamentally distinct from conventional cross-effects --- and clarify the conceptual and physical differences between these two effects. Genuine ICC emerges as a novel feature of coupled transport~\cite{wang2020inverse,ghosh2026inverse}, which is achieved by breaking the symmetry between the energy and particle excitation correspondence through an attractive interaction between the QDs mediated by spin-polarized fermionic reservoirs. While a detailed quantitative analysis is discussed elsewhere~\cite {ghosh2026inverse}, the model provides a promising platform for exploring autonomous thermoelectric heat engines and refrigerators powered by ICC phenomena.

We conclude this summary with the aim of clarifying the key concepts and thermodynamic principles underlying coupled QD thermal transport. Beyond consolidating existing knowledge, we hope that the review presented here --- particularly the role of inverse currents --- will inspire future efforts to design more efficient autonomous quantum engines and refrigerators. In this way, this review seeks to make a modest, yet meaningful contribution to both the theoretical understanding and practical realization of quantum coupled transport.

\section{Outlook}\label{outlook}

Throughout this review, we have analyzed several variants of two-and three-terminal CQD architectures to demonstrate well-known thermodynamic cross-effects, such as the Seebeck and Peltier effects, as well as the more recently discovered ICC phenomena, from a rigorous thermodynamic perspective. Although the present review focuses exclusively on positive-temperature regimes, exploring the effects of negative temperatures would be an interesting direction for future research. We further emphasize that, from a practical point of view, it is highly possible to experimentally realize ICC effects in QD devices. This is primarily because an analogous three-terminal CQD setup has already been realized experimentally by Thierschmann \textit{et al.} as a model of a nanoscale thermoelectric energy harvester~\cite{thierschmann2015three}. While two-terminal setups are used to analyze thermoelectricity at the nano-scale, they pose serious challenges in \textit{resolving} the heat current from the particle current. By spatially separating the heat reservoir from the conductor channel, the three-terminal setup allows us to independently \textit{control} the direction of particle current without affecting the heat current direction. In Ref.~\cite{thierschmann2015three}, the device was fabricated from a GaAs/AlGaAs wafer containing a 2D electron gas (2DEG). To establish reliable electrical connections with the 2DEG, annealed Au/Ge pads were used inside the electronic contacts. Though an isolated electron produces a repulsive Coulomb interaction, embedding it in an appropriately engineered environment that reverses the sign of its Coulomb potential can make the interaction attractive between electrons~\cite{hamo2016electron,tabatabaei2018charge,little1964possibility,prawiroatmodjo2017negativeU}. Recently, such an attractive interaction was demonstrated by Hamo \textit{et al.} in a carbon-nanotube double QD coupled to a charge qubit~\cite{hamo2016electron}. In this setup, the coupled QDs and the qubit were fabricated on separate microchips and arranged in a perpendicular geometry. When the qubit is brought sufficiently close to the QDs, it induces an attractive interaction between the electrons in the dots. The oscillating polarization field of the qubit effectively ``dresses'' the electronic potentials, thereby favoring doubly occupied states over singly occupied ones~\cite{hamo2016electron,tabatabaei2018charge,little1964possibility,prawiroatmodjo2017negativeU}, as schematically illustrated in Fig.~\ref{CQD_ICC Model}(c). We expect that the high degree of tunability of such platforms makes it particularly suitable for realizing ICC in coupled QDs.

Moreover, similar setups could potentially be used to demonstrate spin-thermoelectric effects in a junction between a magnetic material and a conductor. Typically, a ferrimagnetic insulator such as yttrium iron garnet (YIG) and a paramagnetic metal such as Pt are used for this purpose due to their thermally excited dynamics and localized magnetic moments~\cite{uchida2021transport}. While circuit-QED magneto-spectroscopy can be utilized to realize spinful states~\cite{contamin2023inhomogeneous}, the resulting spin-bias voltage can be measured via the Hanle method~\cite{breton2011thermal} or converted into an electrical voltage through the inverse Hall effect~\cite{uchida2010spin}. Quantum transport through spin-polarized reservoirs highlights the potential for more versatile spin-thermoelectric devices beyond conventional ones, with their novelty arising from the counterintuitive thermodynamic behavior of ICC. Interestingly, an autonomous circular heat engine inspired by ICC in classical systems~\cite{wang2020inverse} has recently been proposed by Benenti et al.~\cite{benenti2022autonomous}. We therefore anticipate that the present findings could contribute to the advancement of ICC-assisted autonomous quantum thermal devices and stimulate further experimental and theoretical investigations in this direction.

However, one natural question arises: are spin-polarized reservoirs essential for the mechanism of ICC, or do they represent only one possible implementation? Here, we address this question. It has been shown by Wang \textit{et al.}~\cite{wang2020inverse} that ICC can occur even in a classical version of the Lieb-Liniger model~\cite{lieb-liniger1963exact,lieb1963exact}. In their study, the effect was attributed to self-organization induced by the applied forces. Their findings established self-organization as one possible mechanism. In the present CQD model, a reversal of the conventional particle-energy excitation symmetry is identified as a key microscopic ingredient underlying ICC. In conventional transport, energy excitation is typically associated with particle excitation, whereas in the ICC regime, particle excitation can instead be accompanied by energy stabilization. This reversal may be viewed as a dynamical breaking of the conventional particle-energy excitation correspondence and provides a microscopic basis for the appearance of ICC. Based on these observations, we \textit{conjecture} that some form of symmetry breaking or
self-organization may constitute a generic mechanism behind ICC in both classical and quantum
systems~\cite{ghosh2026inverse}. Establishing whether such a connection is universal, however, requires a systematic investigation
of a wider class of interacting transport models. Thus, spin-polarized reservoirs should not be regarded as a fundamental requirement. Rather, they provide one possible physical implementation for realizing ICC in quantum systems.

From a thermodynamic perspective, however, the possibility of ICC can be understood directly from Eq.~\eqref{Onsager}. In the linear response regime, the diagonal response alone can't generate a current opposite to its conjugate thermodynamic force, and it requires the coupling between the two transport channels to generate ICC. If we take energy and particle transports, then the second law only restricts
$\dot{\Sigma}=J_{\mathrm{E}} \mathcal{F}_{\rm{E}}+J_{\mathrm{N}} \mathcal{F}_{\rm{N}}\geq 0$ and does not impose each individual contribution, $J_{\mathrm{E}}\mathcal{F}_{\mathrm{E}}$ and $J_{\mathrm{N}}\mathcal{F}_{\mathrm{N}}$, to be positive. Consequently, one current can become negative, i.e., flowing against its conjugate thermodynamic force, provided the other transport channel generates a sufficiently large positive entropy production to compensate for the negative contribution. This result is model-independent, and it is the \textit{only} thermodynamic constraint of fundamental significance that must be satisfied for the emergence of ICC. More generally, ICC exhibits the same richness in both classical and quantum systems: one induced current must flow against two mutually parallel thermodynamic forces, without violating the laws of thermodynamics.

In the present CQD model, the above condition is satisfied by the attractive interdot interaction and thereby provides the microscopic mechanism responsible for the realization
of ICC. This observation suggests that similar interaction-induced mechanisms may also govern the emergence of ICC in other CQD architectures. Interestingly, several studies have reported ICC-like
behavior in alternative QD geometries, particularly in ``two-terminal'' setups in which each
dot is simultaneously coupled to multiple reservoirs, often including the same leads, in addition to being strongly and capacitively coupled to each other~\cite{zhang2021inverse}. Although such configurations might provide an alternative route to inverse transport, their dynamics are more complex than those of the present three-terminal model. More importantly, the inverse currents reported in these two-terminal setups are characterized by temperature and chemical-potential differences. Since these gradients are not, in general, identical to the thermodynamic affinities used to define the entropy production rate, an inverse response with respect to $\Delta T$ or $\Delta\mu$ alone does not necessarily establish genuine ICC. A systematic analysis based on thermodynamic forces and entropy production rate, together with an examination of the underlying nonequilibrium dynamics, is therefore necessary to determine whether the same symmetry-principle or self-organization is responsible for the observed inverse transport in such two-terminal configurations. Such an investigation lies beyond the scope of the present review and is left as an open question for future studies on ICC.

As a last point, let us make some remarks on the robustness of the proposed ICC mechanism beyond the weak-coupling, Markovian, sequential-tunneling approximation. In realistic QD devices, cotunneling and finite level broadening may introduce additional transport channels, while coherence, non-Markovian effects, and stronger system--reservoir coupling can modify the dynamics involving energy and particle transport. Such corrections could change the parameter regime of ICC, although they do not necessarily preclude it. Likewise, nonlinear responses need not destroy ICC; in fact, ICC has been reported beyond the linear-response regime in interacting classical Hamiltonian systems~\cite{wang2020inverse}. Therefore, in realistic QD devices, a systematic treatment of these effects is required to determine the robustness of the ICC mechanism, and this cannot be solely determined from the non-negativity of the entropy production rate.

The proposed quantum models could be useful in several futuristic quantum-thermodynamic devices, such as quantum and topological batteries~\cite{alicki2013entanglement,lu2025topological}, quantum
energy-storage devices and the study of macroscopic particle transport in dissipative quantum systems~\cite{li2026macroscopic}. In particular, the CQD configuration considered here can operate as an efficient thermal diode and switch, with ideal rectification behavior even at very low temperatures~\cite {shuvadip2022univarsal}. This feature suggests a route toward programmable and experimentally realizable quantum thermal logic gates~\cite{ghosh2026quantum} in which logic operations are encoded and processed using heat currents rather than charge currents. One could expect that, by appropriately combining multiple thermal diodes, the basic logic gate operations can, in principle, be extended to cascading architectures, opening the possibility of more complex quantum thermal circuits and heat-based computation. The same CQD platform also provides a natural connection to information-driven transport. With realistic measurement noise and finite feedback delay incorporated into the control protocol, the setup can be extended to realize an information-powered Maxwell-demon refrigerator~\cite{ghosh2026information}. Furthermore, incorporating genuine quantum resources, such as quantum coherence and entanglement, could enable additional functionalities and open avenues toward quantum thermal interferometric devices~\cite{satrya2025heat}. Such extensions underscore the broader potential of interacting CQD systems as versatile platforms for developing advanced quantum thermal management devices and energy-efficient quantum technologies.

\section*{Acknowledgments}
AG acknowledges IITK for infrastructure and partial financial support. SG acknowledges the Ministry of Education, Government of India, for the Prime Minister Research Fellowship (PMRF). SB is thankful to IITK for providing institute fellowship. We sincerely acknowledge the valuable suggestions and constructive comments provided by the anonymous reviewers, which have significantly improved the quality of our paper.

\onecolumngrid 
\appendix

\section{Derivation of the Lindblad Master Equation for general Quantum Dot--Fermionic Reservoir Model}
\label{Appendix-A}

The dynamical evolution of the system is described using the quantum master equation formulated in the interaction picture. The tunneling Hamiltonian between $\alpha$-th QD, and a $\lambda$-th bath, introduced in the main text [cf.~Eq.~\eqref{H-tunn}], defined as
\begin{equation}\label{A1}
\begin{split}
H_{{\rm{T}}}^{\lambda}\equiv H_{{\rm{T}}}^{\alpha\lambda}=\hbar\sum_k[t^{\alpha\lambda}_k c_{\lambda k}^\dagger d_{\alpha}+t^{\alpha\lambda*}_ kd_{\alpha}^{\dagger}c_{\lambda k}].
\end{split}
\end{equation}
Hence, the total tunneling Hamiltonian between the system and the baths can be defined as $H_{\rm{T}}=\sum_{{\lambda}} H_{\rm{T}}^{{\lambda}}$. To derive the quantum master equation, we start with the von Neumann equation for the total density operator $\rho_{\rm tot}(t)$ in the interaction picture,
\begin{equation}\label{A2}
\frac{d}{dt}\rho_{\rm tot}(t)
=
-\frac{i}{\hbar}
\big[ H_{\rm T}(t), \rho_{\rm tot}(t) \big].
\end{equation}
Integrating Eq.~\eqref{A2} and tracing the degrees of freedom of  the reservoir, we find
\begin{equation}\label{A3}
\begin{split}
\frac{d}{dt}\rho_{\rm s}(t)
=
\frac{1}{(i\hbar)^2}
\int_0^t dt'\,
\Tr_{\lambda}
\Big[
H_{\rm T}(t),
\big[
H_{\rm T}(t-t'),
\rho_{\rm tot}(t')
\big]
\Big],
\end{split}
\end{equation}
where $\rho_{\rm s}(t)=\Tr_{\lambda}\{\rho_{\rm tot}(t)\}$ is the reduced system density operator. We further assume
$\Tr_{\lambda}\big[H_{\rm T}(t),\rho_{\rm tot}(0)\big]=0$. Under the Born-Markov approximation and initially uncorrelated system-bath states, Eq.~\eqref{A3} can be rewritten as~\cite{breuer2002book,gupt2022PRE,shuvadip2022univarsal}
\begin{equation}\label{A4}
\begin{split}
\dot{\rho}_{\rm s}(t)
=
\frac{1}{(i\hbar)^2}
\sum_{\lambda,\lambda^\prime}
\int_0^\infty dt'\,
\Tr_{\lambda}
\Big[
H_{\rm T}^{\lambda}(t),
\big[
H_{\rm T}^{\lambda^\prime}(t-t'),
\rho_{\rm s}(t)\otimes\prod_{\lambda}\rho_\lambda
\big]
\Big].
\end{split}
\end{equation}
We assume that $\rho_{\lambda}$ is the initial equilibrium density matrix of the $\lambda$-th bath and we use the fact that ${\rm Tr}_{\lambda}\{{c}_{\lambda}(t)\rho_{\lambda}\}=0={\rm Tr}_{\lambda}\{{c}^\dagger_{\lambda}(t)\rho_{\lambda}\}$
and ${\rm Tr}_{\lambda}\{ [H^{\lambda}_{\rm{T}}(t),[H^{\lambda^{\prime}}_{\rm{T}}(t-t'),{\rho_{\rm{s}}}(t)\otimes \prod_{\lambda}\rho_{\lambda}]]\}=0;{\lambda}\ne{\lambda^{\prime}}$. Here, ${\rm Tr}_{\lambda}$ refers to the trace over each bath degree of freedom and ${\rm Tr}_{\lambda}\{ \rho_{\rm{tot}}(t)\}=\rho_{\rm{s}}(t)$, being the reduced density operator of the system. We further assume that ${\rm Tr}_{\lambda}[H_{\rm{T}}(t),\rho_{\rm{tot}}(0)]=0$. Finally, eliminating the high-frequency oscillating terms by the standard procedure of secular approximation, we arrive at the Lindblad master equation (LME) of the following form
\begin{equation}\label{A6}
\dot{\rho}_{\rm s}(t)
=
\sum_{\lambda}
\mathcal{L}_{\lambda}[\rho_{\rm s}(t)],
\end{equation}
where the Lindblad superoperator $\mathcal{L}_{\lambda}$ is given by
\begin{eqnarray}\label{A7}
\mathcal{L}_{\lambda}[\rho_{\rm s}]
&=&
\sum_{\omega_\alpha>0}
\mathcal{G}_{\lambda}(\omega_\alpha)
\left[
d_\alpha^\dagger(\omega_\alpha)\rho_{\rm s} d_\alpha(\omega_\alpha)
-\frac{1}{2}
\big\{\rho_{\rm s},
d_\alpha(\omega_\alpha)d_\alpha^\dagger(\omega_\alpha)\big\}
\right]
\nonumber\\
&&+
\mathcal{G}_{\lambda}(-\omega_\alpha)
\left[
d_\alpha(\omega_\alpha)\rho_{\rm s} d_\alpha^\dagger(\omega_\alpha)
-\frac{1}{2}
\big\{\rho_{\rm s},
d_\alpha^\dagger(\omega_\alpha)d_\alpha(\omega_\alpha)\big\}
\right].
\end{eqnarray}
Here, the interaction-picture system and bath operators are defined as
\begin{eqnarray}\label{A5}
d_\alpha(t)
=
\sum_{\omega_{\mathbb{ij}}>0}
e^{-i\omega_{\mathbb{ij}} t/\hbar}\,
d_\alpha(\omega_{\mathbb{ij}})\qquad;\qquad 
c_{\lambda k}(t)
=
e^{-i(\epsilon_k^\lambda-\mu_\lambda)t/\hbar}\,
c_{\lambda k},
\end{eqnarray}
where,
$\omega_{\mathbb{ij}}$ is the transition energy between eigenstates
$|\mathbb{i}\rangle$ and $|\mathbb{j}\rangle$. $\mathcal{G}_{\lambda}(\pm \omega_\alpha)
=
\gamma_\lambda(\omega_\alpha)\, f_\lambda^{\pm}(\omega_\alpha),$ is the temperature-dependent bath spectral functions, where $\gamma_\lambda(\omega_\alpha)=2\pi\sum_k
|t_k^{\alpha\lambda}|^2
\delta(\omega_\alpha-\epsilon_k^\lambda).$ denotes the bare tunnelling rate between reservoir $\lambda$ and  $\mathrm{QD}_\alpha$. The functions $f_\lambda^{\pm}(\omega_{\mathbb{ij}})$ represent the Fermi--Dirac distribution functions, corresponding to transitions between
$|\mathbb{i}\rangle\rightarrow|\mathbb{j}\rangle$.

\section{Derivation of the Steady-State Currents}
\label{Appendix-B}

We assume that the system is initially prepared in equilibrium, i.e.,
$\rho_{\rm s}(0)=\rho_{\rm s}^{\rm eq}$ and the equilibrium density operator of the system is~\cite{strasberg2022quantum}
\begin{equation}
\rho_{\rm s}^{\rm eq}
=
\frac{e^{-\bar{\beta}(H_{\rm s}-\bar{\mu}\mathcal{N})}}
{\mathcal{Z}(\bar{\beta},\bar{\mu})},
\end{equation}
where the grand-canonical partition function is given by
\[
\mathcal{Z}(\bar{\beta},\bar{\mu})
=
\Tr\!\left[e^{-\bar{\beta}(H_{\rm s}-\bar{\mu}\mathcal{N})}\right].
\]
Here,
$\bar{\beta} = \sum_\lambda \beta_\lambda, \;
\bar{\mu} = \sum_\lambda \mu_\lambda$, are effective thermodynamic parameters in the presence of multiple reservoirs. Due to interaction with the reservoirs, the system evolves to a time-dependent state $\rho_{\rm s}(t)$, which deviates from equilibrium by an infinitesimally small amount, 
$\delta\rho_{\rm s}(t)
=
\rho_{\rm s}(t)-\rho_{\rm s}(0)
=
\mathcal{O}(\xi)$,
where $\xi$ is a small expansion parameter. As long as the system remains close to equilibrium, the von-Neumann entropy is identified with the thermodynamic entropy,
\begin{equation}\label{S1}
\mathcal{S}_{\rm s}(t)
=
- k_B \Tr_{\rm s}
\!\left[\rho_{\rm s}(t)\ln\rho_{\rm s}(t)\right].
\end{equation}
Following Ref.~\cite{strasberg2022quantum} and keeping terms up to first order in $\xi$, the change in entropy is given ny
\begin{equation}\label{S3}
\begin{split}
\Delta\mathcal{S}_{\rm s}(t)
&=
\mathcal{S}_{\rm s}(t)-\mathcal{S}_{\rm s}(0)
\simeq
- k_B \Tr_{\rm s}
\!\left[\delta\rho_{\rm s}(t)\ln\rho_{\rm s}^{\rm eq}\right]
\\
&=
k_B\beta \Tr_{\rm s}
\!\left[\delta\rho_{\rm s}(t)H_{\rm s}\right]
-
k_B\beta\mu \Tr_{\rm s}
\!\left[\delta\rho_{\rm s}(t)\mathcal{N}\right].
\end{split}
\end{equation}
In the absence of mechanical work, the entropy change is given by
\begin{equation}\label{S4}
\begin{split}
\Delta Q = \Delta E - \mu \Delta N \qquad;\qquad
\Delta S = k_B\beta\,\Delta E - k_B\beta\mu\,\Delta N.
\end{split}
\end{equation}
A comparison between Eqs.~\eqref{S3} and~\eqref{S4}, yields
\begin{equation}\label{S5}
\Delta E
=
\Tr_{\rm s}\!\left[\delta\rho_{\rm s}(t)H_{\rm s}\right]\qquad;\qquad
\Delta N
=
\Tr_{\rm s}\!\left[\delta\rho_{\rm s}(t)\mathcal{N}\right].
\end{equation}
Since $H_{\rm s}$ and $\mathcal{N}$ are time-independent, energy and particle currents flowing into the system are given by
\begin{equation}\label{S6}
\begin{split}
J_{\rm E}(t)
=
\Tr_{\rm s}\!\left[\dot{\rho}_{\rm s}(t)H_{\rm s}\right]\qquad;\qquad
J_{\rm N}(t)
=
\Tr_{\rm s}\!\left[\dot{\rho}_{\rm s}(t)\mathcal{N}\right].
\end{split}
\end{equation}
With the help of the LME,
$\dot{\rho}_{\rm s}(t)=\sum_\lambda\mathcal{L}_\lambda[\rho_{\rm s}(t)]$,
the currents can be expressed as
\begin{equation}\label{S7}
\begin{split}
J_{\rm E}(t)
=
\sum_\lambda
\Tr_{\rm s}\left[
\mathcal{L}_\lambda[\rho_{\rm s}(t)]H_{\rm s}
\right]\qquad;\qquad
J_{\rm N}(t)
=
\sum_\lambda
\Tr_{\rm s}\left[
\mathcal{L}_\lambda[\rho_{\rm s}(t)]\mathcal{N}
\right],
\end{split}
\end{equation}
while the total currents can be decomposed into reservoir-resolved contributions,
\begin{equation}\label{S8}
J_{\rm E}(t)=\sum_\lambda J_{\rm E}^\lambda(t)\qquad;\qquad
J_{\rm N}(t)=\sum_\lambda J_{\rm N}^\lambda(t).
\end{equation}
Here, $J_{\rm E}^\lambda$ and $J_{\rm N}^\lambda$ are taken as positive if energy or particles flow from the reservoir $\lambda$ into the system. Comparing Eqs.~\eqref{S7} and~\eqref{S8}, the reservoir-resolved currents are identified as
\begin{equation}\label{S9}
\begin{split}
J_{\rm E}^\lambda(t)
=
\Tr_{\rm s}\left[
\mathcal{L}_\lambda[\rho_{\rm s}(t)]H_{\rm S}
\right]\qquad;\qquad
J_{\rm N}^\lambda(t)
=
\Tr_{\rm s}\left[
\mathcal{L}_\lambda[\rho_{\rm s}(t)]\mathcal{N}
\right].
\end{split}
\end{equation}
The heat current is defined using the thermodynamic relation $J_{\rm Q}(t)
=
J_{\rm E}(t)-\mu J_{\rm N}(t).$
Decomposing this relation into reservoir contributions 
\begin{equation}\label{S11}
J_{\rm Q}(t)
=
\sum_\lambda J_{\rm Q}^\lambda(t)
=
\sum_\lambda
\Big[
J_{\rm E}^\lambda(t)
-
\mu_\lambda J_{\rm N}^\lambda(t)
\Big],
\end{equation}
yields the reservoir-resolved heat current as
\begin{equation}\label{S12}
J_{\rm Q}^\lambda(t)
=
J_{\rm E}^\lambda(t)
-
\mu_\lambda J_{\rm N}^\lambda(t).
\end{equation}
At steady state, $J_{\rm Q}^\lambda(t)$ becomes time-independent, and Eq.~\eqref{S12} reduces to
\begin{equation}\label{S13}
J_{\rm Q}^\lambda
=
J_{\rm E}^\lambda
-
\mu_\lambda J_{\rm N}^\lambda,
\end{equation}
where steady state $J_{\rm E}^\lambda$ and $J_{\rm N}^\lambda$ are given by
\begin{equation}\label{S14}
\begin{split}
J_{\rm E}^\lambda
=
\Tr_{\rm s}\left[
\mathcal{L}_\lambda[\rho_{ss}]H_{\rm s}
\right]\qquad;\qquad
J_{\rm N}^\lambda
=
\Tr_{\rm s}\left[
\mathcal{L}_\lambda[\rho_{ss}]\mathcal{N}
\right].
\end{split}
\end{equation}


\begin{thebibliography}{189}%
	\makeatletter
	\providecommand \@ifxundefined [1]{%
		\@ifx{#1\undefined}
	}%
	\providecommand \@ifnum [1]{%
		\ifnum #1\expandafter \@firstoftwo
		\else \expandafter \@secondoftwo
		\fi
	}%
	\providecommand \@ifx [1]{%
		\ifx #1\expandafter \@firstoftwo
		\else \expandafter \@secondoftwo
		\fi
	}%
	\providecommand \natexlab [1]{#1}%
	\providecommand \enquote  [1]{``#1''}%
	\providecommand \bibnamefont  [1]{#1}%
	\providecommand \bibfnamefont [1]{#1}%
	\providecommand \citenamefont [1]{#1}%
	\providecommand \href@noop [0]{\@secondoftwo}%
	\providecommand \href [0]{\begingroup \@sanitize@url \@href}%
	\providecommand \@href[1]{\@@startlink{#1}\@@href}%
	\providecommand \@@href[1]{\endgroup#1\@@endlink}%
	\providecommand \@sanitize@url [0]{\catcode `\\12\catcode `\$12\catcode
		`\&12\catcode `\#12\catcode `\^12\catcode `\_12\catcode `\%12\relax}%
	\providecommand \@@startlink[1]{}%
	\providecommand \@@endlink[0]{}%
	\providecommand \url  [0]{\begingroup\@sanitize@url \@url }%
	\providecommand \@url [1]{\endgroup\@href {#1}{\urlprefix }}%
	\providecommand \urlprefix  [0]{URL }%
	\providecommand \Eprint [0]{\href }%
	\providecommand \doibase [0]{https://doi.org/}%
	\providecommand \selectlanguage [0]{\@gobble}%
	\providecommand \bibinfo  [0]{\@secondoftwo}%
	\providecommand \bibfield  [0]{\@secondoftwo}%
	\providecommand \translation [1]{[#1]}%
	\providecommand \BibitemOpen [0]{}%
	\providecommand \bibitemStop [0]{}%
	\providecommand \bibitemNoStop [0]{.\EOS\space}%
	\providecommand \EOS [0]{\spacefactor3000\relax}%
	\providecommand \BibitemShut  [1]{\csname bibitem#1\endcsname}%
	\let\auto@bib@innerbib\@empty
	\bibitem [{\citenamefont {Callen}(1985)}]{callen1985book}%
	\BibitemOpen
	\bibfield  {author} {\bibinfo {author} {\bibfnamefont {H.~B.}\ \bibnamefont
			{Callen}},\ }\href@noop {} {\emph {\bibinfo {title} {{Thermodynamics and an
					Introduction to Thermostatistics}}}},\ \bibinfo {edition} {2nd}\ ed.\
	(\bibinfo  {publisher} {John Wiley \& Sons, Inc.},\ \bibinfo {address} {New
		York},\ \bibinfo {year} {1985})\BibitemShut {NoStop}%
	\bibitem [{\citenamefont {Kondepudi}\ and\ \citenamefont
		{Prigogine}(2015)}]{kondepudi2015book}%
	\BibitemOpen
	\bibfield  {author} {\bibinfo {author} {\bibfnamefont {D.}~\bibnamefont
			{Kondepudi}}\ and\ \bibinfo {author} {\bibfnamefont {I.}~\bibnamefont
			{Prigogine}},\ }\href@noop {} {\emph {\bibinfo {title} {Modern
				Thermodynamics}}},\ \bibinfo {edition} {2nd}\ ed.\ (\bibinfo  {publisher}
	{John Wiley \& Sons Ltd},\ \bibinfo {address} {Chichester},\ \bibinfo {year}
	{2015})\BibitemShut {NoStop}%
	\bibitem [{\citenamefont {Gemmer}\ \emph {et~al.}(2009)\citenamefont {Gemmer},
		\citenamefont {Michel},\ and\ \citenamefont {Mahler}}]{gemmer2009book}%
	\BibitemOpen
	\bibfield  {author} {\bibinfo {author} {\bibfnamefont {J.}~\bibnamefont
			{Gemmer}}, \bibinfo {author} {\bibfnamefont {M.}~\bibnamefont {Michel}},\
		and\ \bibinfo {author} {\bibfnamefont {G.}~\bibnamefont {Mahler}},\ }\href
	{https://doi.org/10.1007/978-3-540-70510-9} {\emph {\bibinfo {title} {Quantum
				Thermodynamics}}},\ \bibinfo {edition} {2nd}\ ed.\ (\bibinfo  {publisher}
	{Springer-Verlag},\ \bibinfo {address} {Berlin Heidelberg},\ \bibinfo {year}
	{2009})\BibitemShut {NoStop}%
	\bibitem [{\citenamefont {Vinjanampathy}\ and\ \citenamefont
		{Anders}(2016)}]{vinjanampathy2016quantum}%
	\BibitemOpen
	\bibfield  {author} {\bibinfo {author} {\bibfnamefont {S.}~\bibnamefont
			{Vinjanampathy}}\ and\ \bibinfo {author} {\bibfnamefont {J.}~\bibnamefont
			{Anders}},\ }\bibfield  {title} {\bibinfo {title} {Quantum thermodynamics},\
	}\href {https://doi.org/10.1080/00107514.2016.1201896} {\bibfield  {journal}
		{\bibinfo  {journal} {Contemp. Phys.}\ }\textbf {\bibinfo {volume} {57}},\
		\bibinfo {pages} {1} (\bibinfo {year} {2016})}\BibitemShut {NoStop}%
	\bibitem [{\citenamefont {Datta}(1997)}]{datta1997electronic}%
	\BibitemOpen
	\bibfield  {author} {\bibinfo {author} {\bibfnamefont {S.}~\bibnamefont
			{Datta}},\ }\href {https://books.google.co.in/books?id=28BC-ofEhvUC} {\emph
		{\bibinfo {title} {Electronic Transport in Mesoscopic Systems}}},\ Cambridge
	Studies in Semiconductor Physi\ (\bibinfo  {publisher} {Cambridge University
		Press},\ \bibinfo {year} {1997})\BibitemShut {NoStop}%
	\bibitem [{\citenamefont {Imry}(2002)}]{imry2002introduction}%
	\BibitemOpen
	\bibfield  {author} {\bibinfo {author} {\bibfnamefont {Y.}~\bibnamefont
			{Imry}},\ }\href@noop {} {\emph {\bibinfo {title} {Introduction to Mesoscopic
				Physics}}}\ (\bibinfo  {publisher} {Oxford University Press},\ \bibinfo
	{year} {2002})\BibitemShut {NoStop}%
	\bibitem [{\citenamefont {de~Groot}\ and\ \citenamefont
		{Mazur}(1962)}]{degroot1962non}%
	\BibitemOpen
	\bibfield  {author} {\bibinfo {author} {\bibfnamefont {S.~R.}\ \bibnamefont
			{de~Groot}}\ and\ \bibinfo {author} {\bibfnamefont {P.}~\bibnamefont
			{Mazur}},\ }\href {https://doi.org/10.1016/C2013-0-02388-7} {\emph {\bibinfo
			{title} {non-equilibrium thermodynamics}}}\ (\bibinfo  {publisher}
	{north-holland},\ \bibinfo {address} {amsterdam},\ \bibinfo {year}
	{1962})\BibitemShut {NoStop}%
	\bibitem [{\citenamefont {Nazarov}\ and\ \citenamefont
		{Blanter}(2009)}]{nazarov2009quantum}%
	\BibitemOpen
	\bibfield  {author} {\bibinfo {author} {\bibfnamefont {Y.}~\bibnamefont
			{Nazarov}}\ and\ \bibinfo {author} {\bibfnamefont {Y.}~\bibnamefont
			{Blanter}},\ }\href {https://books.google.co.in/books?id=bjmXJOFmqZIC} {\emph
		{\bibinfo {title} {Quantum Transport: Introduction to Nanoscience}}}\
	(\bibinfo  {publisher} {Cambridge University Press},\ \bibinfo {year}
	{2009})\BibitemShut {NoStop}%
	\bibitem [{\citenamefont {Di~Ventra}(2008)}]{diventra2008electrical}%
	\BibitemOpen
	\bibfield  {author} {\bibinfo {author} {\bibfnamefont {M.}~\bibnamefont
			{Di~Ventra}},\ }\href@noop {} {\emph {\bibinfo {title} {Electrical Transport
				in Nanoscale Systems}}}\ (\bibinfo  {publisher} {Cambridge University
		Press},\ \bibinfo {year} {2008})\BibitemShut {NoStop}%
	\bibitem [{\citenamefont {Seifert}(2012)}]{seifert2012stochastic}%
	\BibitemOpen
	\bibfield  {author} {\bibinfo {author} {\bibfnamefont {U.}~\bibnamefont
			{Seifert}},\ }\bibfield  {title} {\bibinfo {title} {Stochastic
			thermodynamics, fluctuation theorems and molecular machines},\ }\href
	{https://doi.org/10.1088/0034-4885/75/12/126001} {\bibfield  {journal}
		{\bibinfo  {journal} {Reports on Progress in Physics}\ }\textbf {\bibinfo
			{volume} {75}},\ \bibinfo {pages} {126001} (\bibinfo {year}
		{2012})}\BibitemShut {NoStop}%
	\bibitem [{\citenamefont {Callen}(1948)}]{callen1948the-application}%
	\BibitemOpen
	\bibfield  {author} {\bibinfo {author} {\bibfnamefont {H.~B.}\ \bibnamefont
			{Callen}},\ }\bibfield  {title} {\bibinfo {title} {The application of
			onsager's reciprocal relations to thermoelectric, thermomagnetic, and
			galvanomagnetic effects},\ }\href {https://doi.org/10.1103/PhysRev.73.1349}
	{\bibfield  {journal} {\bibinfo  {journal} {Phys. Rev.}\ }\textbf {\bibinfo
			{volume} {73}},\ \bibinfo {pages} {1349} (\bibinfo {year}
		{1948})}\BibitemShut {NoStop}%
	\bibitem [{\citenamefont {Prigogine}(1961)}]{prigogine1961book}%
	\BibitemOpen
	\bibfield  {author} {\bibinfo {author} {\bibfnamefont {I.}~\bibnamefont
			{Prigogine}},\ }\href@noop {} {\emph {\bibinfo {title} {Introduction to
				Thermodynamics of Irreversible Processes}}},\ \bibinfo {edition} {2nd}\ ed.\
	(\bibinfo  {publisher} {Interscience Publishers},\ \bibinfo {address} {New
		York London},\ \bibinfo {year} {1961})\BibitemShut {NoStop}%
	\bibitem [{\citenamefont
		{Onsager}(1931{\natexlab{a}})}]{onsager1931reciprocal-I}%
	\BibitemOpen
	\bibfield  {author} {\bibinfo {author} {\bibfnamefont {L.}~\bibnamefont
			{Onsager}},\ }\bibfield  {title} {\bibinfo {title} {Reciprocal relations in
			irreversible processes. i.},\ }\href {https://doi.org/10.1103/PhysRev.37.405}
	{\bibfield  {journal} {\bibinfo  {journal} {Phys. Rev.}\ }\textbf {\bibinfo
			{volume} {37}},\ \bibinfo {pages} {405} (\bibinfo {year}
		{1931}{\natexlab{a}})}\BibitemShut {NoStop}%
	\bibitem [{\citenamefont
		{Onsager}(1931{\natexlab{b}})}]{onsager1931reciprocal-II}%
	\BibitemOpen
	\bibfield  {author} {\bibinfo {author} {\bibfnamefont {L.}~\bibnamefont
			{Onsager}},\ }\bibfield  {title} {\bibinfo {title} {Reciprocal relations in
			irreversible processes. ii.},\ }\href
	{https://doi.org/10.1103/PhysRev.38.2265} {\bibfield  {journal} {\bibinfo
			{journal} {Phys. Rev.}\ }\textbf {\bibinfo {volume} {38}},\ \bibinfo {pages}
		{2265} (\bibinfo {year} {1931}{\natexlab{b}})}\BibitemShut {NoStop}%
	\bibitem [{\citenamefont {Monsel}\ \emph {et~al.}(2022)\citenamefont {Monsel},
		\citenamefont {Schulenborg}, \citenamefont {Baquet},\ and\ \citenamefont
		{Splettstoesser}}]{monsel2022geometric}%
	\BibitemOpen
	\bibfield  {author} {\bibinfo {author} {\bibfnamefont {J.}~\bibnamefont
			{Monsel}}, \bibinfo {author} {\bibfnamefont {J.}~\bibnamefont {Schulenborg}},
		\bibinfo {author} {\bibfnamefont {T.}~\bibnamefont {Baquet}},\ and\ \bibinfo
		{author} {\bibfnamefont {J.}~\bibnamefont {Splettstoesser}},\ }\bibfield
	{title} {\bibinfo {title} {Geometric energy transport and refrigeration with
			driven quantum dots},\ }\href {https://doi.org/10.1103/PhysRevB.106.035405}
	{\bibfield  {journal} {\bibinfo  {journal} {Phys. Rev. B}\ }\textbf {\bibinfo
			{volume} {106}},\ \bibinfo {pages} {035405} (\bibinfo {year}
		{2022})}\BibitemShut {NoStop}%
	\bibitem [{\citenamefont {Josefsson}\ \emph {et~al.}(2018)\citenamefont
		{Josefsson}, \citenamefont {Svilans}, \citenamefont {Burke}, \citenamefont
		{Hoffmann}, \citenamefont {Fahlvik~Svensson}, \citenamefont {Leijnse},\ and\
		\citenamefont {Linke}}]{josefsson2018a}%
	\BibitemOpen
	\bibfield  {author} {\bibinfo {author} {\bibfnamefont {M.}~\bibnamefont
			{Josefsson}}, \bibinfo {author} {\bibfnamefont {A.}~\bibnamefont {Svilans}},
		\bibinfo {author} {\bibfnamefont {A.~M.}\ \bibnamefont {Burke}}, \bibinfo
		{author} {\bibfnamefont {E.~A.}\ \bibnamefont {Hoffmann}}, \bibinfo {author}
		{\bibfnamefont {S.}~\bibnamefont {Fahlvik~Svensson}}, \bibinfo {author}
		{\bibfnamefont {M.}~\bibnamefont {Leijnse}},\ and\ \bibinfo {author}
		{\bibfnamefont {H.}~\bibnamefont {Linke}},\ }\bibfield  {title} {\bibinfo
		{title} {A quantum-dot heat engine operating close to the thermodynamic
			efficiency limits},\ }\href@noop {} {\bibfield  {journal} {\bibinfo
			{journal} {Nature Nanotechnology}\ }\textbf {\bibinfo {volume} {13}},\
		\bibinfo {pages} {920} (\bibinfo {year} {2018})}\BibitemShut {NoStop}%
	\bibitem [{\citenamefont {Prete}\ \emph {et~al.}(2019)\citenamefont {Prete},
		\citenamefont {Erdman}, \citenamefont {Demesmaeker},\ and\ \citenamefont
		{Sothmann}}]{prete2019thermoelectric}%
	\BibitemOpen
	\bibfield  {author} {\bibinfo {author} {\bibfnamefont {D.}~\bibnamefont
			{Prete}}, \bibinfo {author} {\bibfnamefont {P.~A.}\ \bibnamefont {Erdman}},
		\bibinfo {author} {\bibfnamefont {S.}~\bibnamefont {Demesmaeker}},\ and\
		\bibinfo {author} {\bibfnamefont {B.}~\bibnamefont {Sothmann}},\ }\bibfield
	{title} {\bibinfo {title} {Thermoelectric conversion at 30 k in inas/inp
			nanowire quantum dots},\ }\href@noop {} {\bibfield  {journal} {\bibinfo
			{journal} {Nano Letters}\ }\textbf {\bibinfo {volume} {19}},\ \bibinfo
		{pages} {3033} (\bibinfo {year} {2019})}\BibitemShut {NoStop}%
	\bibitem [{\citenamefont {S\'anchez}\ and\ \citenamefont
		{B\"uttiker}(2011)}]{sanchez2011optimal}%
	\BibitemOpen
	\bibfield  {author} {\bibinfo {author} {\bibfnamefont {R.}~\bibnamefont
			{S\'anchez}}\ and\ \bibinfo {author} {\bibfnamefont {M.}~\bibnamefont
			{B\"uttiker}},\ }\bibfield  {title} {\bibinfo {title} {Optimal energy quanta
			to current conversion},\ }\href {https://doi.org/10.1103/PhysRevB.83.085428}
	{\bibfield  {journal} {\bibinfo  {journal} {Phys. Rev. B}\ }\textbf {\bibinfo
			{volume} {83}},\ \bibinfo {pages} {085428} (\bibinfo {year}
		{2011})}\BibitemShut {NoStop}%
	\bibitem [{\citenamefont {Zhang}\ \emph {et~al.}(2015)\citenamefont {Zhang},
		\citenamefont {Lin},\ and\ \citenamefont {Chen}}]{zhang2015three}%
	\BibitemOpen
	\bibfield  {author} {\bibinfo {author} {\bibfnamefont {Y.}~\bibnamefont
			{Zhang}}, \bibinfo {author} {\bibfnamefont {G.}~\bibnamefont {Lin}},\ and\
		\bibinfo {author} {\bibfnamefont {J.}~\bibnamefont {Chen}},\ }\bibfield
	{title} {\bibinfo {title} {Three-terminal quantum-dot refrigerators},\ }\href
	{https://doi.org/10.1103/PhysRevE.91.052118} {\bibfield  {journal} {\bibinfo
			{journal} {Phys. Rev. E}\ }\textbf {\bibinfo {volume} {91}},\ \bibinfo
		{pages} {052118} (\bibinfo {year} {2015})}\BibitemShut {NoStop}%
	\bibitem [{\citenamefont {Thierschmann}\ \emph {et~al.}(2015)\citenamefont
		{Thierschmann}, \citenamefont {S{\'a}nchez}, \citenamefont {Sothmann},
		\citenamefont {Arnold}, \citenamefont {Heyn}, \citenamefont {Hansen},
		\citenamefont {Buhmann},\ and\ \citenamefont
		{Molenkamp}}]{thierschmann2015three}%
	\BibitemOpen
	\bibfield  {author} {\bibinfo {author} {\bibfnamefont {H.}~\bibnamefont
			{Thierschmann}}, \bibinfo {author} {\bibfnamefont {R.}~\bibnamefont
			{S{\'a}nchez}}, \bibinfo {author} {\bibfnamefont {B.}~\bibnamefont
			{Sothmann}}, \bibinfo {author} {\bibfnamefont {F.}~\bibnamefont {Arnold}},
		\bibinfo {author} {\bibfnamefont {C.}~\bibnamefont {Heyn}}, \bibinfo {author}
		{\bibfnamefont {W.}~\bibnamefont {Hansen}}, \bibinfo {author} {\bibfnamefont
			{H.}~\bibnamefont {Buhmann}},\ and\ \bibinfo {author} {\bibfnamefont {L.~W.}\
			\bibnamefont {Molenkamp}},\ }\bibfield  {title} {\bibinfo {title}
		{Three-terminal energy harvester with coupled quantum dots},\ }\href
	{https://doi.org/10.1038/nnano.2015.176} {\bibfield  {journal} {\bibinfo
			{journal} {Nature Nanotechnology}\ }\textbf {\bibinfo {volume} {10}},\
		\bibinfo {pages} {854} (\bibinfo {year} {2015})}\BibitemShut {NoStop}%
	\bibitem [{\citenamefont {Thierschmann}\ \emph {et~al.}(2016)\citenamefont
		{Thierschmann}, \citenamefont {S\'anchez}, \citenamefont {Sothmann},
		\citenamefont {Buhmann},\ and\ \citenamefont
		{Molenkamp}}]{thierschmann2016thermoelectrics}%
	\BibitemOpen
	\bibfield  {author} {\bibinfo {author} {\bibfnamefont {H.}~\bibnamefont
			{Thierschmann}}, \bibinfo {author} {\bibfnamefont {R.}~\bibnamefont
			{S\'anchez}}, \bibinfo {author} {\bibfnamefont {B.}~\bibnamefont {Sothmann}},
		\bibinfo {author} {\bibfnamefont {H.}~\bibnamefont {Buhmann}},\ and\ \bibinfo
		{author} {\bibfnamefont {L.~W.}\ \bibnamefont {Molenkamp}},\ }\bibfield
	{title} {\bibinfo {title} {Thermoelectrics with coulomb-coupled quantum
			dots},\ }\href {https://doi.org/10.1016/j.crhy.2016.08.001} {\bibfield
		{journal} {\bibinfo  {journal} {Comptes Rendus. Physique}\ }\textbf {\bibinfo
			{volume} {17}},\ \bibinfo {pages} {1109} (\bibinfo {year}
		{2016})}\BibitemShut {NoStop}%
	\bibitem [{\citenamefont {Erdman}\ \emph {et~al.}(2017)\citenamefont {Erdman},
		\citenamefont {Mazza}, \citenamefont {Bosisio}, \citenamefont {Benenti},
		\citenamefont {Fazio},\ and\ \citenamefont
		{Taddei}}]{erdman2017thermoelectric}%
	\BibitemOpen
	\bibfield  {author} {\bibinfo {author} {\bibfnamefont {P.~A.}\ \bibnamefont
			{Erdman}}, \bibinfo {author} {\bibfnamefont {F.}~\bibnamefont {Mazza}},
		\bibinfo {author} {\bibfnamefont {R.}~\bibnamefont {Bosisio}}, \bibinfo
		{author} {\bibfnamefont {G.}~\bibnamefont {Benenti}}, \bibinfo {author}
		{\bibfnamefont {R.}~\bibnamefont {Fazio}},\ and\ \bibinfo {author}
		{\bibfnamefont {F.}~\bibnamefont {Taddei}},\ }\bibfield  {title} {\bibinfo
		{title} {Thermoelectric properties of an interacting quantum dot based heat
			engine},\ }\href {https://doi.org/10.1103/PhysRevB.95.245432} {\bibfield
		{journal} {\bibinfo  {journal} {Phys. Rev. B}\ }\textbf {\bibinfo {volume}
			{95}},\ \bibinfo {pages} {245432} (\bibinfo {year} {2017})}\BibitemShut
	{NoStop}%
	\bibitem [{\citenamefont {Jurgens}\ \emph {et~al.}(2013)\citenamefont
		{Jurgens}, \citenamefont {Haupt}, \citenamefont {Moskalets},\ and\
		\citenamefont {Splettstoesser}}]{jurgens2013thermoelectric}%
	\BibitemOpen
	\bibfield  {author} {\bibinfo {author} {\bibfnamefont {S.}~\bibnamefont
			{Jurgens}}, \bibinfo {author} {\bibfnamefont {F.}~\bibnamefont {Haupt}},
		\bibinfo {author} {\bibfnamefont {M.}~\bibnamefont {Moskalets}},\ and\
		\bibinfo {author} {\bibfnamefont {J.}~\bibnamefont {Splettstoesser}},\
	}\bibfield  {title} {\bibinfo {title} {Thermoelectric performance of a driven
			double quantum dot},\ }\href {https://doi.org/10.1103/PhysRevB.87.245423}
	{\bibfield  {journal} {\bibinfo  {journal} {Physical Review B}\ }\textbf
		{\bibinfo {volume} {87}},\ \bibinfo {pages} {245423} (\bibinfo {year}
		{2013})}\BibitemShut {NoStop}%
	\bibitem [{\citenamefont {Seebeck}(1822)}]{Seebeck1822magnetische}%
	\BibitemOpen
	\bibfield  {author} {\bibinfo {author} {\bibfnamefont {T.~J.}\ \bibnamefont
			{Seebeck}},\ }\bibfield  {title} {\bibinfo {title} {Magnetische polarisation
			der metalle und erze durch temperatur-differenz},\ }\href@noop {} {\bibfield
		{journal} {\bibinfo  {journal} {Abhandlungen der Königlichen Akademie der
				Wissenschaften zu Berlin}\ ,\ \bibinfo {pages} {265}} (\bibinfo {year}
		{1822})}\BibitemShut {NoStop}%
	\bibitem [{\citenamefont {Thomson}(1856)}]{Kelvin1856on}%
	\BibitemOpen
	\bibfield  {author} {\bibinfo {author} {\bibfnamefont {W.}~\bibnamefont
			{Thomson}},\ }\bibfield  {title} {\bibinfo {title} {On the dynamical theory
			of heat},\ }\href@noop {} {\bibfield  {journal} {\bibinfo  {journal}
			{Transactions of the Royal Society of Edinburgh}\ }\textbf {\bibinfo {volume}
			{21}},\ \bibinfo {pages} {123} (\bibinfo {year} {1856})}\BibitemShut
	{NoStop}%
	\bibitem [{\citenamefont {de~Groot}\ and\ \citenamefont
		{Mazur}(1984)}]{degroot1984non}%
	\BibitemOpen
	\bibfield  {author} {\bibinfo {author} {\bibfnamefont {S.~R.}\ \bibnamefont
			{de~Groot}}\ and\ \bibinfo {author} {\bibfnamefont {P.}~\bibnamefont
			{Mazur}},\ }\href@noop {} {\emph {\bibinfo {title} {Non-Equilibrium
				Thermodynamics}}}\ (\bibinfo  {publisher} {Dover Publications},\ \bibinfo
	{address} {New York},\ \bibinfo {year} {1984})\BibitemShut {NoStop}%
	\bibitem [{\citenamefont {Mahan}\ and\ \citenamefont
		{Sofo}(1996)}]{Mahan1996the}%
	\BibitemOpen
	\bibfield  {author} {\bibinfo {author} {\bibfnamefont {G.~D.}\ \bibnamefont
			{Mahan}}\ and\ \bibinfo {author} {\bibfnamefont {J.~O.}\ \bibnamefont
			{Sofo}},\ }\bibfield  {title} {\bibinfo {title} {The best thermoelectric},\
	}\href@noop {} {\bibfield  {journal} {\bibinfo  {journal} {Proceedings of the
				National Academy of Sciences}\ }\textbf {\bibinfo {volume} {93}},\ \bibinfo
		{pages} {7436} (\bibinfo {year} {1996})}\BibitemShut {NoStop}%
	\bibitem [{\citenamefont {Esposito}\ \emph
		{et~al.}(2010{\natexlab{a}})\citenamefont {Esposito}, \citenamefont
		{Lindenberg},\ and\ \citenamefont {Van~den
			Broeck}}]{esposito2009thermoelectric}%
	\BibitemOpen
	\bibfield  {author} {\bibinfo {author} {\bibfnamefont {M.}~\bibnamefont
			{Esposito}}, \bibinfo {author} {\bibfnamefont {K.}~\bibnamefont
			{Lindenberg}},\ and\ \bibinfo {author} {\bibfnamefont {C.}~\bibnamefont
			{Van~den Broeck}},\ }\bibfield  {title} {\bibinfo {title} {Thermoelectric
			efficiency at maximum power in a quantum dot},\ }\href
	{https://doi.org/10.1088/1367-2630/12/1/013013} {\bibfield  {journal}
		{\bibinfo  {journal} {New Journal of Physics}\ }\textbf {\bibinfo {volume}
			{12}},\ \bibinfo {pages} {013013} (\bibinfo {year}
		{2010}{\natexlab{a}})}\BibitemShut {NoStop}%
	\bibitem [{\citenamefont {Svensson}\ \emph {et~al.}(2012)\citenamefont
		{Svensson}, \citenamefont {Persson}, \citenamefont {Hoffmann}, \citenamefont
		{Nakpathomkun}, \citenamefont {Nilsson}, \citenamefont {Xu}, \citenamefont
		{Samuelson},\ and\ \citenamefont {Linke}}]{svensson2012lineshape}%
	\BibitemOpen
	\bibfield  {author} {\bibinfo {author} {\bibfnamefont {S.~F.}\ \bibnamefont
			{Svensson}}, \bibinfo {author} {\bibfnamefont {A.~I.}\ \bibnamefont
			{Persson}}, \bibinfo {author} {\bibfnamefont {E.~A.}\ \bibnamefont
			{Hoffmann}}, \bibinfo {author} {\bibfnamefont {N.}~\bibnamefont
			{Nakpathomkun}}, \bibinfo {author} {\bibfnamefont {H.~A.}\ \bibnamefont
			{Nilsson}}, \bibinfo {author} {\bibfnamefont {H.~Q.}\ \bibnamefont {Xu}},
		\bibinfo {author} {\bibfnamefont {L.}~\bibnamefont {Samuelson}},\ and\
		\bibinfo {author} {\bibfnamefont {H.}~\bibnamefont {Linke}},\ }\bibfield
	{title} {\bibinfo {title} {Lineshape of the thermopower of quantum dots},\
	}\href {https://doi.org/10.1088/1367-2630/14/3/033041} {\bibfield  {journal}
		{\bibinfo  {journal} {New Journal of Physics}\ }\textbf {\bibinfo {volume}
			{14}},\ \bibinfo {pages} {033041} (\bibinfo {year} {2012})}\BibitemShut
	{NoStop}%
	\bibitem [{\citenamefont {Whitney}(2014)}]{whitney2014most}%
	\BibitemOpen
	\bibfield  {author} {\bibinfo {author} {\bibfnamefont {R.~S.}\ \bibnamefont
			{Whitney}},\ }\bibfield  {title} {\bibinfo {title} {Most efficient quantum
			thermoelectric at finite power output},\ }\href@noop {} {\bibfield  {journal}
		{\bibinfo  {journal} {Physical Review Letters}\ }\textbf {\bibinfo {volume}
			{112}},\ \bibinfo {pages} {130601} (\bibinfo {year} {2014})}\BibitemShut
	{NoStop}%
	\bibitem [{\citenamefont {Peltier}(1834)}]{peltier1834nou}%
	\BibitemOpen
	\bibfield  {author} {\bibinfo {author} {\bibfnamefont {J.~C.~A.}\
			\bibnamefont {Peltier}},\ }\bibfield  {title} {\bibinfo {title} {Nouvelles
			exp{\'e}riences sur la caloricit{\'e} des courants {\'e}lectriques},\
	}\href@noop {} {\bibfield  {journal} {\bibinfo  {journal} {Annales de Chimie
				et de Physique}\ }\textbf {\bibinfo {volume} {56}},\ \bibinfo {pages} {371}
		(\bibinfo {year} {1834})}\BibitemShut {NoStop}%
	\bibitem [{\citenamefont {Jordan}\ \emph {et~al.}(2013)\citenamefont {Jordan},
		\citenamefont {Sothmann}, \citenamefont {S\'anchez},\ and\ \citenamefont
		{B\"uttiker}}]{jordan2013powerful}%
	\BibitemOpen
	\bibfield  {author} {\bibinfo {author} {\bibfnamefont {A.~N.}\ \bibnamefont
			{Jordan}}, \bibinfo {author} {\bibfnamefont {B.}~\bibnamefont {Sothmann}},
		\bibinfo {author} {\bibfnamefont {R.}~\bibnamefont {S\'anchez}},\ and\
		\bibinfo {author} {\bibfnamefont {M.}~\bibnamefont {B\"uttiker}},\ }\bibfield
	{title} {\bibinfo {title} {Powerful and efficient energy harvester with
			resonant-tunneling quantum dots},\ }\href
	{https://doi.org/10.1103/PhysRevB.87.075312} {\bibfield  {journal} {\bibinfo
			{journal} {Phys. Rev. B}\ }\textbf {\bibinfo {volume} {87}},\ \bibinfo
		{pages} {075312} (\bibinfo {year} {2013})}\BibitemShut {NoStop}%
	\bibitem [{\citenamefont {Sothmann}\ \emph {et~al.}(2012)\citenamefont
		{Sothmann}, \citenamefont {S{\'a}nchez},\ and\ \citenamefont
		{Jordan}}]{sothmann2012quantum}%
	\BibitemOpen
	\bibfield  {author} {\bibinfo {author} {\bibfnamefont {B.}~\bibnamefont
			{Sothmann}}, \bibinfo {author} {\bibfnamefont {R.}~\bibnamefont
			{S{\'a}nchez}},\ and\ \bibinfo {author} {\bibfnamefont {A.~N.}\ \bibnamefont
			{Jordan}},\ }\bibfield  {title} {\bibinfo {title} {Quantum dots as thermal
			rectifiers},\ }\href@noop {} {\bibfield  {journal} {\bibinfo  {journal}
			{Nanotechnology}\ }\textbf {\bibinfo {volume} {23}},\ \bibinfo {pages}
		{365101} (\bibinfo {year} {2012})}\BibitemShut {NoStop}%
	\bibitem [{\citenamefont {Donsa}\ \emph {et~al.}(2014)\citenamefont {Donsa},
		\citenamefont {Andergassen},\ and\ \citenamefont {Held}}]{donsa2014double}%
	\BibitemOpen
	\bibfield  {author} {\bibinfo {author} {\bibfnamefont {S.}~\bibnamefont
			{Donsa}}, \bibinfo {author} {\bibfnamefont {S.}~\bibnamefont {Andergassen}},\
		and\ \bibinfo {author} {\bibfnamefont {K.}~\bibnamefont {Held}},\ }\bibfield
	{title} {\bibinfo {title} {Double quantum dot as a minimal thermoelectric
			generator},\ }\href {https://doi.org/10.1103/PhysRevB.89.125103} {\bibfield
		{journal} {\bibinfo  {journal} {Phys. Rev. B}\ }\textbf {\bibinfo {volume}
			{89}},\ \bibinfo {pages} {125103} (\bibinfo {year} {2014})}\BibitemShut
	{NoStop}%
	\bibitem [{\citenamefont {Dar\'e}\ and\ \citenamefont
		{Lombardo}(2017)}]{dare2017powerful}%
	\BibitemOpen
	\bibfield  {author} {\bibinfo {author} {\bibfnamefont {A.-M.}\ \bibnamefont
			{Dar\'e}}\ and\ \bibinfo {author} {\bibfnamefont {P.}~\bibnamefont
			{Lombardo}},\ }\bibfield  {title} {\bibinfo {title} {Powerful coulomb-drag
			thermoelectric engine},\ }\href {https://doi.org/10.1103/PhysRevB.96.115414}
	{\bibfield  {journal} {\bibinfo  {journal} {Phys. Rev. B}\ }\textbf {\bibinfo
			{volume} {96}},\ \bibinfo {pages} {115414} (\bibinfo {year}
		{2017})}\BibitemShut {NoStop}%
	\bibitem [{\citenamefont {Singha}(2020)}]{singha2020realistic}%
	\BibitemOpen
	\bibfield  {author} {\bibinfo {author} {\bibfnamefont {A.}~\bibnamefont
			{Singha}},\ }\bibfield  {title} {\bibinfo {title} {{A realistic non-local
				heat engine based on Coulomb-coupled systems}},\ }\href
	{https://doi.org/10.1063/5.0007347} {\bibfield  {journal} {\bibinfo
			{journal} {Journal of Applied Physics}\ }\textbf {\bibinfo {volume} {127}},\
		\bibinfo {pages} {234903} (\bibinfo {year} {2020})}\BibitemShut {NoStop}%
	\bibitem [{\citenamefont {Erdman}\ \emph {et~al.}(2018)\citenamefont {Erdman},
		\citenamefont {Bhandari}, \citenamefont {Fazio}, \citenamefont {Pekola},\
		and\ \citenamefont {Taddei}}]{erdman2018absorption}%
	\BibitemOpen
	\bibfield  {author} {\bibinfo {author} {\bibfnamefont {P.~A.}\ \bibnamefont
			{Erdman}}, \bibinfo {author} {\bibfnamefont {B.}~\bibnamefont {Bhandari}},
		\bibinfo {author} {\bibfnamefont {R.}~\bibnamefont {Fazio}}, \bibinfo
		{author} {\bibfnamefont {J.~P.}\ \bibnamefont {Pekola}},\ and\ \bibinfo
		{author} {\bibfnamefont {F.}~\bibnamefont {Taddei}},\ }\bibfield  {title}
	{\bibinfo {title} {Absorption refrigerators based on coulomb-coupled
			single-electron systems},\ }\href
	{https://doi.org/10.1103/PhysRevB.98.045433} {\bibfield  {journal} {\bibinfo
			{journal} {Phys. Rev. B}\ }\textbf {\bibinfo {volume} {98}},\ \bibinfo
		{pages} {045433} (\bibinfo {year} {2018})}\BibitemShut {NoStop}%
	\bibitem [{\citenamefont {Dar\'e}(2019)}]{dare2019comparative}%
	\BibitemOpen
	\bibfield  {author} {\bibinfo {author} {\bibfnamefont {A.-M.}\ \bibnamefont
			{Dar\'e}},\ }\bibfield  {title} {\bibinfo {title} {Comparative study of
			heat-driven and power-driven refrigerators with coulomb-coupled quantum
			dots},\ }\href {https://doi.org/10.1103/PhysRevB.100.195427} {\bibfield
		{journal} {\bibinfo  {journal} {Phys. Rev. B}\ }\textbf {\bibinfo {volume}
			{100}},\ \bibinfo {pages} {195427} (\bibinfo {year} {2019})}\BibitemShut
	{NoStop}%
	\bibitem [{\citenamefont {Mukherjee}\ \emph {et~al.}(2020)\citenamefont
		{Mukherjee}, \citenamefont {De},\ and\ \citenamefont
		{Muralidharan}}]{mukherjee2020three}%
	\BibitemOpen
	\bibfield  {author} {\bibinfo {author} {\bibfnamefont {S.}~\bibnamefont
			{Mukherjee}}, \bibinfo {author} {\bibfnamefont {B.}~\bibnamefont {De}},\ and\
		\bibinfo {author} {\bibfnamefont {B.}~\bibnamefont {Muralidharan}},\
	}\bibfield  {title} {\bibinfo {title} {{Three-terminal vibron-coupled hybrid
				quantum dot thermoelectric refrigeration}},\ }\href
	{https://doi.org/10.1063/5.0032215} {\bibfield  {journal} {\bibinfo
			{journal} {Journal of Applied Physics}\ }\textbf {\bibinfo {volume} {128}},\
		\bibinfo {pages} {234303} (\bibinfo {year} {2020})}\BibitemShut {NoStop}%
	\bibitem [{\citenamefont {Barman}\ \emph {et~al.}(2021)\citenamefont {Barman},
		\citenamefont {Halder}, \citenamefont {Varshney}, \citenamefont {Dutta},\
		and\ \citenamefont {Singha}}]{barman2021realistic}%
	\BibitemOpen
	\bibfield  {author} {\bibinfo {author} {\bibfnamefont {A.}~\bibnamefont
			{Barman}}, \bibinfo {author} {\bibfnamefont {S.}~\bibnamefont {Halder}},
		\bibinfo {author} {\bibfnamefont {S.~K.}\ \bibnamefont {Varshney}}, \bibinfo
		{author} {\bibfnamefont {G.}~\bibnamefont {Dutta}},\ and\ \bibinfo {author}
		{\bibfnamefont {A.}~\bibnamefont {Singha}},\ }\bibfield  {title} {\bibinfo
		{title} {Realistic nonlocal refrigeration engine based on coulomb-coupled
			systems},\ }\href {https://doi.org/10.1103/PhysRevE.103.012131} {\bibfield
		{journal} {\bibinfo  {journal} {Phys. Rev. E}\ }\textbf {\bibinfo {volume}
			{103}},\ \bibinfo {pages} {012131} (\bibinfo {year} {2021})}\BibitemShut
	{NoStop}%
	\bibitem [{\citenamefont {Mayrhofer}\ \emph {et~al.}(2021)\citenamefont
		{Mayrhofer}, \citenamefont {Elouard}, \citenamefont {Splettstoesser},\ and\
		\citenamefont {Jordan}}]{mayrhofer2021stochastic}%
	\BibitemOpen
	\bibfield  {author} {\bibinfo {author} {\bibfnamefont {R.~D.}\ \bibnamefont
			{Mayrhofer}}, \bibinfo {author} {\bibfnamefont {C.}~\bibnamefont {Elouard}},
		\bibinfo {author} {\bibfnamefont {J.}~\bibnamefont {Splettstoesser}},\ and\
		\bibinfo {author} {\bibfnamefont {A.~N.}\ \bibnamefont {Jordan}},\ }\bibfield
	{title} {\bibinfo {title} {Stochastic thermodynamic cycles of a mesoscopic
			thermoelectric engine},\ }\href {https://doi.org/10.1103/PhysRevB.103.075404}
	{\bibfield  {journal} {\bibinfo  {journal} {Phys. Rev. B}\ }\textbf {\bibinfo
			{volume} {103}},\ \bibinfo {pages} {075404} (\bibinfo {year}
		{2021})}\BibitemShut {NoStop}%
	\bibitem [{\citenamefont {Wang}\ \emph {et~al.}(2020)\citenamefont {Wang},
		\citenamefont {Casati},\ and\ \citenamefont {Benenti}}]{wang2020inverse}%
	\BibitemOpen
	\bibfield  {author} {\bibinfo {author} {\bibfnamefont {J.}~\bibnamefont
			{Wang}}, \bibinfo {author} {\bibfnamefont {G.}~\bibnamefont {Casati}},\ and\
		\bibinfo {author} {\bibfnamefont {G.}~\bibnamefont {Benenti}},\ }\bibfield
	{title} {\bibinfo {title} {Inverse currents in hamiltonian coupled
			transport},\ }\href {https://doi.org/10.1103/PhysRevLett.124.110607}
	{\bibfield  {journal} {\bibinfo  {journal} {Phys. Rev. Lett.}\ }\textbf
		{\bibinfo {volume} {124}},\ \bibinfo {pages} {110607} (\bibinfo {year}
		{2020})}\BibitemShut {NoStop}%
	\bibitem [{\citenamefont {Ghosh}\ \emph
		{et~al.}(2026{\natexlab{a}})\citenamefont {Ghosh}, \citenamefont {Gupt},\
		and\ \citenamefont {Ghosh}}]{ghosh2026inverse}%
	\BibitemOpen
	\bibfield  {author} {\bibinfo {author} {\bibfnamefont {S.}~\bibnamefont
			{Ghosh}}, \bibinfo {author} {\bibfnamefont {N.}~\bibnamefont {Gupt}},\ and\
		\bibinfo {author} {\bibfnamefont {A.}~\bibnamefont {Ghosh}},\ }\bibfield
	{title} {\bibinfo {title} {Inverse current in coupled transport: A quantum
			thermodynamic model},\ }\href {https://doi.org/10.1103/6r7j-j3n1} {\bibfield
		{journal} {\bibinfo  {journal} {Phys. Rev. Res.}\ }\textbf {\bibinfo {volume}
			{8}},\ \bibinfo {pages} {023166} (\bibinfo {year}
		{2026}{\natexlab{a}})}\BibitemShut {NoStop}%
	\bibitem [{\citenamefont {Fourier}(1822)}]{fourier1822the}%
	\BibitemOpen
	\bibfield  {author} {\bibinfo {author} {\bibfnamefont {J.}~\bibnamefont
			{Fourier}},\ }\href {https://doi.org/10.1017/CBO9780511693146} {\emph
		{\bibinfo {title} {the analytical theory of heat}}}\ (\bibinfo  {publisher}
	{cambridge university press},\ \bibinfo {address} {cambridge},\ \bibinfo
	{year} {1822})\BibitemShut {NoStop}%
	\bibitem [{\citenamefont {Fick}(1855)}]{fick1855on}%
	\BibitemOpen
	\bibfield  {author} {\bibinfo {author} {\bibfnamefont {A.}~\bibnamefont
			{Fick}},\ }\bibfield  {title} {\bibinfo {title} {on liquid diffusion},\
	}\href {https://doi.org/10.1080/14786445508641925} {\bibfield  {journal}
		{\bibinfo  {journal} {philosophical magazine}\ }\textbf {\bibinfo {volume}
			{10}},\ \bibinfo {pages} {30} (\bibinfo {year} {1855})}\BibitemShut {NoStop}%
	\bibitem [{\citenamefont {Ohm}(1827)}]{ohm1827die}%
	\BibitemOpen
	\bibfield  {author} {\bibinfo {author} {\bibfnamefont {G.~S.}\ \bibnamefont
			{Ohm}},\ }\href {https://doi.org/10.1007/978-3-642-61520-3} {\emph {\bibinfo
			{title} {die galvanische kette mathematisch bearbeitet}}}\ (\bibinfo
	{publisher} {riegel und wie{\ss}ner},\ \bibinfo {address} {berlin},\ \bibinfo
	{year} {1827})\BibitemShut {NoStop}%
	\bibitem [{\citenamefont {Ashcroft}\ and\ \citenamefont
		{Mermin}(1976)}]{ashcroft1976solid}%
	\BibitemOpen
	\bibfield  {author} {\bibinfo {author} {\bibfnamefont {N.~W.}\ \bibnamefont
			{Ashcroft}}\ and\ \bibinfo {author} {\bibfnamefont {N.~D.}\ \bibnamefont
			{Mermin}},\ }\href {https://doi.org/10.1002/9780470544846} {\emph {\bibinfo
			{title} {solid state physics}}}\ (\bibinfo  {publisher} {holt, rinehart and
		winston},\ \bibinfo {address} {new york},\ \bibinfo {year}
	{1976})\BibitemShut {NoStop}%
	\bibitem [{\citenamefont {Ziman}(1972)}]{ziman1972principles}%
	\BibitemOpen
	\bibfield  {author} {\bibinfo {author} {\bibfnamefont {J.~M.}\ \bibnamefont
			{Ziman}},\ }\href {https://doi.org/10.1017/CBO9780511813214} {\emph {\bibinfo
			{title} {Principles of the Theory of Solids}}}\ (\bibinfo  {publisher}
	{Cambridge University Press},\ \bibinfo {year} {1972})\BibitemShut {NoStop}%
	\bibitem [{\citenamefont {Datta}(1995)}]{datta1995electronic}%
	\BibitemOpen
	\bibfield  {author} {\bibinfo {author} {\bibfnamefont {S.}~\bibnamefont
			{Datta}},\ }\href {https://doi.org/10.1017/CBO9780511805776} {\emph {\bibinfo
			{title} {Electronic Transport in Mesoscopic Systems}}}\ (\bibinfo
	{publisher} {Cambridge University Press},\ \bibinfo {year}
	{1995})\BibitemShut {NoStop}%
	\bibitem [{\citenamefont {Žutić}\ \emph {et~al.}(2004)\citenamefont
		{Žutić}, \citenamefont {Fabian},\ and\ \citenamefont
		{Das~Sarma}}]{zutzic2004spintronics}%
	\BibitemOpen
	\bibfield  {author} {\bibinfo {author} {\bibfnamefont {I.}~\bibnamefont
			{Žutić}}, \bibinfo {author} {\bibfnamefont {J.}~\bibnamefont {Fabian}},\
		and\ \bibinfo {author} {\bibfnamefont {S.}~\bibnamefont {Das~Sarma}},\
	}\bibfield  {title} {\bibinfo {title} {spintronics: fundamentals and
			applications},\ }\href@noop {} {\bibfield  {journal} {\bibinfo  {journal}
			{reviews of modern physics}\ }\textbf {\bibinfo {volume} {76}},\ \bibinfo
		{pages} {323} (\bibinfo {year} {2004})}\BibitemShut {NoStop}%
	\bibitem [{\citenamefont {Landau}\ and\ \citenamefont
		{Lifshitz}(1987)}]{landau1987fluid}%
	\BibitemOpen
	\bibfield  {author} {\bibinfo {author} {\bibfnamefont {L.~D.}\ \bibnamefont
			{Landau}}\ and\ \bibinfo {author} {\bibfnamefont {E.~M.}\ \bibnamefont
			{Lifshitz}},\ }\href {https://doi.org/10.1016/C2009-0-02002-2} {\emph
		{\bibinfo {title} {fluid mechanics}}},\ \bibinfo {series} {course of
		theoretical physics}, Vol.~\bibinfo {volume} {6}\ (\bibinfo  {publisher}
	{pergamon press},\ \bibinfo {address} {oxford},\ \bibinfo {year}
	{1987})\BibitemShut {NoStop}%
	\bibitem [{\citenamefont {Machura}\ \emph {et~al.}(2007)\citenamefont
		{Machura}, \citenamefont {Kostur}, \citenamefont {Talkner}, \citenamefont
		{\L{}uczka},\ and\ \citenamefont {H\"anggi}}]{machura2007absolute}%
	\BibitemOpen
	\bibfield  {author} {\bibinfo {author} {\bibfnamefont {L.}~\bibnamefont
			{Machura}}, \bibinfo {author} {\bibfnamefont {M.}~\bibnamefont {Kostur}},
		\bibinfo {author} {\bibfnamefont {P.}~\bibnamefont {Talkner}}, \bibinfo
		{author} {\bibfnamefont {J.}~\bibnamefont {\L{}uczka}},\ and\ \bibinfo
		{author} {\bibfnamefont {P.}~\bibnamefont {H\"anggi}},\ }\bibfield  {title}
	{\bibinfo {title} {Absolute negative mobility induced by thermal equilibrium
			fluctuations},\ }\href {https://doi.org/10.1103/PhysRevLett.98.040601}
	{\bibfield  {journal} {\bibinfo  {journal} {Phys. Rev. Lett.}\ }\textbf
		{\bibinfo {volume} {98}},\ \bibinfo {pages} {040601} (\bibinfo {year}
		{2007})}\BibitemShut {NoStop}%
	\bibitem [{\citenamefont {Reimann}(2002)}]{reimann2002brownian}%
	\BibitemOpen
	\bibfield  {author} {\bibinfo {author} {\bibfnamefont {P.}~\bibnamefont
			{Reimann}},\ }\bibfield  {title} {\bibinfo {title} {Brownian motors: noisy
			transport far from equilibrium},\ }\href@noop {} {\bibfield  {journal}
		{\bibinfo  {journal} {Physics Reports}\ }\textbf {\bibinfo {volume} {361}},\
		\bibinfo {pages} {57} (\bibinfo {year} {2002})}\BibitemShut {NoStop}%
	\bibitem [{\citenamefont {Eichhorn}\ \emph {et~al.}(2002)\citenamefont
		{Eichhorn}, \citenamefont {Reimann},\ and\ \citenamefont
		{H\"anggi}}]{eichhorn2002brownian}%
	\BibitemOpen
	\bibfield  {author} {\bibinfo {author} {\bibfnamefont {R.}~\bibnamefont
			{Eichhorn}}, \bibinfo {author} {\bibfnamefont {P.}~\bibnamefont {Reimann}},\
		and\ \bibinfo {author} {\bibfnamefont {P.}~\bibnamefont {H\"anggi}},\
	}\bibfield  {title} {\bibinfo {title} {Brownian motion exhibiting absolute
			negative mobility},\ }\href {https://doi.org/10.1103/PhysRevLett.88.190601}
	{\bibfield  {journal} {\bibinfo  {journal} {Phys. Rev. Lett.}\ }\textbf
		{\bibinfo {volume} {88}},\ \bibinfo {pages} {190601} (\bibinfo {year}
		{2002})}\BibitemShut {NoStop}%
	\bibitem [{\citenamefont {Nagel}\ \emph {et~al.}(2008)\citenamefont {Nagel},
		\citenamefont {Speer}, \citenamefont {Gaber}, \citenamefont {Sterck},
		\citenamefont {Eichhorn}, \citenamefont {Reimann}, \citenamefont {Ilin},
		\citenamefont {Siegel}, \citenamefont {Koelle},\ and\ \citenamefont
		{Kleiner}}]{nagel2008observation}%
	\BibitemOpen
	\bibfield  {author} {\bibinfo {author} {\bibfnamefont {J.}~\bibnamefont
			{Nagel}}, \bibinfo {author} {\bibfnamefont {D.}~\bibnamefont {Speer}},
		\bibinfo {author} {\bibfnamefont {T.}~\bibnamefont {Gaber}}, \bibinfo
		{author} {\bibfnamefont {A.}~\bibnamefont {Sterck}}, \bibinfo {author}
		{\bibfnamefont {R.}~\bibnamefont {Eichhorn}}, \bibinfo {author}
		{\bibfnamefont {P.}~\bibnamefont {Reimann}}, \bibinfo {author} {\bibfnamefont
			{K.}~\bibnamefont {Ilin}}, \bibinfo {author} {\bibfnamefont {M.}~\bibnamefont
			{Siegel}}, \bibinfo {author} {\bibfnamefont {D.}~\bibnamefont {Koelle}},\
		and\ \bibinfo {author} {\bibfnamefont {R.}~\bibnamefont {Kleiner}},\
	}\bibfield  {title} {\bibinfo {title} {Observation of negative absolute
			resistance in a josephson junction},\ }\href
	{https://doi.org/10.1103/PhysRevLett.100.217001} {\bibfield  {journal}
		{\bibinfo  {journal} {Phys. Rev. Lett.}\ }\textbf {\bibinfo {volume} {100}},\
		\bibinfo {pages} {217001} (\bibinfo {year} {2008})}\BibitemShut {NoStop}%
	\bibitem [{\citenamefont {Zhu}\ and\ \citenamefont
		{Balatsky}(2004)}]{zhu2004absolute}%
	\BibitemOpen
	\bibfield  {author} {\bibinfo {author} {\bibfnamefont {J.-X.}\ \bibnamefont
			{Zhu}}\ and\ \bibinfo {author} {\bibfnamefont {A.~V.}\ \bibnamefont
			{Balatsky}},\ }\bibfield  {title} {\bibinfo {title} {Absolute negative
			mobility in quantum transport},\ }\href@noop {} {\bibfield  {journal}
		{\bibinfo  {journal} {Physical Review B}\ }\textbf {\bibinfo {volume} {70}},\
		\bibinfo {pages} {195107} (\bibinfo {year} {2004})}\BibitemShut {NoStop}%
	\bibitem [{\citenamefont {H\"opfel}\ \emph {et~al.}(1986)\citenamefont
		{H\"opfel}, \citenamefont {Shah}, \citenamefont {Wolff},\ and\ \citenamefont
		{Gossard}}]{hopfel1986negative}%
	\BibitemOpen
	\bibfield  {author} {\bibinfo {author} {\bibfnamefont {R.~A.}\ \bibnamefont
			{H\"opfel}}, \bibinfo {author} {\bibfnamefont {J.}~\bibnamefont {Shah}},
		\bibinfo {author} {\bibfnamefont {P.~A.}\ \bibnamefont {Wolff}},\ and\
		\bibinfo {author} {\bibfnamefont {A.~C.}\ \bibnamefont {Gossard}},\
	}\bibfield  {title} {\bibinfo {title} {Negative absolute mobility of minority
			electrons in gaas quantum wells},\ }\href
	{https://doi.org/10.1103/PhysRevLett.56.2736} {\bibfield  {journal} {\bibinfo
			{journal} {Phys. Rev. Lett.}\ }\textbf {\bibinfo {volume} {56}},\ \bibinfo
		{pages} {2736} (\bibinfo {year} {1986})}\BibitemShut {NoStop}%
	\bibitem [{\citenamefont {Keay}\ \emph {et~al.}(1995)\citenamefont {Keay},
		\citenamefont {Zeuner}, \citenamefont {Allen}, \citenamefont {Maranowski},
		\citenamefont {Gossard}, \citenamefont {Bhattacharya},\ and\ \citenamefont
		{Rodwell}}]{keay1995dynamic}%
	\BibitemOpen
	\bibfield  {author} {\bibinfo {author} {\bibfnamefont {B.~J.}\ \bibnamefont
			{Keay}}, \bibinfo {author} {\bibfnamefont {S.}~\bibnamefont {Zeuner}},
		\bibinfo {author} {\bibfnamefont {S.~J.}\ \bibnamefont {Allen}}, \bibinfo
		{author} {\bibfnamefont {K.~D.}\ \bibnamefont {Maranowski}}, \bibinfo
		{author} {\bibfnamefont {A.~C.}\ \bibnamefont {Gossard}}, \bibinfo {author}
		{\bibfnamefont {U.}~\bibnamefont {Bhattacharya}},\ and\ \bibinfo {author}
		{\bibfnamefont {M.~J.~W.}\ \bibnamefont {Rodwell}},\ }\bibfield  {title}
	{\bibinfo {title} {Dynamic localization, absolute negative conductance, and
			stimulated, multiphoton emission in sequential resonant tunneling
			semiconductor superlattices},\ }\href
	{https://doi.org/10.1103/PhysRevLett.75.4102} {\bibfield  {journal} {\bibinfo
			{journal} {Phys. Rev. Lett.}\ }\textbf {\bibinfo {volume} {75}},\ \bibinfo
		{pages} {4102} (\bibinfo {year} {1995})}\BibitemShut {NoStop}%
	\bibitem [{\citenamefont {Sarracino}\ \emph {et~al.}(2016)\citenamefont
		{Sarracino}, \citenamefont {Cecconi}, \citenamefont {Puglisi},\ and\
		\citenamefont {Vulpiani}}]{sarracino2016nonlinear}%
	\BibitemOpen
	\bibfield  {author} {\bibinfo {author} {\bibfnamefont {A.}~\bibnamefont
			{Sarracino}}, \bibinfo {author} {\bibfnamefont {F.}~\bibnamefont {Cecconi}},
		\bibinfo {author} {\bibfnamefont {A.}~\bibnamefont {Puglisi}},\ and\ \bibinfo
		{author} {\bibfnamefont {A.}~\bibnamefont {Vulpiani}},\ }\bibfield  {title}
	{\bibinfo {title} {Nonlinear response of inertial tracers in steady laminar
			flows: Differential and absolute negative mobility},\ }\href
	{https://doi.org/10.1103/PhysRevLett.117.174501} {\bibfield  {journal}
		{\bibinfo  {journal} {Phys. Rev. Lett.}\ }\textbf {\bibinfo {volume} {117}},\
		\bibinfo {pages} {174501} (\bibinfo {year} {2016})}\BibitemShut {NoStop}%
	\bibitem [{\citenamefont {Ghosh}\ \emph {et~al.}(2014)\citenamefont {Ghosh},
		\citenamefont {H\"anggi}, \citenamefont {Marchesoni},\ and\ \citenamefont
		{Nori}}]{ghosh2014giant}%
	\BibitemOpen
	\bibfield  {author} {\bibinfo {author} {\bibfnamefont {P.~K.}\ \bibnamefont
			{Ghosh}}, \bibinfo {author} {\bibfnamefont {P.}~\bibnamefont {H\"anggi}},
		\bibinfo {author} {\bibfnamefont {F.}~\bibnamefont {Marchesoni}},\ and\
		\bibinfo {author} {\bibfnamefont {F.}~\bibnamefont {Nori}},\ }\bibfield
	{title} {\bibinfo {title} {Giant negative mobility of janus particles in a
			corrugated channel},\ }\href {https://doi.org/10.1103/PhysRevE.89.062115}
	{\bibfield  {journal} {\bibinfo  {journal} {Phys. Rev. E}\ }\textbf {\bibinfo
			{volume} {89}},\ \bibinfo {pages} {062115} (\bibinfo {year}
		{2014})}\BibitemShut {NoStop}%
	\bibitem [{\citenamefont {Reguera}\ \emph {et~al.}(2012)\citenamefont
		{Reguera}, \citenamefont {Luque}, \citenamefont {Burada}, \citenamefont
		{Schmid}, \citenamefont {Rub\'{\i}},\ and\ \citenamefont
		{H\"anggi}}]{rugueara2012entropic}%
	\BibitemOpen
	\bibfield  {author} {\bibinfo {author} {\bibfnamefont {D.}~\bibnamefont
			{Reguera}}, \bibinfo {author} {\bibfnamefont {A.}~\bibnamefont {Luque}},
		\bibinfo {author} {\bibfnamefont {P.~S.}\ \bibnamefont {Burada}}, \bibinfo
		{author} {\bibfnamefont {G.}~\bibnamefont {Schmid}}, \bibinfo {author}
		{\bibfnamefont {J.~M.}\ \bibnamefont {Rub\'{\i}}},\ and\ \bibinfo {author}
		{\bibfnamefont {P.}~\bibnamefont {H\"anggi}},\ }\bibfield  {title} {\bibinfo
		{title} {Entropic splitter for particle separation},\ }\href
	{https://doi.org/10.1103/PhysRevLett.108.020604} {\bibfield  {journal}
		{\bibinfo  {journal} {Phys. Rev. Lett.}\ }\textbf {\bibinfo {volume} {108}},\
		\bibinfo {pages} {020604} (\bibinfo {year} {2012})}\BibitemShut {NoStop}%
	\bibitem [{\citenamefont {S\l{}apik}\ \emph {et~al.}(2019)\citenamefont
		{S\l{}apik}, \citenamefont {\L{}uczka}, \citenamefont {H\"anggi},\ and\
		\citenamefont {Spiechowicz}}]{slapik2019tunable}%
	\BibitemOpen
	\bibfield  {author} {\bibinfo {author} {\bibfnamefont {A.}~\bibnamefont
			{S\l{}apik}}, \bibinfo {author} {\bibfnamefont {J.}~\bibnamefont
			{\L{}uczka}}, \bibinfo {author} {\bibfnamefont {P.}~\bibnamefont
			{H\"anggi}},\ and\ \bibinfo {author} {\bibfnamefont {J.}~\bibnamefont
			{Spiechowicz}},\ }\bibfield  {title} {\bibinfo {title} {Tunable mass
			separation via negative mobility},\ }\href
	{https://doi.org/10.1103/PhysRevLett.122.070602} {\bibfield  {journal}
		{\bibinfo  {journal} {Phys. Rev. Lett.}\ }\textbf {\bibinfo {volume} {122}},\
		\bibinfo {pages} {070602} (\bibinfo {year} {2019})}\BibitemShut {NoStop}%
	\bibitem [{\citenamefont {Cleuren}\ and\ \citenamefont {den
			Broeck}(2001)}]{cleuren2001ising}%
	\BibitemOpen
	\bibfield  {author} {\bibinfo {author} {\bibfnamefont {B.}~\bibnamefont
			{Cleuren}}\ and\ \bibinfo {author} {\bibfnamefont {C.~V.}\ \bibnamefont {den
				Broeck}},\ }\bibfield  {title} {\bibinfo {title} {Ising model for a brownian
			donkey},\ }\href {https://doi.org/10.1209/epl/i2001-00274-6} {\bibfield
		{journal} {\bibinfo  {journal} {Europhysics Letters}\ }\textbf {\bibinfo
			{volume} {54}},\ \bibinfo {pages} {1} (\bibinfo {year} {2001})}\BibitemShut
	{NoStop}%
	\bibitem [{\citenamefont {Cleuren}\ and\ \citenamefont {Van~den
			Broeck}(2002)}]{cleuren2002random}%
	\BibitemOpen
	\bibfield  {author} {\bibinfo {author} {\bibfnamefont {B.}~\bibnamefont
			{Cleuren}}\ and\ \bibinfo {author} {\bibfnamefont {C.}~\bibnamefont {Van~den
				Broeck}},\ }\bibfield  {title} {\bibinfo {title} {Random walks with absolute
			negative mobility},\ }\href {https://doi.org/10.1103/PhysRevE.65.030101}
	{\bibfield  {journal} {\bibinfo  {journal} {Phys. Rev. E}\ }\textbf {\bibinfo
			{volume} {65}},\ \bibinfo {pages} {030101} (\bibinfo {year}
		{2002})}\BibitemShut {NoStop}%
	\bibitem [{\citenamefont {Dandogbessi}\ and\ \citenamefont
		{Kenfack}(2015)}]{dandigbessi2015absolute}%
	\BibitemOpen
	\bibfield  {author} {\bibinfo {author} {\bibfnamefont {B.~S.}\ \bibnamefont
			{Dandogbessi}}\ and\ \bibinfo {author} {\bibfnamefont {A.}~\bibnamefont
			{Kenfack}},\ }\bibfield  {title} {\bibinfo {title} {Absolute negative
			mobility induced by potential phase modulation},\ }\href
	{https://doi.org/10.1103/PhysRevE.92.062903} {\bibfield  {journal} {\bibinfo
			{journal} {Phys. Rev. E}\ }\textbf {\bibinfo {volume} {92}},\ \bibinfo
		{pages} {062903} (\bibinfo {year} {2015})}\BibitemShut {NoStop}%
	\bibitem [{\citenamefont {Spiechowicz}\ \emph {et~al.}(2019)\citenamefont
		{Spiechowicz}, \citenamefont {Hänggi},\ and\ \citenamefont
		{Łuczka}}]{spiechowicz2019coexistence}%
	\BibitemOpen
	\bibfield  {author} {\bibinfo {author} {\bibfnamefont {J.}~\bibnamefont
			{Spiechowicz}}, \bibinfo {author} {\bibfnamefont {P.}~\bibnamefont
			{Hänggi}},\ and\ \bibinfo {author} {\bibfnamefont {J.}~\bibnamefont
			{Łuczka}},\ }\bibfield  {title} {\bibinfo {title} {Coexistence of absolute
			negative mobility and anomalous diffusion},\ }\href
	{https://doi.org/10.1088/1367-2630/ab3764} {\bibfield  {journal} {\bibinfo
			{journal} {New Journal of Physics}\ }\textbf {\bibinfo {volume} {21}},\
		\bibinfo {pages} {083029} (\bibinfo {year} {2019})}\BibitemShut {NoStop}%
	\bibitem [{\citenamefont {Ros}\ \emph {et~al.}(2005)\citenamefont {Ros},
		\citenamefont {Eichhorn}, \citenamefont {Regtmeier}, \citenamefont {Duong},
		\citenamefont {Reimann},\ and\ \citenamefont {Anselmetti}}]{ros2005absolute}%
	\BibitemOpen
	\bibfield  {author} {\bibinfo {author} {\bibfnamefont {A.}~\bibnamefont
			{Ros}}, \bibinfo {author} {\bibfnamefont {R.}~\bibnamefont {Eichhorn}},
		\bibinfo {author} {\bibfnamefont {J.}~\bibnamefont {Regtmeier}}, \bibinfo
		{author} {\bibfnamefont {T.~T.}\ \bibnamefont {Duong}}, \bibinfo {author}
		{\bibfnamefont {P.}~\bibnamefont {Reimann}},\ and\ \bibinfo {author}
		{\bibfnamefont {D.}~\bibnamefont {Anselmetti}},\ }\bibfield  {title}
	{\bibinfo {title} {Absolute negative particle mobility},\ }\href
	{https://doi.org/10.1038/436928a} {\bibfield  {journal} {\bibinfo  {journal}
			{Nature}\ }\textbf {\bibinfo {volume} {436}},\ \bibinfo {pages} {928}
		(\bibinfo {year} {2005})}\BibitemShut {NoStop}%
	\bibitem [{\citenamefont {{Casimir}}(1945)}]{casimir1945on}%
	\BibitemOpen
	\bibfield  {author} {\bibinfo {author} {\bibfnamefont {H.~B.}\ \bibnamefont
			{{Casimir}}},\ }\bibfield  {title} {\bibinfo {title} {{On Onsager's Principle
				of Microscopic Reversibility}},\ }\href
	{https://doi.org/10.1103/RevModPhys.17.343} {\bibfield  {journal} {\bibinfo
			{journal} {Reviews of Modern Physics}\ }\textbf {\bibinfo {volume} {17}},\
		\bibinfo {pages} {343} (\bibinfo {year} {1945})}\BibitemShut {NoStop}%
	\bibitem [{\citenamefont {Benenti}\ \emph
		{et~al.}(2017{\natexlab{a}})\citenamefont {Benenti}, \citenamefont {Casati},
		\citenamefont {Saito},\ and\ \citenamefont
		{Whitney}}]{benenti2017fundamental}%
	\BibitemOpen
	\bibfield  {author} {\bibinfo {author} {\bibfnamefont {G.}~\bibnamefont
			{Benenti}}, \bibinfo {author} {\bibfnamefont {G.}~\bibnamefont {Casati}},
		\bibinfo {author} {\bibfnamefont {K.}~\bibnamefont {Saito}},\ and\ \bibinfo
		{author} {\bibfnamefont {R.}~\bibnamefont {Whitney}},\ }\bibfield  {title}
	{\bibinfo {title} {Fundamental aspects of steady-state conversion of heat to
			work at the nanoscale},\ }\href
	{https://doi.org/https://doi.org/10.1016/j.physrep.2017.05.008} {\bibfield
		{journal} {\bibinfo  {journal} {Physics Reports}\ }\textbf {\bibinfo {volume}
			{694}},\ \bibinfo {pages} {1} (\bibinfo {year} {2017}{\natexlab{a}})},\
	\bibinfo {note} {fundamental aspects of steady-state conversion of heat to
		work at the nanoscale}\BibitemShut {NoStop}%
	\bibitem [{\citenamefont {Pekola}\ and\ \citenamefont
		{Karimi}(2021)}]{pekola2021colloquium}%
	\BibitemOpen
	\bibfield  {author} {\bibinfo {author} {\bibfnamefont {J.~P.}\ \bibnamefont
			{Pekola}}\ and\ \bibinfo {author} {\bibfnamefont {B.}~\bibnamefont
			{Karimi}},\ }\bibfield  {title} {\bibinfo {title} {Colloquium: Quantum heat
			transport in condensed matter systems},\ }\href
	{https://doi.org/10.1103/RevModPhys.93.041001} {\bibfield  {journal}
		{\bibinfo  {journal} {Rev. Mod. Phys.}\ }\textbf {\bibinfo {volume} {93}},\
		\bibinfo {pages} {041001} (\bibinfo {year} {2021})}\BibitemShut {NoStop}%
	\bibitem [{\citenamefont {S\'anchez}\ and\ \citenamefont
		{Sothmann}(2019)}]{sanchez2019thermoelectric}%
	\BibitemOpen
	\bibfield  {author} {\bibinfo {author} {\bibfnamefont {R.}~\bibnamefont
			{S\'anchez}}\ and\ \bibinfo {author} {\bibfnamefont {B.}~\bibnamefont
			{Sothmann}},\ }\bibfield  {title} {\bibinfo {title} {Thermoelectric transport
			of mesoscopic conductors coupled to voltage and temperature probes},\
	}\href@noop {} {\bibfield  {journal} {\bibinfo  {journal} {Physical Review
				B}\ }\textbf {\bibinfo {volume} {99}},\ \bibinfo {pages} {245304} (\bibinfo
		{year} {2019})}\BibitemShut {NoStop}%
	\bibitem [{\citenamefont {Kouwenhoven}\ \emph {et~al.}(1997)\citenamefont
		{Kouwenhoven}, \citenamefont {Marcus}, \citenamefont {McEuen}, \citenamefont
		{Tarucha}, \citenamefont {Westervelt},\ and\ \citenamefont
		{Wingreen}}]{kouwenhoven1997electron}%
	\BibitemOpen
	\bibfield  {author} {\bibinfo {author} {\bibfnamefont {L.~P.}\ \bibnamefont
			{Kouwenhoven}}, \bibinfo {author} {\bibfnamefont {C.~M.}\ \bibnamefont
			{Marcus}}, \bibinfo {author} {\bibfnamefont {P.~L.}\ \bibnamefont {McEuen}},
		\bibinfo {author} {\bibfnamefont {S.}~\bibnamefont {Tarucha}}, \bibinfo
		{author} {\bibfnamefont {R.~M.}\ \bibnamefont {Westervelt}},\ and\ \bibinfo
		{author} {\bibfnamefont {N.~S.}\ \bibnamefont {Wingreen}},\ }\bibfield
	{title} {\bibinfo {title} {Electron transport in quantum dots},\ }\href@noop
	{} {\bibfield  {journal} {\bibinfo  {journal} {NATO ASI Series E}\ }\textbf
		{\bibinfo {volume} {345}},\ \bibinfo {pages} {105} (\bibinfo {year}
		{1997})}\BibitemShut {NoStop}%
	\bibitem [{\citenamefont {van~der Wiel}\ \emph {et~al.}(2002)\citenamefont
		{van~der Wiel}, \citenamefont {De~Franceschi}, \citenamefont {Elzerman},
		\citenamefont {Fujisawa}, \citenamefont {Tarucha},\ and\ \citenamefont
		{Kouwenhoven}}]{vanDerWiel2002electron}%
	\BibitemOpen
	\bibfield  {author} {\bibinfo {author} {\bibfnamefont {W.~G.}\ \bibnamefont
			{van~der Wiel}}, \bibinfo {author} {\bibfnamefont {S.}~\bibnamefont
			{De~Franceschi}}, \bibinfo {author} {\bibfnamefont {J.~M.}\ \bibnamefont
			{Elzerman}}, \bibinfo {author} {\bibfnamefont {T.}~\bibnamefont {Fujisawa}},
		\bibinfo {author} {\bibfnamefont {S.}~\bibnamefont {Tarucha}},\ and\ \bibinfo
		{author} {\bibfnamefont {L.~P.}\ \bibnamefont {Kouwenhoven}},\ }\bibfield
	{title} {\bibinfo {title} {Electron transport through double quantum dots},\
	}\href@noop {} {\bibfield  {journal} {\bibinfo  {journal} {Reviews of Modern
				Physics}\ }\textbf {\bibinfo {volume} {75}},\ \bibinfo {pages} {1} (\bibinfo
		{year} {2002})}\BibitemShut {NoStop}%
	\bibitem [{\citenamefont {Linke}\ and\ \citenamefont
		{Humphrey}(2015)}]{linke2015harnessing}%
	\BibitemOpen
	\bibfield  {author} {\bibinfo {author} {\bibfnamefont {H.}~\bibnamefont
			{Linke}}\ and\ \bibinfo {author} {\bibfnamefont {T.~E.}\ \bibnamefont
			{Humphrey}},\ }\bibfield  {title} {\bibinfo {title} {Harnessing
			non-equilibrium fluctuations for thermoelectric devices},\ }\href@noop {}
	{\bibfield  {journal} {\bibinfo  {journal} {Physics Today}\ }\textbf
		{\bibinfo {volume} {68}},\ \bibinfo {pages} {32} (\bibinfo {year}
		{2015})}\BibitemShut {NoStop}%
	\bibitem [{\citenamefont {Breuer}\ and\ \citenamefont
		{Petruccione}(2007)}]{breuer2002book}%
	\BibitemOpen
	\bibfield  {author} {\bibinfo {author} {\bibfnamefont {H.-P.}\ \bibnamefont
			{Breuer}}\ and\ \bibinfo {author} {\bibfnamefont {F.}~\bibnamefont
			{Petruccione}},\ }\href
	{https://doi.org/10.1093/acprof:oso/9780199213900.001.0001} {\emph {\bibinfo
			{title} {{The Theory of Open Quantum Systems}}}}\ (\bibinfo  {publisher}
	{Oxford University Press},\ \bibinfo {year} {2007})\BibitemShut {NoStop}%
	\bibitem [{\citenamefont {Louisell}(1990)}]{louisell1990book}%
	\BibitemOpen
	\bibfield  {author} {\bibinfo {author} {\bibfnamefont {W.~H.}\ \bibnamefont
			{Louisell}},\ }\href {https://books.google.co.in/books?id=NRlRAAAAMAAJ}
	{\emph {\bibinfo {title} {Quantum Statistical Properties of Radiation}}},\ A
	Wiley-Interscience publication\ (\bibinfo  {publisher} {Wiley},\ \bibinfo
	{address} {New York},\ \bibinfo {year} {1990})\BibitemShut {NoStop}%
	\bibitem [{\citenamefont {Carmichael}(1993)}]{carmichael1993book}%
	\BibitemOpen
	\bibfield  {author} {\bibinfo {author} {\bibfnamefont {H.}~\bibnamefont
			{Carmichael}},\ }\href@noop {} {\emph {\bibinfo {title} {An Open Systems
				Approach to Quantum Optics}}}\ (\bibinfo  {publisher} {Springer-Verlag},\
	\bibinfo {address} {Berlin Heidelberg},\ \bibinfo {year} {1993})\BibitemShut
	{NoStop}%
	\bibitem [{\citenamefont {Gelbwaser-Klimovsky}\ \emph
		{et~al.}(2015)\citenamefont {Gelbwaser-Klimovsky}, \citenamefont {Niedenzu},\
		and\ \citenamefont {Kurizki}}]{gelbwaser2015thermodynamics}%
	\BibitemOpen
	\bibfield  {author} {\bibinfo {author} {\bibfnamefont {D.}~\bibnamefont
			{Gelbwaser-Klimovsky}}, \bibinfo {author} {\bibfnamefont {W.}~\bibnamefont
			{Niedenzu}},\ and\ \bibinfo {author} {\bibfnamefont {G.}~\bibnamefont
			{Kurizki}},\ }\bibfield  {title} {\bibinfo {title} {Thermodynamics of quantum
			systems under dynamical control},\ }\href
	{https://doi.org/10.1016/bs.aamop.2015.07.002} {\bibfield  {journal}
		{\bibinfo  {journal} {Adv. At. Mol. Opt. Phys.}\ }\textbf {\bibinfo {volume}
			{64}},\ \bibinfo {pages} {329} (\bibinfo {year} {2015})}\BibitemShut
	{NoStop}%
	\bibitem [{\citenamefont {Kurizki}\ and\ \citenamefont
		{Kofman}(2022)}]{kurizki_kofman_2022}%
	\BibitemOpen
	\bibfield  {author} {\bibinfo {author} {\bibfnamefont {G.}~\bibnamefont
			{Kurizki}}\ and\ \bibinfo {author} {\bibfnamefont {A.~G.}\ \bibnamefont
			{Kofman}},\ }\href {https://doi.org/10.1017/9781316798454} {\emph {\bibinfo
			{title} {Thermodynamics and Control of Open Quantum Systems}}}\ (\bibinfo
	{publisher} {Cambridge University Press},\ \bibinfo {year}
	{2022})\BibitemShut {NoStop}%
	\bibitem [{\citenamefont {Ghosh}\ \emph {et~al.}(2017)\citenamefont {Ghosh},
		\citenamefont {Latune}, \citenamefont {Davidovich},\ and\ \citenamefont
		{Kurizki}}]{ghosh2017catalysis}%
	\BibitemOpen
	\bibfield  {author} {\bibinfo {author} {\bibfnamefont {A.}~\bibnamefont
			{Ghosh}}, \bibinfo {author} {\bibfnamefont {C.~L.}\ \bibnamefont {Latune}},
		\bibinfo {author} {\bibfnamefont {L.}~\bibnamefont {Davidovich}},\ and\
		\bibinfo {author} {\bibfnamefont {G.}~\bibnamefont {Kurizki}},\ }\bibfield
	{title} {\bibinfo {title} {Catalysis of heat-to-work conversion in quantum
			machines},\ }\href {https://doi.org/10.1073/pnas.1711381114} {\bibfield
		{journal} {\bibinfo  {journal} {Proc. Natl. Acad. Sci. U.S.A.}\ }\textbf
		{\bibinfo {volume} {114}},\ \bibinfo {pages} {12156} (\bibinfo {year}
		{2017})}\BibitemShut {NoStop}%
	\bibitem [{\citenamefont {Gupt}\ \emph {et~al.}(2022)\citenamefont {Gupt},
		\citenamefont {Bhattacharyya}, \citenamefont {Das}, \citenamefont {Datta},
		\citenamefont {Mukherjee},\ and\ \citenamefont {Ghosh}}]{gupt2022PRE}%
	\BibitemOpen
	\bibfield  {author} {\bibinfo {author} {\bibfnamefont {N.}~\bibnamefont
			{Gupt}}, \bibinfo {author} {\bibfnamefont {S.}~\bibnamefont {Bhattacharyya}},
		\bibinfo {author} {\bibfnamefont {B.}~\bibnamefont {Das}}, \bibinfo {author}
		{\bibfnamefont {S.}~\bibnamefont {Datta}}, \bibinfo {author} {\bibfnamefont
			{V.}~\bibnamefont {Mukherjee}},\ and\ \bibinfo {author} {\bibfnamefont
			{A.}~\bibnamefont {Ghosh}},\ }\bibfield  {title} {\bibinfo {title} {Floquet
			quantum thermal transistor},\ }\href
	{https://doi.org/10.1103/PhysRevE.106.024110} {\bibfield  {journal} {\bibinfo
			{journal} {Phys. Rev. E}\ }\textbf {\bibinfo {volume} {106}},\ \bibinfo
		{pages} {024110} (\bibinfo {year} {2022})}\BibitemShut {NoStop}%
	\bibitem [{\citenamefont {Gupt}\ \emph {et~al.}(2023)\citenamefont {Gupt},
		\citenamefont {Ghosh},\ and\ \citenamefont {Ghosh}}]{gupt2023topranked}%
	\BibitemOpen
	\bibfield  {author} {\bibinfo {author} {\bibfnamefont {N.}~\bibnamefont
			{Gupt}}, \bibinfo {author} {\bibfnamefont {S.}~\bibnamefont {Ghosh}},\ and\
		\bibinfo {author} {\bibfnamefont {A.}~\bibnamefont {Ghosh}},\ }\bibfield
	{title} {\bibinfo {title} {Top-ranked cycle flux network analysis of
			molecular photocells},\ }\href {https://doi.org/10.1103/PhysRevE.108.034305}
	{\bibfield  {journal} {\bibinfo  {journal} {Phys. Rev. E}\ }\textbf {\bibinfo
			{volume} {108}},\ \bibinfo {pages} {034305} (\bibinfo {year}
		{2023})}\BibitemShut {NoStop}%
	\bibitem [{\citenamefont {Gupt}\ \emph {et~al.}(2024)\citenamefont {Gupt},
		\citenamefont {Ghosh},\ and\ \citenamefont {Ghosh}}]{gupt2024graph}%
	\BibitemOpen
	\bibfield  {author} {\bibinfo {author} {\bibfnamefont {N.}~\bibnamefont
			{Gupt}}, \bibinfo {author} {\bibfnamefont {S.}~\bibnamefont {Ghosh}},\ and\
		\bibinfo {author} {\bibfnamefont {A.}~\bibnamefont {Ghosh}},\ }\bibfield
	{title} {\bibinfo {title} {Graph theoretic analysis of three-terminal quantum
			dot thermocouples: Onsager relations and spin-thermoelectric effects},\
	}\href {https://doi.org/10.1103/PhysRevB.109.125124} {\bibfield  {journal}
		{\bibinfo  {journal} {Phys. Rev. B}\ }\textbf {\bibinfo {volume} {109}},\
		\bibinfo {pages} {125124} (\bibinfo {year} {2024})}\BibitemShut {NoStop}%
	\bibitem [{\citenamefont {Ghosh}\ \emph {et~al.}(2022)\citenamefont {Ghosh},
		\citenamefont {Gupt},\ and\ \citenamefont {Ghosh}}]{shuvadip2022univarsal}%
	\BibitemOpen
	\bibfield  {author} {\bibinfo {author} {\bibfnamefont {S.}~\bibnamefont
			{Ghosh}}, \bibinfo {author} {\bibfnamefont {N.}~\bibnamefont {Gupt}},\ and\
		\bibinfo {author} {\bibfnamefont {A.}~\bibnamefont {Ghosh}},\ }\bibfield
	{title} {\bibinfo {title} {Universal behavior of the coulomb-coupled
			fermionic thermal diode},\ }\href {https://www.mdpi.com/1099-4300/24/12/1810}
	{\bibfield  {journal} {\bibinfo  {journal} {Entropy}\ }\textbf {\bibinfo
			{volume} {24}} (\bibinfo {year} {2022})}\BibitemShut {NoStop}%
	\bibitem [{\citenamefont {Jaliel}\ \emph {et~al.}(2019)\citenamefont {Jaliel},
		\citenamefont {Demesmaeker},\ and\ \citenamefont
		{Sothmann}}]{jaliel2019experimental}%
	\BibitemOpen
	\bibfield  {author} {\bibinfo {author} {\bibfnamefont {G.}~\bibnamefont
			{Jaliel}}, \bibinfo {author} {\bibfnamefont {S.}~\bibnamefont
			{Demesmaeker}},\ and\ \bibinfo {author} {\bibfnamefont {B.}~\bibnamefont
			{Sothmann}},\ }\bibfield  {title} {\bibinfo {title} {Experimental realization
			of a quantum-dot energy harvester},\ }\href@noop {} {\bibfield  {journal}
		{\bibinfo  {journal} {Physical Review Letters}\ }\textbf {\bibinfo {volume}
			{123}},\ \bibinfo {pages} {117701} (\bibinfo {year} {2019})}\BibitemShut
	{NoStop}%
	\bibitem [{\citenamefont {Nakpathomkun}\ \emph {et~al.}(2010)\citenamefont
		{Nakpathomkun}, \citenamefont {Xu},\ and\ \citenamefont
		{Linke}}]{nakpathomkun2010thermoelctric}%
	\BibitemOpen
	\bibfield  {author} {\bibinfo {author} {\bibfnamefont {N.}~\bibnamefont
			{Nakpathomkun}}, \bibinfo {author} {\bibfnamefont {H.~Q.}\ \bibnamefont
			{Xu}},\ and\ \bibinfo {author} {\bibfnamefont {H.}~\bibnamefont {Linke}},\
	}\bibfield  {title} {\bibinfo {title} {Thermoelectric efficiency at maximum
			power in low-dimensional systems},\ }\href
	{https://doi.org/10.1103/PhysRevB.82.235428} {\bibfield  {journal} {\bibinfo
			{journal} {Phys. Rev. B}\ }\textbf {\bibinfo {volume} {82}},\ \bibinfo
		{pages} {235428} (\bibinfo {year} {2010})}\BibitemShut {NoStop}%
	\bibitem [{\citenamefont {Josefsson}\ \emph {et~al.}(2019)\citenamefont
		{Josefsson}, \citenamefont {Sothmann}, \citenamefont {Leijnse},\ and\
		\citenamefont {Linke}}]{josefsson2019optimal}%
	\BibitemOpen
	\bibfield  {author} {\bibinfo {author} {\bibfnamefont {M.}~\bibnamefont
			{Josefsson}}, \bibinfo {author} {\bibfnamefont {B.}~\bibnamefont {Sothmann}},
		\bibinfo {author} {\bibfnamefont {M.}~\bibnamefont {Leijnse}},\ and\ \bibinfo
		{author} {\bibfnamefont {H.}~\bibnamefont {Linke}},\ }\bibfield  {title}
	{\bibinfo {title} {Optimal power and efficiency of single quantum dot heat
			engines: theory and experiment},\ }\href@noop {} {\bibfield  {journal}
		{\bibinfo  {journal} {Physical Review B}\ }\textbf {\bibinfo {volume} {99}},\
		\bibinfo {pages} {235432} (\bibinfo {year} {2019})}\BibitemShut {NoStop}%
	\bibitem [{\citenamefont {McConnell}\ and\ \citenamefont
		{Nazir}(2022)}]{mcConnell2022strong}%
	\BibitemOpen
	\bibfield  {author} {\bibinfo {author} {\bibfnamefont {C.}~\bibnamefont
			{McConnell}}\ and\ \bibinfo {author} {\bibfnamefont {A.}~\bibnamefont
			{Nazir}},\ }\bibfield  {title} {\bibinfo {title} {Strong coupling in
			thermoelectric nanojunctions: A reaction coordinate framework},\ }\href
	{https://doi.org/10.1088/1367-2630/ac5795} {\bibfield  {journal} {\bibinfo
			{journal} {New Journal of Physics}\ }\textbf {\bibinfo {volume} {24}},\
		\bibinfo {pages} {025002} (\bibinfo {year} {2022})}\BibitemShut {NoStop}%
	\bibitem [{\citenamefont {Strasberg}\ \emph {et~al.}(2013)\citenamefont
		{Strasberg}, \citenamefont {Schaller}, \citenamefont {Brandes},\ and\
		\citenamefont {Esposito}}]{strasberg2013thermodynamic}%
	\BibitemOpen
	\bibfield  {author} {\bibinfo {author} {\bibfnamefont {P.}~\bibnamefont
			{Strasberg}}, \bibinfo {author} {\bibfnamefont {G.}~\bibnamefont {Schaller}},
		\bibinfo {author} {\bibfnamefont {T.}~\bibnamefont {Brandes}},\ and\ \bibinfo
		{author} {\bibfnamefont {M.}~\bibnamefont {Esposito}},\ }\bibfield  {title}
	{\bibinfo {title} {Thermodynamics of a physical model implementing a maxwell
			demon},\ }\href {https://doi.org/10.1103/PhysRevLett.110.040601} {\bibfield
		{journal} {\bibinfo  {journal} {Phys. Rev. Lett.}\ }\textbf {\bibinfo
			{volume} {110}},\ \bibinfo {pages} {040601} (\bibinfo {year}
		{2013})}\BibitemShut {NoStop}%
	\bibitem [{\citenamefont {Esposito}\ and\ \citenamefont
		{Schaller}(2012)}]{esposito2012stochastic}%
	\BibitemOpen
	\bibfield  {author} {\bibinfo {author} {\bibfnamefont {M.}~\bibnamefont
			{Esposito}}\ and\ \bibinfo {author} {\bibfnamefont {G.}~\bibnamefont
			{Schaller}},\ }\bibfield  {title} {\bibinfo {title} {Stochastic
			thermodynamics for feedback-controlled systems},\ }\href@noop {} {\bibfield
		{journal} {\bibinfo  {journal} {Europhysics Letters}\ }\textbf {\bibinfo
			{volume} {99}},\ \bibinfo {pages} {30003} (\bibinfo {year}
		{2012})}\BibitemShut {NoStop}%
	\bibitem [{\citenamefont {Annby-Andersson}\ \emph {et~al.}(2022)\citenamefont
		{Annby-Andersson}, \citenamefont {Bakhshinezhad}, \citenamefont
		{Bhattacharyya}, \citenamefont {De~Sousa}, \citenamefont {Jarzynski},
		\citenamefont {Samuelsson},\ and\ \citenamefont {Potts}}]{annby2022quantum}%
	\BibitemOpen
	\bibfield  {author} {\bibinfo {author} {\bibfnamefont {B.}~\bibnamefont
			{Annby-Andersson}}, \bibinfo {author} {\bibfnamefont {F.}~\bibnamefont
			{Bakhshinezhad}}, \bibinfo {author} {\bibfnamefont {D.}~\bibnamefont
			{Bhattacharyya}}, \bibinfo {author} {\bibfnamefont {G.}~\bibnamefont
			{De~Sousa}}, \bibinfo {author} {\bibfnamefont {C.}~\bibnamefont {Jarzynski}},
		\bibinfo {author} {\bibfnamefont {P.}~\bibnamefont {Samuelsson}},\ and\
		\bibinfo {author} {\bibfnamefont {P.~P.}\ \bibnamefont {Potts}},\ }\bibfield
	{title} {\bibinfo {title} {Quantum fokker-planck master equation for
			continuous feedback control},\ }\href
	{https://doi.org/10.1103/PhysRevLett.129.050401} {\bibfield  {journal}
		{\bibinfo  {journal} {Phys. Rev. Lett.}\ }\textbf {\bibinfo {volume} {129}},\
		\bibinfo {pages} {050401} (\bibinfo {year} {2022})}\BibitemShut {NoStop}%
	\bibitem [{\citenamefont {Annby-Andersson}\ \emph {et~al.}(2024)\citenamefont
		{Annby-Andersson}, \citenamefont {Bhattacharyya}, \citenamefont
		{Bakhshinezhad}, \citenamefont {Holst}, \citenamefont {De~Sousa},
		\citenamefont {Jarzynski}, \citenamefont {Samuelsson},\ and\ \citenamefont
		{Potts}}]{annby2024maxwell}%
	\BibitemOpen
	\bibfield  {author} {\bibinfo {author} {\bibfnamefont {B.}~\bibnamefont
			{Annby-Andersson}}, \bibinfo {author} {\bibfnamefont {D.}~\bibnamefont
			{Bhattacharyya}}, \bibinfo {author} {\bibfnamefont {P.}~\bibnamefont
			{Bakhshinezhad}}, \bibinfo {author} {\bibfnamefont {D.}~\bibnamefont
			{Holst}}, \bibinfo {author} {\bibfnamefont {G.}~\bibnamefont {De~Sousa}},
		\bibinfo {author} {\bibfnamefont {C.}~\bibnamefont {Jarzynski}}, \bibinfo
		{author} {\bibfnamefont {P.}~\bibnamefont {Samuelsson}},\ and\ \bibinfo
		{author} {\bibfnamefont {P.~P.}\ \bibnamefont {Potts}},\ }\bibfield  {title}
	{\bibinfo {title} {Maxwell's demon across the quantum-to-classical
			transition},\ }\href {https://doi.org/10.1103/PhysRevResearch.6.043216}
	{\bibfield  {journal} {\bibinfo  {journal} {Phys. Rev. Res.}\ }\textbf
		{\bibinfo {volume} {6}},\ \bibinfo {pages} {043216} (\bibinfo {year}
		{2024})}\BibitemShut {NoStop}%
	\bibitem [{\citenamefont {Kutvonen}\ \emph {et~al.}(2016)\citenamefont
		{Kutvonen}, \citenamefont {Sagawa},\ and\ \citenamefont
		{Ala-Nissila}}]{kutvonen2016thermodynamics}%
	\BibitemOpen
	\bibfield  {author} {\bibinfo {author} {\bibfnamefont {A.}~\bibnamefont
			{Kutvonen}}, \bibinfo {author} {\bibfnamefont {T.}~\bibnamefont {Sagawa}},\
		and\ \bibinfo {author} {\bibfnamefont {T.}~\bibnamefont {Ala-Nissila}},\
	}\bibfield  {title} {\bibinfo {title} {Thermodynamics of information exchange
			between two coupled quantum dots},\ }\href
	{https://doi.org/10.1103/PhysRevE.93.032147} {\bibfield  {journal} {\bibinfo
			{journal} {Phys. Rev. E}\ }\textbf {\bibinfo {volume} {93}},\ \bibinfo
		{pages} {032147} (\bibinfo {year} {2016})}\BibitemShut {NoStop}%
	\bibitem [{\citenamefont {Ptaszy\ifmmode~\acute{n}\else \'{n}\fi{}ski}\ and\
		\citenamefont {Esposito}(2019)}]{esposito2019thermodynamics}%
	\BibitemOpen
	\bibfield  {author} {\bibinfo {author} {\bibfnamefont {K.}~\bibnamefont
			{Ptaszy\ifmmode~\acute{n}\else \'{n}\fi{}ski}}\ and\ \bibinfo {author}
		{\bibfnamefont {M.}~\bibnamefont {Esposito}},\ }\bibfield  {title} {\bibinfo
		{title} {Thermodynamics of quantum information flows},\ }\href
	{https://doi.org/10.1103/PhysRevLett.122.150603} {\bibfield  {journal}
		{\bibinfo  {journal} {Phys. Rev. Lett.}\ }\textbf {\bibinfo {volume} {122}},\
		\bibinfo {pages} {150603} (\bibinfo {year} {2019})}\BibitemShut {NoStop}%
	\bibitem [{\citenamefont {Leggett}\ \emph
		{et~al.}(1987{\natexlab{a}})\citenamefont {Leggett}, \citenamefont
		{Chakravarty}, \citenamefont {Dorsey}, \citenamefont {Fisher}, \citenamefont
		{Garg},\ and\ \citenamefont {Zwerger}}]{leggett1987dynamics}%
	\BibitemOpen
	\bibfield  {author} {\bibinfo {author} {\bibfnamefont {A.~J.}\ \bibnamefont
			{Leggett}}, \bibinfo {author} {\bibfnamefont {S.}~\bibnamefont
			{Chakravarty}}, \bibinfo {author} {\bibfnamefont {A.~T.}\ \bibnamefont
			{Dorsey}}, \bibinfo {author} {\bibfnamefont {M.~P.~A.}\ \bibnamefont
			{Fisher}}, \bibinfo {author} {\bibfnamefont {A.}~\bibnamefont {Garg}},\ and\
		\bibinfo {author} {\bibfnamefont {W.}~\bibnamefont {Zwerger}},\ }\bibfield
	{title} {\bibinfo {title} {Dynamics of the dissipative two-state system},\
	}\href@noop {} {\bibfield  {journal} {\bibinfo  {journal} {Reviews of Modern
				Physics}\ }\textbf {\bibinfo {volume} {59}},\ \bibinfo {pages} {1} (\bibinfo
		{year} {1987}{\natexlab{a}})}\BibitemShut {NoStop}%
	\bibitem [{\citenamefont {Weiss}(2012)}]{weiss2012quantum}%
	\BibitemOpen
	\bibfield  {author} {\bibinfo {author} {\bibfnamefont {U.}~\bibnamefont
			{Weiss}},\ }\href@noop {} {\emph {\bibinfo {title} {quantum dissipative
				systems}}}\ (\bibinfo  {publisher} {world scientific},\ \bibinfo {year}
	{2012})\BibitemShut {NoStop}%
	\bibitem [{\citenamefont {Agarwalla}\ and\ \citenamefont
		{Segal}(2017)}]{agarwalla2017energy}%
	\BibitemOpen
	\bibfield  {author} {\bibinfo {author} {\bibfnamefont {B.~K.}\ \bibnamefont
			{Agarwalla}}\ and\ \bibinfo {author} {\bibfnamefont {D.}~\bibnamefont
			{Segal}},\ }\bibfield  {title} {\bibinfo {title} {Energy current and its
			statistics in the nonequilibrium spin-boson model: Majorana fermion
			representation},\ }\href {https://doi.org/10.1088/1367-2630/aa6657}
	{\bibfield  {journal} {\bibinfo  {journal} {New Journal of Physics}\ }\textbf
		{\bibinfo {volume} {19}},\ \bibinfo {pages} {043030} (\bibinfo {year}
		{2017})}\BibitemShut {NoStop}%
	\bibitem [{\citenamefont {Strasberg}\ \emph {et~al.}(2017)\citenamefont
		{Strasberg}, \citenamefont {Schaller}, \citenamefont {Brandes},\ and\
		\citenamefont {Esposito}}]{strasberg2016quantum}%
	\BibitemOpen
	\bibfield  {author} {\bibinfo {author} {\bibfnamefont {P.}~\bibnamefont
			{Strasberg}}, \bibinfo {author} {\bibfnamefont {G.}~\bibnamefont {Schaller}},
		\bibinfo {author} {\bibfnamefont {T.}~\bibnamefont {Brandes}},\ and\ \bibinfo
		{author} {\bibfnamefont {M.}~\bibnamefont {Esposito}},\ }\bibfield  {title}
	{\bibinfo {title} {Quantum and information thermodynamics: A unifying
			framework based on repeated interactions},\ }\href@noop {} {\bibfield
		{journal} {\bibinfo  {journal} {Physical Review X}\ }\textbf {\bibinfo
			{volume} {7}},\ \bibinfo {pages} {021003} (\bibinfo {year}
		{2017})}\BibitemShut {NoStop}%
	\bibitem [{\citenamefont {Gelbwaser-Klimovsky}\ and\ \citenamefont
		{Kurizki}(2014)}]{gelbwaser2014heat}%
	\BibitemOpen
	\bibfield  {author} {\bibinfo {author} {\bibfnamefont {D.}~\bibnamefont
			{Gelbwaser-Klimovsky}}\ and\ \bibinfo {author} {\bibfnamefont
			{G.}~\bibnamefont {Kurizki}},\ }\bibfield  {title} {\bibinfo {title}
		{Heat-machine control by quantum-state preparation: From quantum engines to
			refrigerators},\ }\href {https://doi.org/10.1103/PhysRevE.90.022102}
	{\bibfield  {journal} {\bibinfo  {journal} {Phys. Rev. E}\ }\textbf {\bibinfo
			{volume} {90}},\ \bibinfo {pages} {022102} (\bibinfo {year}
		{2014})}\BibitemShut {NoStop}%
	\bibitem [{\citenamefont {Segal}(2006)}]{segal2006heat}%
	\BibitemOpen
	\bibfield  {author} {\bibinfo {author} {\bibfnamefont {D.}~\bibnamefont
			{Segal}},\ }\bibfield  {title} {\bibinfo {title} {Heat flow in nonlinear
			molecular junctions: Master equation analysis},\ }\href@noop {} {\bibfield
		{journal} {\bibinfo  {journal} {Physical Review B}\ }\textbf {\bibinfo
			{volume} {73}},\ \bibinfo {pages} {205415} (\bibinfo {year}
		{2006})}\BibitemShut {NoStop}%
	\bibitem [{\citenamefont {Caldeira}\ and\ \citenamefont
		{Leggett}(1981)}]{calderia1981influence}%
	\BibitemOpen
	\bibfield  {author} {\bibinfo {author} {\bibfnamefont {A.~O.}\ \bibnamefont
			{Caldeira}}\ and\ \bibinfo {author} {\bibfnamefont {A.~J.}\ \bibnamefont
			{Leggett}},\ }\bibfield  {title} {\bibinfo {title} {Influence of dissipation
			on quantum tunneling in macroscopic systems},\ }\href
	{https://doi.org/10.1103/PhysRevLett.46.211} {\bibfield  {journal} {\bibinfo
			{journal} {Phys. Rev. Lett.}\ }\textbf {\bibinfo {volume} {46}},\ \bibinfo
		{pages} {211} (\bibinfo {year} {1981})}\BibitemShut {NoStop}%
	\bibitem [{\citenamefont {Caldeira}\ and\ \citenamefont
		{Leggett}(1983)}]{caldeira1983path}%
	\BibitemOpen
	\bibfield  {author} {\bibinfo {author} {\bibfnamefont {A.~O.}\ \bibnamefont
			{Caldeira}}\ and\ \bibinfo {author} {\bibfnamefont {A.~J.}\ \bibnamefont
			{Leggett}},\ }\bibfield  {title} {\bibinfo {title} {Path integral approach to
			quantum brownian motion},\ }\href@noop {} {\bibfield  {journal} {\bibinfo
			{journal} {Physica A}\ }\textbf {\bibinfo {volume} {121}},\ \bibinfo {pages}
		{587} (\bibinfo {year} {1983})}\BibitemShut {NoStop}%
	\bibitem [{\citenamefont {Leggett}\ \emph
		{et~al.}(1987{\natexlab{b}})\citenamefont {Leggett}, \citenamefont
		{Chakravarty}, \citenamefont {Dorsey}, \citenamefont {Fisher}, \citenamefont
		{Garg},\ and\ \citenamefont {Zwerger}}]{calderia1987dynamics}%
	\BibitemOpen
	\bibfield  {author} {\bibinfo {author} {\bibfnamefont {A.~J.}\ \bibnamefont
			{Leggett}}, \bibinfo {author} {\bibfnamefont {S.}~\bibnamefont
			{Chakravarty}}, \bibinfo {author} {\bibfnamefont {A.~T.}\ \bibnamefont
			{Dorsey}}, \bibinfo {author} {\bibfnamefont {M.~P.~A.}\ \bibnamefont
			{Fisher}}, \bibinfo {author} {\bibfnamefont {A.}~\bibnamefont {Garg}},\ and\
		\bibinfo {author} {\bibfnamefont {W.}~\bibnamefont {Zwerger}},\ }\bibfield
	{title} {\bibinfo {title} {Dynamics of the dissipative two-state system},\
	}\href {https://doi.org/10.1103/RevModPhys.59.1} {\bibfield  {journal}
		{\bibinfo  {journal} {Rev. Mod. Phys.}\ }\textbf {\bibinfo {volume} {59}},\
		\bibinfo {pages} {1} (\bibinfo {year} {1987}{\natexlab{b}})}\BibitemShut
	{NoStop}%
	\bibitem [{\citenamefont {Kosloff}\ and\ \citenamefont
		{Levy}(2014)}]{kosloff2014quantum}%
	\BibitemOpen
	\bibfield  {author} {\bibinfo {author} {\bibfnamefont {R.}~\bibnamefont
			{Kosloff}}\ and\ \bibinfo {author} {\bibfnamefont {A.}~\bibnamefont {Levy}},\
	}\bibfield  {title} {\bibinfo {title} {Quantum heat engines and
			refrigerators: Continuous devices},\ }\href
	{https://doi.org/10.1146/annurev-physchem-040513-103724} {\bibfield
		{journal} {\bibinfo  {journal} {Annu. Rev. Phys. Chem.}\ }\textbf {\bibinfo
			{volume} {65}},\ \bibinfo {pages} {365} (\bibinfo {year} {2014})}\BibitemShut
	{NoStop}%
	\bibitem [{\citenamefont {Gupt}\ \emph {et~al.}(2021)\citenamefont {Gupt},
		\citenamefont {Bhattacharyya},\ and\ \citenamefont
		{Ghosh}}]{nikhil2021statistical}%
	\BibitemOpen
	\bibfield  {author} {\bibinfo {author} {\bibfnamefont {N.}~\bibnamefont
			{Gupt}}, \bibinfo {author} {\bibfnamefont {S.}~\bibnamefont
			{Bhattacharyya}},\ and\ \bibinfo {author} {\bibfnamefont {A.}~\bibnamefont
			{Ghosh}},\ }\bibfield  {title} {\bibinfo {title} {Statistical generalization
			of regenerative bosonic and fermionic stirling cycles},\ }\href
	{https://doi.org/10.1103/PhysRevE.104.054130} {\bibfield  {journal} {\bibinfo
			{journal} {Phys. Rev. E}\ }\textbf {\bibinfo {volume} {104}},\ \bibinfo
		{pages} {054130} (\bibinfo {year} {2021})}\BibitemShut {NoStop}%
	\bibitem [{\citenamefont {Kumar}\ \emph {et~al.}(2023)\citenamefont {Kumar},
		\citenamefont {Gupt}, \citenamefont {Ghosh},\ and\ \citenamefont
		{Ghosh}}]{samarth2023introduction}%
	\BibitemOpen
	\bibfield  {author} {\bibinfo {author} {\bibfnamefont {S.}~\bibnamefont
			{Kumar}}, \bibinfo {author} {\bibfnamefont {N.}~\bibnamefont {Gupt}},
		\bibinfo {author} {\bibfnamefont {S.}~\bibnamefont {Ghosh}},\ and\ \bibinfo
		{author} {\bibfnamefont {A.}~\bibnamefont {Ghosh}},\ }\bibfield  {title}
	{\bibinfo {title} {Introduction to quantum thermodynamic cycles},\ }\href
	{https://doi.org/10.1007/s12039-023-02154-5} {\bibfield  {journal} {\bibinfo
			{journal} {Journal of Chemical Sciences}\ }\textbf {\bibinfo {volume}
			{135}},\ \bibinfo {pages} {36} (\bibinfo {year} {2023})}\BibitemShut
	{NoStop}%
	\bibitem [{\citenamefont {Bhardwaj}\ \emph {et~al.}(2017)\citenamefont
		{Bhardwaj}, \citenamefont {Kaur}, \citenamefont {Wuest},\ and\ \citenamefont
		{Wuest}}]{bhardwaj2017in}%
	\BibitemOpen
	\bibfield  {author} {\bibinfo {author} {\bibfnamefont {A.}~\bibnamefont
			{Bhardwaj}}, \bibinfo {author} {\bibfnamefont {J.}~\bibnamefont {Kaur}},
		\bibinfo {author} {\bibfnamefont {M.}~\bibnamefont {Wuest}},\ and\ \bibinfo
		{author} {\bibfnamefont {F.}~\bibnamefont {Wuest}},\ }\bibfield  {title}
	{\bibinfo {title} {In situ click chemistry generation of cyclooxygenase-2
			inhibitors},\ }\href {https://doi.org/10.1038/s41467-017-00001-x} {\bibfield
		{journal} {\bibinfo  {journal} {Nature Communications}\ }\textbf {\bibinfo
			{volume} {8}},\ \bibinfo {pages} {1} (\bibinfo {year} {2017})}\BibitemShut
	{NoStop}%
	\bibitem [{\citenamefont {Levy}\ and\ \citenamefont
		{Kosloff}(2012)}]{levy2012quantum}%
	\BibitemOpen
	\bibfield  {author} {\bibinfo {author} {\bibfnamefont {A.}~\bibnamefont
			{Levy}}\ and\ \bibinfo {author} {\bibfnamefont {R.}~\bibnamefont {Kosloff}},\
	}\bibfield  {title} {\bibinfo {title} {Quantum absorption refrigerator},\
	}\href {https://doi.org/10.1103/PhysRevLett.108.070604} {\bibfield  {journal}
		{\bibinfo  {journal} {Phys. Rev. Lett.}\ }\textbf {\bibinfo {volume} {108}},\
		\bibinfo {pages} {070604} (\bibinfo {year} {2012})}\BibitemShut {NoStop}%
	\bibitem [{\citenamefont {Hofer}\ \emph {et~al.}(2016)\citenamefont {Hofer},
		\citenamefont {Perarnau-Llobet}, \citenamefont {Brask}, \citenamefont
		{Silva}, \citenamefont {Huber},\ and\ \citenamefont
		{Brunner}}]{hofer2016autonomous}%
	\BibitemOpen
	\bibfield  {author} {\bibinfo {author} {\bibfnamefont {P.~P.}\ \bibnamefont
			{Hofer}}, \bibinfo {author} {\bibfnamefont {M.}~\bibnamefont
			{Perarnau-Llobet}}, \bibinfo {author} {\bibfnamefont {J.~B.}\ \bibnamefont
			{Brask}}, \bibinfo {author} {\bibfnamefont {R.}~\bibnamefont {Silva}},
		\bibinfo {author} {\bibfnamefont {M.}~\bibnamefont {Huber}},\ and\ \bibinfo
		{author} {\bibfnamefont {N.}~\bibnamefont {Brunner}},\ }\bibfield  {title}
	{\bibinfo {title} {Autonomous quantum refrigerator in a circuit qed
			architecture based on a josephson junction},\ }\href
	{https://doi.org/10.1103/PhysRevB.94.235420} {\bibfield  {journal} {\bibinfo
			{journal} {Phys. Rev. B}\ }\textbf {\bibinfo {volume} {94}},\ \bibinfo
		{pages} {235420} (\bibinfo {year} {2016})}\BibitemShut {NoStop}%
	\bibitem [{\citenamefont {Chen}\ \emph {et~al.}(2023)\citenamefont {Chen},
		\citenamefont {Wang}, \citenamefont {Chen},\ and\ \citenamefont
		{Su}}]{chen2023optimal}%
	\BibitemOpen
	\bibfield  {author} {\bibinfo {author} {\bibfnamefont {J.}~\bibnamefont
			{Chen}}, \bibinfo {author} {\bibfnamefont {Y.}~\bibnamefont {Wang}}, \bibinfo
		{author} {\bibfnamefont {J.}~\bibnamefont {Chen}},\ and\ \bibinfo {author}
		{\bibfnamefont {S.}~\bibnamefont {Su}},\ }\bibfield  {title} {\bibinfo
		{title} {Optimal figure of merit of low-dissipation quantum refrigerators},\
	}\href {https://doi.org/10.1103/PhysRevE.107.044118} {\bibfield  {journal}
		{\bibinfo  {journal} {Physical Review E}\ }\textbf {\bibinfo {volume}
			{107}},\ \bibinfo {pages} {044118} (\bibinfo {year} {2023})}\BibitemShut
	{NoStop}%
	\bibitem [{\citenamefont {Bhattacharyya}\ \emph {et~al.}(2025)\citenamefont
		{Bhattacharyya}, \citenamefont {Ghoshal},\ and\ \citenamefont
		{Sen}}]{bhattacharyya2025transient}%
	\BibitemOpen
	\bibfield  {author} {\bibinfo {author} {\bibfnamefont {A.}~\bibnamefont
			{Bhattacharyya}}, \bibinfo {author} {\bibfnamefont {A.}~\bibnamefont
			{Ghoshal}},\ and\ \bibinfo {author} {\bibfnamefont {U.}~\bibnamefont {Sen}},\
	}\bibfield  {title} {\bibinfo {title} {Transient effects in quantum
			refrigerators with finite environments},\ }\href
	{https://doi.org/10.1103/PhysRevA.111.012209} {\bibfield  {journal} {\bibinfo
			{journal} {Physical Review A}\ }\textbf {\bibinfo {volume} {111}},\ \bibinfo
		{pages} {012209} (\bibinfo {year} {2025})}\BibitemShut {NoStop}%
	\bibitem [{\citenamefont {Li}\ \emph {et~al.}(2012)\citenamefont {Li},
		\citenamefont {Ren}, \citenamefont {Wang}, \citenamefont {Zhang},
		\citenamefont {H\"anggi},\ and\ \citenamefont {Li}}]{li2012colloquium}%
	\BibitemOpen
	\bibfield  {author} {\bibinfo {author} {\bibfnamefont {N.}~\bibnamefont
			{Li}}, \bibinfo {author} {\bibfnamefont {J.}~\bibnamefont {Ren}}, \bibinfo
		{author} {\bibfnamefont {L.}~\bibnamefont {Wang}}, \bibinfo {author}
		{\bibfnamefont {G.}~\bibnamefont {Zhang}}, \bibinfo {author} {\bibfnamefont
			{P.}~\bibnamefont {H\"anggi}},\ and\ \bibinfo {author} {\bibfnamefont
			{B.}~\bibnamefont {Li}},\ }\bibfield  {title} {\bibinfo {title} {Colloquium:
			Phononics: Manipulating heat flow with electronic analogs and beyond},\
	}\href {https://doi.org/10.1103/RevModPhys.84.1045} {\bibfield  {journal}
		{\bibinfo  {journal} {Rev. Mod. Phys.}\ }\textbf {\bibinfo {volume} {84}},\
		\bibinfo {pages} {1045} (\bibinfo {year} {2012})}\BibitemShut {NoStop}%
	\bibitem [{\citenamefont {S{\'{a}}nchez}\ \emph {et~al.}(2015)\citenamefont
		{S{\'{a}}nchez}, \citenamefont {Sothmann},\ and\ \citenamefont
		{Jordan}}]{sanchez2015heat}%
	\BibitemOpen
	\bibfield  {author} {\bibinfo {author} {\bibfnamefont {R.}~\bibnamefont
			{S{\'{a}}nchez}}, \bibinfo {author} {\bibfnamefont {B.}~\bibnamefont
			{Sothmann}},\ and\ \bibinfo {author} {\bibfnamefont {A.~N.}\ \bibnamefont
			{Jordan}},\ }\bibfield  {title} {\bibinfo {title} {Heat diode and engine
			based on quantum hall edge states},\ }\href
	{https://doi.org/10.1088/1367-2630/17/7/075006} {\bibfield  {journal}
		{\bibinfo  {journal} {New Journal of Physics}\ }\textbf {\bibinfo {volume}
			{17}},\ \bibinfo {pages} {075006} (\bibinfo {year} {2015})}\BibitemShut
	{NoStop}%
	\bibitem [{\citenamefont {Li}\ \emph {et~al.}(2004)\citenamefont {Li},
		\citenamefont {Wang},\ and\ \citenamefont {Casati}}]{li2004thermal}%
	\BibitemOpen
	\bibfield  {author} {\bibinfo {author} {\bibfnamefont {B.}~\bibnamefont
			{Li}}, \bibinfo {author} {\bibfnamefont {L.}~\bibnamefont {Wang}},\ and\
		\bibinfo {author} {\bibfnamefont {G.}~\bibnamefont {Casati}},\ }\bibfield
	{title} {\bibinfo {title} {Thermal diode: Rectification of heat flux},\
	}\href {https://doi.org/10.1103/PhysRevLett.93.184301} {\bibfield  {journal}
		{\bibinfo  {journal} {Phys. Rev. Lett.}\ }\textbf {\bibinfo {volume} {93}},\
		\bibinfo {pages} {184301} (\bibinfo {year} {2004})}\BibitemShut {NoStop}%
	\bibitem [{\citenamefont {Scheibner}\ \emph {et~al.}(2008)\citenamefont
		{Scheibner}, \citenamefont {König}, \citenamefont {Reuter}, \citenamefont
		{Wieck}, \citenamefont {Gould}, \citenamefont {Buhmann},\ and\ \citenamefont
		{Molenkamp}}]{scheibner2008quantum}%
	\BibitemOpen
	\bibfield  {author} {\bibinfo {author} {\bibfnamefont {R.}~\bibnamefont
			{Scheibner}}, \bibinfo {author} {\bibfnamefont {M.}~\bibnamefont {König}},
		\bibinfo {author} {\bibfnamefont {D.}~\bibnamefont {Reuter}}, \bibinfo
		{author} {\bibfnamefont {A.~D.}\ \bibnamefont {Wieck}}, \bibinfo {author}
		{\bibfnamefont {C.}~\bibnamefont {Gould}}, \bibinfo {author} {\bibfnamefont
			{H.}~\bibnamefont {Buhmann}},\ and\ \bibinfo {author} {\bibfnamefont {L.~W.}\
			\bibnamefont {Molenkamp}},\ }\bibfield  {title} {\bibinfo {title} {Quantum
			dot as thermal rectifier},\ }\href
	{https://doi.org/10.1088/1367-2630/10/8/083016} {\bibfield  {journal}
		{\bibinfo  {journal} {New Journal of Physics}\ }\textbf {\bibinfo {volume}
			{10}},\ \bibinfo {pages} {083016} (\bibinfo {year} {2008})}\BibitemShut
	{NoStop}%
	\bibitem [{\citenamefont {Ordonez-Miranda}\ \emph {et~al.}(2017)\citenamefont
		{Ordonez-Miranda}, \citenamefont {Ezzahri},\ and\ \citenamefont
		{Joulain}}]{ordonez2017quantum}%
	\BibitemOpen
	\bibfield  {author} {\bibinfo {author} {\bibfnamefont {J.}~\bibnamefont
			{Ordonez-Miranda}}, \bibinfo {author} {\bibfnamefont {Y.}~\bibnamefont
			{Ezzahri}},\ and\ \bibinfo {author} {\bibfnamefont {K.}~\bibnamefont
			{Joulain}},\ }\bibfield  {title} {\bibinfo {title} {Quantum thermal diode
			based on two interacting spinlike systems under different excitations},\
	}\href {https://doi.org/10.1103/PhysRevE.95.022128} {\bibfield  {journal}
		{\bibinfo  {journal} {Phys. Rev. E}\ }\textbf {\bibinfo {volume} {95}},\
		\bibinfo {pages} {022128} (\bibinfo {year} {2017})}\BibitemShut {NoStop}%
	\bibitem [{\citenamefont {Wang}\ \emph {et~al.}(2019)\citenamefont {Wang},
		\citenamefont {Xu}, \citenamefont {Liu},\ and\ \citenamefont
		{Gao}}]{wang2019thermal}%
	\BibitemOpen
	\bibfield  {author} {\bibinfo {author} {\bibfnamefont {C.}~\bibnamefont
			{Wang}}, \bibinfo {author} {\bibfnamefont {D.}~\bibnamefont {Xu}}, \bibinfo
		{author} {\bibfnamefont {H.}~\bibnamefont {Liu}},\ and\ \bibinfo {author}
		{\bibfnamefont {X.}~\bibnamefont {Gao}},\ }\bibfield  {title} {\bibinfo
		{title} {Thermal rectification and heat amplification in a nonequilibrium
			v-type three-level system},\ }\href
	{https://doi.org/10.1103/PhysRevE.99.042102} {\bibfield  {journal} {\bibinfo
			{journal} {Physical Review E}\ }\textbf {\bibinfo {volume} {99}},\ \bibinfo
		{pages} {042102} (\bibinfo {year} {2019})}\BibitemShut {NoStop}%
	\bibitem [{\citenamefont {Kalantar}\ \emph {et~al.}(2021)\citenamefont
		{Kalantar}, \citenamefont {Agarwalla},\ and\ \citenamefont
		{Segal}}]{kalantar2021harmonic}%
	\BibitemOpen
	\bibfield  {author} {\bibinfo {author} {\bibfnamefont {N.}~\bibnamefont
			{Kalantar}}, \bibinfo {author} {\bibfnamefont {B.~K.}\ \bibnamefont
			{Agarwalla}},\ and\ \bibinfo {author} {\bibfnamefont {D.}~\bibnamefont
			{Segal}},\ }\bibfield  {title} {\bibinfo {title} {Harmonic chains and the
			thermal diode effect},\ }\href {https://doi.org/10.1103/PhysRevE.103.052130}
	{\bibfield  {journal} {\bibinfo  {journal} {Phys. Rev. E}\ }\textbf {\bibinfo
			{volume} {103}},\ \bibinfo {pages} {052130} (\bibinfo {year}
		{2021})}\BibitemShut {NoStop}%
	\bibitem [{\citenamefont {Marcos-Vicioso}\ \emph {et~al.}(2018)\citenamefont
		{Marcos-Vicioso}, \citenamefont {L\'opez-Jurado}, \citenamefont
		{Ruiz-Garcia},\ and\ \citenamefont {S\'anchez}}]{marcos2018thermal}%
	\BibitemOpen
	\bibfield  {author} {\bibinfo {author} {\bibfnamefont {A.}~\bibnamefont
			{Marcos-Vicioso}}, \bibinfo {author} {\bibfnamefont {C.}~\bibnamefont
			{L\'opez-Jurado}}, \bibinfo {author} {\bibfnamefont {M.}~\bibnamefont
			{Ruiz-Garcia}},\ and\ \bibinfo {author} {\bibfnamefont {R.}~\bibnamefont
			{S\'anchez}},\ }\bibfield  {title} {\bibinfo {title} {Thermal rectification
			with interacting electronic channels: Exploiting degeneracy, quantum
			superpositions, and interference},\ }\href
	{https://doi.org/10.1103/PhysRevB.98.035414} {\bibfield  {journal} {\bibinfo
			{journal} {Phys. Rev. B}\ }\textbf {\bibinfo {volume} {98}},\ \bibinfo
		{pages} {035414} (\bibinfo {year} {2018})}\BibitemShut {NoStop}%
	\bibitem [{\citenamefont {Terraneo}\ \emph {et~al.}(2002)\citenamefont
		{Terraneo}, \citenamefont {Peyrard},\ and\ \citenamefont
		{Casati}}]{terraneo2002controlling}%
	\BibitemOpen
	\bibfield  {author} {\bibinfo {author} {\bibfnamefont {M.}~\bibnamefont
			{Terraneo}}, \bibinfo {author} {\bibfnamefont {M.}~\bibnamefont {Peyrard}},\
		and\ \bibinfo {author} {\bibfnamefont {G.}~\bibnamefont {Casati}},\
	}\bibfield  {title} {\bibinfo {title} {Controlling the energy flow in
			nonlinear lattices: A model for a thermal rectifier},\ }\href
	{https://doi.org/10.1103/PhysRevLett.88.094302} {\bibfield  {journal}
		{\bibinfo  {journal} {Phys. Rev. Lett.}\ }\textbf {\bibinfo {volume} {88}},\
		\bibinfo {pages} {094302} (\bibinfo {year} {2002})}\BibitemShut {NoStop}%
	\bibitem [{\citenamefont {Neumeier}\ \emph {et~al.}(2013)\citenamefont
		{Neumeier}, \citenamefont {Leib},\ and\ \citenamefont
		{Hartmann}}]{neumeier2013single}%
	\BibitemOpen
	\bibfield  {author} {\bibinfo {author} {\bibfnamefont {L.}~\bibnamefont
			{Neumeier}}, \bibinfo {author} {\bibfnamefont {M.}~\bibnamefont {Leib}},\
		and\ \bibinfo {author} {\bibfnamefont {M.~J.}\ \bibnamefont {Hartmann}},\
	}\bibfield  {title} {\bibinfo {title} {Single-photon transistor in circuit
			quantum electrodynamics},\ }\href
	{https://doi.org/10.1103/PhysRevLett.111.063601} {\bibfield  {journal}
		{\bibinfo  {journal} {Phys. Rev. Lett.}\ }\textbf {\bibinfo {volume} {111}},\
		\bibinfo {pages} {063601} (\bibinfo {year} {2013})}\BibitemShut {NoStop}%
	\bibitem [{\citenamefont {Joulain}\ \emph {et~al.}(2016)\citenamefont
		{Joulain}, \citenamefont {Drevillon}, \citenamefont {Ezzahri},\ and\
		\citenamefont {Ordonez-Miranda}}]{joulain2016quantum}%
	\BibitemOpen
	\bibfield  {author} {\bibinfo {author} {\bibfnamefont {K.}~\bibnamefont
			{Joulain}}, \bibinfo {author} {\bibfnamefont {J.}~\bibnamefont {Drevillon}},
		\bibinfo {author} {\bibfnamefont {Y.}~\bibnamefont {Ezzahri}},\ and\ \bibinfo
		{author} {\bibfnamefont {J.}~\bibnamefont {Ordonez-Miranda}},\ }\bibfield
	{title} {\bibinfo {title} {Quantum thermal transistor},\ }\href
	{https://doi.org/10.1103/PhysRevLett.116.200601} {\bibfield  {journal}
		{\bibinfo  {journal} {Phys. Rev. Lett.}\ }\textbf {\bibinfo {volume} {116}},\
		\bibinfo {pages} {200601} (\bibinfo {year} {2016})}\BibitemShut {NoStop}%
	\bibitem [{\citenamefont {Guo}\ \emph {et~al.}(2018)\citenamefont {Guo},
		\citenamefont {Liu},\ and\ \citenamefont {Yu}}]{guo2018quantum}%
	\BibitemOpen
	\bibfield  {author} {\bibinfo {author} {\bibfnamefont {B.-q.}\ \bibnamefont
			{Guo}}, \bibinfo {author} {\bibfnamefont {T.}~\bibnamefont {Liu}},\ and\
		\bibinfo {author} {\bibfnamefont {C.-s.}\ \bibnamefont {Yu}},\ }\bibfield
	{title} {\bibinfo {title} {Quantum thermal transistor based on qubit-qutrit
			coupling},\ }\href@noop {} {\bibfield  {journal} {\bibinfo  {journal}
			{Physical Review E}\ }\textbf {\bibinfo {volume} {98}},\ \bibinfo {pages}
		{022118} (\bibinfo {year} {2018})}\BibitemShut {NoStop}%
	\bibitem [{\citenamefont {Wijesekara}\ \emph {et~al.}(2021)\citenamefont
		{Wijesekara}, \citenamefont {Gunapala},\ and\ \citenamefont
		{Premaratne}}]{wijesekara2021darlington}%
	\BibitemOpen
	\bibfield  {author} {\bibinfo {author} {\bibfnamefont {R.~T.}\ \bibnamefont
			{Wijesekara}}, \bibinfo {author} {\bibfnamefont {S.~D.}\ \bibnamefont
			{Gunapala}},\ and\ \bibinfo {author} {\bibfnamefont {M.}~\bibnamefont
			{Premaratne}},\ }\bibfield  {title} {\bibinfo {title} {Darlington pair of
			quantum thermal transistors},\ }\href
	{https://doi.org/10.1103/PhysRevB.104.045405} {\bibfield  {journal} {\bibinfo
			{journal} {Phys. Rev. B}\ }\textbf {\bibinfo {volume} {104}},\ \bibinfo
		{pages} {045405} (\bibinfo {year} {2021})}\BibitemShut {NoStop}%
	\bibitem [{\citenamefont {Ekanayake}\ \emph {et~al.}(2023)\citenamefont
		{Ekanayake}, \citenamefont {Gunapala},\ and\ \citenamefont
		{Premaratne}}]{ekanayake2023stochastic}%
	\BibitemOpen
	\bibfield  {author} {\bibinfo {author} {\bibfnamefont {U.~N.}\ \bibnamefont
			{Ekanayake}}, \bibinfo {author} {\bibfnamefont {S.~D.}\ \bibnamefont
			{Gunapala}},\ and\ \bibinfo {author} {\bibfnamefont {M.}~\bibnamefont
			{Premaratne}},\ }\bibfield  {title} {\bibinfo {title} {Stochastic model of
			noise for a quantum thermal transistor},\ }\href
	{https://doi.org/10.1103/PhysRevB.108.235421} {\bibfield  {journal} {\bibinfo
			{journal} {Physical Review B}\ }\textbf {\bibinfo {volume} {108}},\ \bibinfo
		{pages} {235421} (\bibinfo {year} {2023})}\BibitemShut {NoStop}%
	\bibitem [{\citenamefont {B{\"u}ttiker}(1986)}]{buttiker1986four}%
	\BibitemOpen
	\bibfield  {author} {\bibinfo {author} {\bibfnamefont {M.}~\bibnamefont
			{B{\"u}ttiker}},\ }\bibfield  {title} {\bibinfo {title} {Four-terminal
			phase-coherent conductance},\ }\href@noop {} {\bibfield  {journal} {\bibinfo
			{journal} {Physical Review Letters}\ }\textbf {\bibinfo {volume} {57}},\
		\bibinfo {pages} {1761} (\bibinfo {year} {1986})}\BibitemShut {NoStop}%
	\bibitem [{\citenamefont {Strasberg}(2022)}]{strasberg2022quantum}%
	\BibitemOpen
	\bibfield  {author} {\bibinfo {author} {\bibfnamefont {P.}~\bibnamefont
			{Strasberg}},\ }\href@noop {} {\emph {\bibinfo {title} {Quantum Stochastic
				Thermodynamics: Foundations and Selected Applications}}}\ (\bibinfo
	{publisher} {Oxford University Press},\ \bibinfo {year} {2022})\BibitemShut
	{NoStop}%
	\bibitem [{\citenamefont {Dorsch}\ \emph {et~al.}(2021)\citenamefont {Dorsch},
		\citenamefont {Svilans}, \citenamefont {Josefsson}, \citenamefont
		{Goldozian}, \citenamefont {Kumar}, \citenamefont {Thelander}, \citenamefont
		{Wacker},\ and\ \citenamefont {Burke}}]{dorsch2021heat}%
	\BibitemOpen
	\bibfield  {author} {\bibinfo {author} {\bibfnamefont {S.}~\bibnamefont
			{Dorsch}}, \bibinfo {author} {\bibfnamefont {A.}~\bibnamefont {Svilans}},
		\bibinfo {author} {\bibfnamefont {M.}~\bibnamefont {Josefsson}}, \bibinfo
		{author} {\bibfnamefont {B.}~\bibnamefont {Goldozian}}, \bibinfo {author}
		{\bibfnamefont {M.}~\bibnamefont {Kumar}}, \bibinfo {author} {\bibfnamefont
			{C.}~\bibnamefont {Thelander}}, \bibinfo {author} {\bibfnamefont
			{A.}~\bibnamefont {Wacker}},\ and\ \bibinfo {author} {\bibfnamefont
			{A.}~\bibnamefont {Burke}},\ }\bibfield  {title} {\bibinfo {title} {Heat
			driven transport in serial double quantum dot devices},\ }\href
	{https://doi.org/10.1021/acs.nanolett.0c04017} {\bibfield  {journal}
		{\bibinfo  {journal} {Nano Letters}\ }\textbf {\bibinfo {volume} {21}},\
		\bibinfo {pages} {988} (\bibinfo {year} {2021})},\ \bibinfo {note} {pMID:
		33459021},\ \Eprint
	{https://arxiv.org/abs/https://doi.org/10.1021/acs.nanolett.0c04017}
	{https://doi.org/10.1021/acs.nanolett.0c04017} \BibitemShut {NoStop}%
	\bibitem [{\citenamefont {Zhang}\ and\ \citenamefont
		{Chen}(2015)}]{zhang2015thermoelectric}%
	\BibitemOpen
	\bibfield  {author} {\bibinfo {author} {\bibfnamefont {Y.}~\bibnamefont
			{Zhang}}\ and\ \bibinfo {author} {\bibfnamefont {G.}~\bibnamefont {Chen}},\
	}\bibfield  {title} {\bibinfo {title} {Thermoelectric effects in serially
			coupled quantum dots},\ }\href@noop {} {\bibfield  {journal} {\bibinfo
			{journal} {Scientific Reports}\ }\textbf {\bibinfo {volume} {5}},\ \bibinfo
		{pages} {16050} (\bibinfo {year} {2015})}\BibitemShut {NoStop}%
	\bibitem [{\citenamefont {Brandes}(2005)}]{brandes2005coherent}%
	\BibitemOpen
	\bibfield  {author} {\bibinfo {author} {\bibfnamefont {T.}~\bibnamefont
			{Brandes}},\ }\bibfield  {title} {\bibinfo {title} {Coherent and collective
			quantum optical effects in mesoscopic systems},\ }\href@noop {} {\bibfield
		{journal} {\bibinfo  {journal} {Physics Reports}\ }\textbf {\bibinfo {volume}
			{408}},\ \bibinfo {pages} {315} (\bibinfo {year} {2005})}\BibitemShut
	{NoStop}%
	\bibitem [{\citenamefont {Chi}\ \emph {et~al.}(2016)\citenamefont {Chi},
		\citenamefont {Zheng},\ and\ \citenamefont {Zhang}}]{chi2016thermoelectric}%
	\BibitemOpen
	\bibfield  {author} {\bibinfo {author} {\bibfnamefont {F.}~\bibnamefont
			{Chi}}, \bibinfo {author} {\bibfnamefont {J.}~\bibnamefont {Zheng}},\ and\
		\bibinfo {author} {\bibfnamefont {K.}~\bibnamefont {Zhang}},\ }\bibfield
	{title} {\bibinfo {title} {Thermoelectric properties of serially coupled
			quantum dots},\ }\href@noop {} {\bibfield  {journal} {\bibinfo  {journal}
			{Nanoscale}\ }\textbf {\bibinfo {volume} {8}},\ \bibinfo {pages} {15927}
		(\bibinfo {year} {2016})}\BibitemShut {NoStop}%
	\bibitem [{\citenamefont {Zhang}\ \emph {et~al.}(2023)\citenamefont {Zhang},
		\citenamefont {Wang}, \citenamefont {Li},\ and\ \citenamefont
		{Yue}}]{zhang2023inverse}%
	\BibitemOpen
	\bibfield  {author} {\bibinfo {author} {\bibfnamefont {Y.}~\bibnamefont
			{Zhang}}, \bibinfo {author} {\bibfnamefont {S.}~\bibnamefont {Wang}},
		\bibinfo {author} {\bibfnamefont {W.}~\bibnamefont {Li}},\ and\ \bibinfo
		{author} {\bibfnamefont {M.}~\bibnamefont {Yue}},\ }\bibfield  {title}
	{\bibinfo {title} {Inverse current induced thermoelectric conversion in a
			parallel-coupled double quantum dot system},\ }\href
	{https://doi.org/10.1088/1402-4896/acfa36} {\bibfield  {journal} {\bibinfo
			{journal} {Physica Scripta}\ }\textbf {\bibinfo {volume} {98}},\ \bibinfo
		{pages} {105245} (\bibinfo {year} {2023})}\BibitemShut {NoStop}%
	\bibitem [{\citenamefont {Purcell}\ and\ \citenamefont
		{Pound}(1951)}]{purcell1951a}%
	\BibitemOpen
	\bibfield  {author} {\bibinfo {author} {\bibfnamefont {E.~M.}\ \bibnamefont
			{Purcell}}\ and\ \bibinfo {author} {\bibfnamefont {R.~V.}\ \bibnamefont
			{Pound}},\ }\bibfield  {title} {\bibinfo {title} {A nuclear spin system at
			negative temperature},\ }\href {https://doi.org/10.1103/PhysRev.81.279}
	{\bibfield  {journal} {\bibinfo  {journal} {Phys. Rev.}\ }\textbf {\bibinfo
			{volume} {81}},\ \bibinfo {pages} {279} (\bibinfo {year} {1951})}\BibitemShut
	{NoStop}%
	\bibitem [{\citenamefont {Ramsey}(1956)}]{ramsey1956thermodynamics}%
	\BibitemOpen
	\bibfield  {author} {\bibinfo {author} {\bibfnamefont {N.~F.}\ \bibnamefont
			{Ramsey}},\ }\bibfield  {title} {\bibinfo {title} {Thermodynamics and
			statistical mechanics at negative absolute temperatures},\ }\href
	{https://doi.org/10.1103/PhysRev.103.20} {\bibfield  {journal} {\bibinfo
			{journal} {Phys. Rev.}\ }\textbf {\bibinfo {volume} {103}},\ \bibinfo {pages}
		{20} (\bibinfo {year} {1956})}\BibitemShut {NoStop}%
	\bibitem [{\citenamefont {Allen}\ and\ \citenamefont
		{Eberly}(1987)}]{allen1987optical}%
	\BibitemOpen
	\bibfield  {author} {\bibinfo {author} {\bibfnamefont {L.}~\bibnamefont
			{Allen}}\ and\ \bibinfo {author} {\bibfnamefont {J.}~\bibnamefont {Eberly}},\
	}\href {https://books.google.co.in/books?id=1q0ae-XNmWwC} {\emph {\bibinfo
			{title} {Optical Resonance and Two-level Atoms}}},\ Dover books on physics
	and chemistry\ (\bibinfo  {publisher} {Dover},\ \bibinfo {year}
	{1987})\BibitemShut {NoStop}%
	\bibitem [{\citenamefont {Ghosh}\ \emph
		{et~al.}(2012{\natexlab{a}})\citenamefont {Ghosh}, \citenamefont {Sinha},\
		and\ \citenamefont {Ray}}]{ghosh2012canonical}%
	\BibitemOpen
	\bibfield  {author} {\bibinfo {author} {\bibfnamefont {A.}~\bibnamefont
			{Ghosh}}, \bibinfo {author} {\bibfnamefont {S.~S.}\ \bibnamefont {Sinha}},\
		and\ \bibinfo {author} {\bibfnamefont {D.~S.}\ \bibnamefont {Ray}},\
	}\bibfield  {title} {\bibinfo {title} {Canonical formulation of quantum
			dissipation and noise in a generalized spin bath},\ }\href
	{https://doi.org/10.1103/PhysRevE.86.011122} {\bibfield  {journal} {\bibinfo
			{journal} {Phys. Rev. E}\ }\textbf {\bibinfo {volume} {86}},\ \bibinfo
		{pages} {011122} (\bibinfo {year} {2012}{\natexlab{a}})}\BibitemShut
	{NoStop}%
	\bibitem [{\citenamefont {Ghosh}\ \emph {et~al.}(2011)\citenamefont {Ghosh},
		\citenamefont {Sinha},\ and\ \citenamefont {Ray}}]{ghosh2011dissipation}%
	\BibitemOpen
	\bibfield  {author} {\bibinfo {author} {\bibfnamefont {A.}~\bibnamefont
			{Ghosh}}, \bibinfo {author} {\bibfnamefont {S.~S.}\ \bibnamefont {Sinha}},\
		and\ \bibinfo {author} {\bibfnamefont {D.~S.}\ \bibnamefont {Ray}},\
	}\bibfield  {title} {\bibinfo {title} {Dissipation in a spin bath: Thermally
			induced coherent intensity and spectral splitting},\ }\href
	{https://doi.org/10.1103/PhysRevE.83.061154} {\bibfield  {journal} {\bibinfo
			{journal} {Phys. Rev. E}\ }\textbf {\bibinfo {volume} {83}},\ \bibinfo
		{pages} {061154} (\bibinfo {year} {2011})}\BibitemShut {NoStop}%
	\bibitem [{\citenamefont {Ghosh}\ \emph
		{et~al.}(2012{\natexlab{b}})\citenamefont {Ghosh}, \citenamefont {Sinha},\
		and\ \citenamefont {Ray}}]{ghosh2012fermionic}%
	\BibitemOpen
	\bibfield  {author} {\bibinfo {author} {\bibfnamefont {A.}~\bibnamefont
			{Ghosh}}, \bibinfo {author} {\bibfnamefont {S.~S.}\ \bibnamefont {Sinha}},\
		and\ \bibinfo {author} {\bibfnamefont {D.~S.}\ \bibnamefont {Ray}},\
	}\bibfield  {title} {\bibinfo {title} {Fermionic oscillator in a fermionic
			bath},\ }\href {https://doi.org/10.1103/PhysRevE.86.011138} {\bibfield
		{journal} {\bibinfo  {journal} {Phys. Rev. E}\ }\textbf {\bibinfo {volume}
			{86}},\ \bibinfo {pages} {011138} (\bibinfo {year}
		{2012}{\natexlab{b}})}\BibitemShut {NoStop}%
	\bibitem [{\citenamefont {Bruus}\ and\ \citenamefont
		{Flensberg}(2004)}]{bruus2004many}%
	\BibitemOpen
	\bibfield  {author} {\bibinfo {author} {\bibfnamefont {H.}~\bibnamefont
			{Bruus}}\ and\ \bibinfo {author} {\bibfnamefont {K.}~\bibnamefont
			{Flensberg}},\ }\href {https://books.google.co.in/books?id=CktuBAAAQBAJ}
	{\emph {\bibinfo {title} {Many-Body Quantum Theory in Condensed Matter
				Physics: An Introduction}}},\ Oxford Graduate Texts\ (\bibinfo  {publisher}
	{OUP Oxford},\ \bibinfo {year} {2004})\BibitemShut {NoStop}%
	\bibitem [{\citenamefont {Datta}(2005)}]{datta2005book}%
	\BibitemOpen
	\bibfield  {author} {\bibinfo {author} {\bibfnamefont {S.}~\bibnamefont
			{Datta}},\ }\href {https://books.google.co.in/books?id=Yj50EJoS224C} {\emph
		{\bibinfo {title} {Quantum Transport: Atom to Transistor}}},\ Ke xue qian yan
	cong shu\ (\bibinfo  {publisher} {Cambridge University Press},\ \bibinfo
	{year} {2005})\BibitemShut {NoStop}%
	\bibitem [{\citenamefont {Wang}\ \emph {et~al.}(2022)\citenamefont {Wang},
		\citenamefont {Wang}, \citenamefont {Wang},\ and\ \citenamefont
		{Ren}}]{wang2022cycleflux}%
	\BibitemOpen
	\bibfield  {author} {\bibinfo {author} {\bibfnamefont {L.}~\bibnamefont
			{Wang}}, \bibinfo {author} {\bibfnamefont {Z.}~\bibnamefont {Wang}}, \bibinfo
		{author} {\bibfnamefont {C.}~\bibnamefont {Wang}},\ and\ \bibinfo {author}
		{\bibfnamefont {J.}~\bibnamefont {Ren}},\ }\bibfield  {title} {\bibinfo
		{title} {Cycle flux ranking of network analysis in quantum thermal devices},\
	}\href {https://doi.org/10.1103/PhysRevLett.128.067701} {\bibfield  {journal}
		{\bibinfo  {journal} {Phys. Rev. Lett.}\ }\textbf {\bibinfo {volume} {128}},\
		\bibinfo {pages} {067701} (\bibinfo {year} {2022})}\BibitemShut {NoStop}%
	\bibitem [{\citenamefont {{Martin}}\ and\ \citenamefont
		{{Schwinger}}(1959)}]{martin1959theory}%
	\BibitemOpen
	\bibfield  {author} {\bibinfo {author} {\bibfnamefont {P.~C.}\ \bibnamefont
			{{Martin}}}\ and\ \bibinfo {author} {\bibfnamefont {J.}~\bibnamefont
			{{Schwinger}}},\ }\bibfield  {title} {\bibinfo {title} {{Theory of
				Many-Particle Systems. I}},\ }\href
	{https://doi.org/10.1103/PhysRev.115.1342} {\bibfield  {journal} {\bibinfo
			{journal} {Physical Review}\ }\textbf {\bibinfo {volume} {115}},\ \bibinfo
		{pages} {1342} (\bibinfo {year} {1959})}\BibitemShut {NoStop}%
	\bibitem [{\citenamefont {Kadanoff}(1989)}]{kadanoff1989quantum}%
	\BibitemOpen
	\bibfield  {author} {\bibinfo {author} {\bibfnamefont {L.~P.}\ \bibnamefont
			{Kadanoff}},\ }\href {https://doi.org/10.1201/9780429493218} {\emph {\bibinfo
			{title} {Quantum Statistical Mechanics}}},\ \bibinfo {edition} {1st}\ ed.\
	(\bibinfo  {publisher} {CRC Press},\ \bibinfo {year} {1989})\BibitemShut
	{NoStop}%
	\bibitem [{\citenamefont {Keldysh}(1965)}]{keldysh1964diagram}%
	\BibitemOpen
	\bibfield  {author} {\bibinfo {author} {\bibfnamefont {L.~V.}\ \bibnamefont
			{Keldysh}},\ }\bibfield  {title} {\bibinfo {title} {{Diagram Technique for
				Nonequilibrium Processes}},\ }\href
	{https://doi.org/10.1142/9789811279461_0007} {\bibfield  {journal} {\bibinfo
			{journal} {Sov. Phys. JETP}\ }\textbf {\bibinfo {volume} {20}},\ \bibinfo
		{pages} {1018} (\bibinfo {year} {1965})}\BibitemShut {NoStop}%
	\bibitem [{\citenamefont {Mahan}(1981)}]{mahan1981many}%
	\BibitemOpen
	\bibfield  {author} {\bibinfo {author} {\bibfnamefont {G.~D.}\ \bibnamefont
			{Mahan}},\ }\href {https://www.osti.gov/biblio/5382466} {\emph {\bibinfo
			{title} {Many-particle physics}}}\ (\bibinfo  {publisher} {Plenum Press,New
		York, NY},\ \bibinfo {year} {1981})\BibitemShut {NoStop}%
	\bibitem [{\citenamefont {Landauer}(1957)}]{landauer1957spatial}%
	\BibitemOpen
	\bibfield  {author} {\bibinfo {author} {\bibfnamefont {R.}~\bibnamefont
			{Landauer}},\ }\bibfield  {title} {\bibinfo {title} {spatial variation of
			currents and fields due to localized scatterers in metallic conduction},\
	}\href@noop {} {\bibfield  {journal} {\bibinfo  {journal} {ibm journal of
				research and development}\ }\textbf {\bibinfo {volume} {1}},\ \bibinfo
		{pages} {223} (\bibinfo {year} {1957})}\BibitemShut {NoStop}%
	\bibitem [{\citenamefont {Camsari}\ \emph {et~al.}(2023)\citenamefont
		{Camsari}, \citenamefont {Chowdhury},\ and\ \citenamefont
		{Datta}}]{Camsari2023}%
	\BibitemOpen
	\bibfield  {author} {\bibinfo {author} {\bibfnamefont {K.~Y.}\ \bibnamefont
			{Camsari}}, \bibinfo {author} {\bibfnamefont {S.}~\bibnamefont {Chowdhury}},\
		and\ \bibinfo {author} {\bibfnamefont {S.}~\bibnamefont {Datta}},\ }\bibinfo
	{title} {The nonequilibrium green function (negf) method},\ in\ \href
	{https://doi.org/10.1007/978-3-030-79827-7\_44} {\emph {\bibinfo {booktitle}
			{Springer Handbook of Semiconductor Devices}}},\ \bibinfo {editor} {edited
		by\ \bibinfo {editor} {\bibfnamefont {M.}~\bibnamefont {Rudan}}, \bibinfo
		{editor} {\bibfnamefont {R.}~\bibnamefont {Brunetti}},\ and\ \bibinfo
		{editor} {\bibfnamefont {S.}~\bibnamefont {Reggiani}}}\ (\bibinfo
	{publisher} {Springer International Publishing},\ \bibinfo {address} {Cham},\
	\bibinfo {year} {2023})\ pp.\ \bibinfo {pages} {1583--1599}\BibitemShut
	{NoStop}%
	\bibitem [{\citenamefont {Majidi}\ \emph {et~al.}(2024)\citenamefont {Majidi},
		\citenamefont {Bergfield}, \citenamefont {Maisi}, \citenamefont {Höfer},
		\citenamefont {Courtois},\ and\ \citenamefont {Winkelmann}}]{majidi2024heat}%
	\BibitemOpen
	\bibfield  {author} {\bibinfo {author} {\bibfnamefont {D.}~\bibnamefont
			{Majidi}}, \bibinfo {author} {\bibfnamefont {J.~P.}\ \bibnamefont
			{Bergfield}}, \bibinfo {author} {\bibfnamefont {V.}~\bibnamefont {Maisi}},
		\bibinfo {author} {\bibfnamefont {J.}~\bibnamefont {Höfer}}, \bibinfo
		{author} {\bibfnamefont {H.}~\bibnamefont {Courtois}},\ and\ \bibinfo
		{author} {\bibfnamefont {C.~B.}\ \bibnamefont {Winkelmann}},\ }\bibfield
	{title} {\bibinfo {title} {Heat transport at the nanoscale and ultralow
			temperatures—implications for quantum technologies},\ }\href
	{https://doi.org/10.1063/5.0204207} {\bibfield  {journal} {\bibinfo
			{journal} {Applied Physics Letters}\ }\textbf {\bibinfo {volume} {124}},\
		\bibinfo {pages} {140504} (\bibinfo {year} {2024})}\BibitemShut {NoStop}%
	\bibitem [{\citenamefont {Annby-Andersson}\ \emph {et~al.}(2020)\citenamefont
		{Annby-Andersson}, \citenamefont {Samuelsson}, \citenamefont {Maisi},\ and\
		\citenamefont {Potts}}]{annby2020maxwell}%
	\BibitemOpen
	\bibfield  {author} {\bibinfo {author} {\bibfnamefont {B.}~\bibnamefont
			{Annby-Andersson}}, \bibinfo {author} {\bibfnamefont {P.}~\bibnamefont
			{Samuelsson}}, \bibinfo {author} {\bibfnamefont {V.~F.}\ \bibnamefont
			{Maisi}},\ and\ \bibinfo {author} {\bibfnamefont {P.~P.}\ \bibnamefont
			{Potts}},\ }\bibfield  {title} {\bibinfo {title} {Maxwell's demon in a double
			quantum dot with continuous charge detection},\ }\href
	{https://doi.org/10.1103/PhysRevB.101.165404} {\bibfield  {journal} {\bibinfo
			{journal} {Phys. Rev. B}\ }\textbf {\bibinfo {volume} {101}},\ \bibinfo
		{pages} {165404} (\bibinfo {year} {2020})}\BibitemShut {NoStop}%
	\bibitem [{\citenamefont {Das}\ \emph {et~al.}(2026)\citenamefont {Das},
		\citenamefont {Mahunta}, \citenamefont {Gupt}, \citenamefont {Mukherjee},\
		and\ \citenamefont {Ghosh}}]{das2026fluctuations}%
	\BibitemOpen
	\bibfield  {author} {\bibinfo {author} {\bibfnamefont {S.}~\bibnamefont
			{Das}}, \bibinfo {author} {\bibfnamefont {S.}~\bibnamefont {Mahunta}},
		\bibinfo {author} {\bibfnamefont {N.}~\bibnamefont {Gupt}}, \bibinfo {author}
		{\bibfnamefont {V.}~\bibnamefont {Mukherjee}},\ and\ \bibinfo {author}
		{\bibfnamefont {A.}~\bibnamefont {Ghosh}},\ }\bibfield  {title} {\bibinfo
		{title} {Fluctuations and optimal control in a floquet quantum thermal
			transistor},\ }\href@noop {} {\bibfield  {journal} {\bibinfo  {journal}
			{Phys. Scr.}\ }\textbf {\bibinfo {volume} {101}},\ \bibinfo {pages} {0335102}
		(\bibinfo {year} {2026})}\BibitemShut {NoStop}%
	\bibitem [{\citenamefont {Binder}\ \emph {et~al.}(2019)\citenamefont {Binder},
		\citenamefont {Correa}, \citenamefont {Gogolin}, \citenamefont {Anders},\
		and\ \citenamefont {Adesso}}]{binder2019thermodynamics}%
	\BibitemOpen
	\bibfield  {author} {\bibinfo {author} {\bibfnamefont {F.}~\bibnamefont
			{Binder}}, \bibinfo {author} {\bibfnamefont {L.}~\bibnamefont {Correa}},
		\bibinfo {author} {\bibfnamefont {C.}~\bibnamefont {Gogolin}}, \bibinfo
		{author} {\bibfnamefont {J.}~\bibnamefont {Anders}},\ and\ \bibinfo {author}
		{\bibfnamefont {G.}~\bibnamefont {Adesso}},\ }\href@noop {} {\emph {\bibinfo
			{title} {Thermodynamics in the Quantum Regime: Fundamental Aspects and New
				Directions}}},\ Fundamental Theories of Physics\ (\bibinfo  {publisher}
	{Springer International Publishing},\ \bibinfo {year} {2019})\BibitemShut
	{NoStop}%
	\bibitem [{\citenamefont {Pyurbeeva}\ and\ \citenamefont
		{Kosloff}(2026)}]{pyurbeeva2026quantum}%
	\BibitemOpen
	\bibfield  {author} {\bibinfo {author} {\bibfnamefont {E.}~\bibnamefont
			{Pyurbeeva}}\ and\ \bibinfo {author} {\bibfnamefont {R.}~\bibnamefont
			{Kosloff}},\ }\bibfield  {title} {\bibinfo {title} {Quantum dot thermal
			machines—a guide to engineering},\ }\bibfield  {journal} {\bibinfo
		{journal} {Entropy}\ }\textbf {\bibinfo {volume} {28}},\ \href
	{https://doi.org/10.3390/e28010002} {10.3390/e28010002} (\bibinfo {year}
	{2026})\BibitemShut {NoStop}%
	\bibitem [{\citenamefont {Zhang}\ and\ \citenamefont
		{Xie}(2021)}]{zhang2021inverse}%
	\BibitemOpen
	\bibfield  {author} {\bibinfo {author} {\bibfnamefont {Y.}~\bibnamefont
			{Zhang}}\ and\ \bibinfo {author} {\bibfnamefont {Z.}~\bibnamefont {Xie}},\
	}\bibfield  {title} {\bibinfo {title} {Inverse currents in coulomb-coupled
			quantum dots},\ }\href {https://doi.org/10.1103/PhysRevE.104.064142}
	{\bibfield  {journal} {\bibinfo  {journal} {Phys. Rev. E}\ }\textbf {\bibinfo
			{volume} {104}},\ \bibinfo {pages} {064142} (\bibinfo {year}
		{2021})}\BibitemShut {NoStop}%
	\bibitem [{\citenamefont {Tesser}\ \emph {et~al.}(2022)\citenamefont {Tesser},
		\citenamefont {Bhandari}, \citenamefont {Erdman}, \citenamefont {Paladino},
		\citenamefont {Fazio},\ and\ \citenamefont {Taddei}}]{tesser2022heat}%
	\BibitemOpen
	\bibfield  {author} {\bibinfo {author} {\bibfnamefont {L.}~\bibnamefont
			{Tesser}}, \bibinfo {author} {\bibfnamefont {B.}~\bibnamefont {Bhandari}},
		\bibinfo {author} {\bibfnamefont {P.~A.}\ \bibnamefont {Erdman}}, \bibinfo
		{author} {\bibfnamefont {E.}~\bibnamefont {Paladino}}, \bibinfo {author}
		{\bibfnamefont {R.}~\bibnamefont {Fazio}},\ and\ \bibinfo {author}
		{\bibfnamefont {F.}~\bibnamefont {Taddei}},\ }\bibfield  {title} {\bibinfo
		{title} {Heat rectification through single and coupled quantum dots},\ }\href
	{https://doi.org/10.1088/1367-2630/ac53b8} {\bibfield  {journal} {\bibinfo
			{journal} {New Journal of Physics}\ }\textbf {\bibinfo {volume} {24}},\
		\bibinfo {pages} {035001} (\bibinfo {year} {2022})}\BibitemShut {NoStop}%
	\bibitem [{\citenamefont {Landi}\ and\ \citenamefont
		{Paternostro}(2021)}]{landi2021irreversible}%
	\BibitemOpen
	\bibfield  {author} {\bibinfo {author} {\bibfnamefont {G.~T.}\ \bibnamefont
			{Landi}}\ and\ \bibinfo {author} {\bibfnamefont {M.}~\bibnamefont
			{Paternostro}},\ }\bibfield  {title} {\bibinfo {title} {Irreversible entropy
			production: From classical to quantum},\ }\href
	{https://doi.org/10.1103/RevModPhys.93.035008} {\bibfield  {journal}
		{\bibinfo  {journal} {Rev. Mod. Phys.}\ }\textbf {\bibinfo {volume} {93}},\
		\bibinfo {pages} {035008} (\bibinfo {year} {2021})}\BibitemShut {NoStop}%
	\bibitem [{\citenamefont {Esposito}\ \emph
		{et~al.}(2010{\natexlab{b}})\citenamefont {Esposito}, \citenamefont
		{Lindenberg},\ and\ \citenamefont {den Broeck}}]{esposito2010entropy}%
	\BibitemOpen
	\bibfield  {author} {\bibinfo {author} {\bibfnamefont {M.}~\bibnamefont
			{Esposito}}, \bibinfo {author} {\bibfnamefont {K.}~\bibnamefont
			{Lindenberg}},\ and\ \bibinfo {author} {\bibfnamefont {C.~V.}\ \bibnamefont
			{den Broeck}},\ }\bibfield  {title} {\bibinfo {title} {Entropy production as
			correlation between system and reservoir},\ }\href
	{https://doi.org/10.1088/1367-2630/12/1/013013} {\bibfield  {journal}
		{\bibinfo  {journal} {New Journal of Physics}\ }\textbf {\bibinfo {volume}
			{12}},\ \bibinfo {pages} {013013} (\bibinfo {year}
		{2010}{\natexlab{b}})}\BibitemShut {NoStop}%
	\bibitem [{\citenamefont {Walldorf}\ \emph {et~al.}(2017)\citenamefont
		{Walldorf}, \citenamefont {Jauho},\ and\ \citenamefont
		{Kaasbjerg}}]{walldorf2017thermoelectrics}%
	\BibitemOpen
	\bibfield  {author} {\bibinfo {author} {\bibfnamefont {N.}~\bibnamefont
			{Walldorf}}, \bibinfo {author} {\bibfnamefont {A.-P.}\ \bibnamefont
			{Jauho}},\ and\ \bibinfo {author} {\bibfnamefont {K.}~\bibnamefont
			{Kaasbjerg}},\ }\bibfield  {title} {\bibinfo {title} {Thermoelectrics in
			coulomb-coupled quantum dots: Cotunneling and energy-dependent lead
			couplings},\ }\href {https://doi.org/10.1103/PhysRevB.96.115415} {\bibfield
		{journal} {\bibinfo  {journal} {Phys. Rev. B}\ }\textbf {\bibinfo {volume}
			{96}},\ \bibinfo {pages} {115415} (\bibinfo {year} {2017})}\BibitemShut
	{NoStop}%
	\bibitem [{\citenamefont {Whitney}\ \emph {et~al.}(2018)\citenamefont
		{Whitney}, \citenamefont {S{\'a}nchez},\ and\ \citenamefont
		{Splettstoesser}}]{whitney2018quantum}%
	\BibitemOpen
	\bibfield  {author} {\bibinfo {author} {\bibfnamefont {R.~S.}\ \bibnamefont
			{Whitney}}, \bibinfo {author} {\bibfnamefont {R.}~\bibnamefont
			{S{\'a}nchez}},\ and\ \bibinfo {author} {\bibfnamefont {J.}~\bibnamefont
			{Splettstoesser}},\ }\href {https://doi.org/10.1007/978-3-319-99046-0_7}
	{\bibinfo {title} {Quantum thermodynamics of nanoscale thermoelectrics and
			electronic devices}} (\bibinfo {year} {2018})\BibitemShut {NoStop}%
	\bibitem [{\citenamefont {DiSalvo}(1999)}]{disalvo1999thermoelectric}%
	\BibitemOpen
	\bibfield  {author} {\bibinfo {author} {\bibfnamefont {F.~J.}\ \bibnamefont
			{DiSalvo}},\ }\bibfield  {title} {\bibinfo {title} {Thermoelectric cooling
			and power generation},\ }\href {https://doi.org/10.1126/science.285.5428.703}
	{\bibfield  {journal} {\bibinfo  {journal} {Science}\ }\textbf {\bibinfo
			{volume} {285}},\ \bibinfo {pages} {703} (\bibinfo {year}
		{1999})}\BibitemShut {NoStop}%
	\bibitem [{\citenamefont {Timm}(2007)}]{timm2007gauge}%
	\BibitemOpen
	\bibfield  {author} {\bibinfo {author} {\bibfnamefont {C.}~\bibnamefont
			{Timm}},\ }\bibfield  {title} {\bibinfo {title} {Gauge theory for the rate
			equations: Electrodynamics on a network},\ }\href
	{https://doi.org/10.1103/PhysRevLett.98.070604} {\bibfield  {journal}
		{\bibinfo  {journal} {Phys. Rev. Lett.}\ }\textbf {\bibinfo {volume} {98}},\
		\bibinfo {pages} {070604} (\bibinfo {year} {2007})}\BibitemShut {NoStop}%
	\bibitem [{\citenamefont {Walschaers}\ \emph {et~al.}(2013)\citenamefont
		{Walschaers}, \citenamefont {Diaz}, \citenamefont {Mulet},\ and\
		\citenamefont {Buchleitner}}]{walschares2013optimally}%
	\BibitemOpen
	\bibfield  {author} {\bibinfo {author} {\bibfnamefont {M.}~\bibnamefont
			{Walschaers}}, \bibinfo {author} {\bibfnamefont {J.~F.-d.-C.}\ \bibnamefont
			{Diaz}}, \bibinfo {author} {\bibfnamefont {R.}~\bibnamefont {Mulet}},\ and\
		\bibinfo {author} {\bibfnamefont {A.}~\bibnamefont {Buchleitner}},\
	}\bibfield  {title} {\bibinfo {title} {Optimally designed quantum transport
			across disordered networks},\ }\href
	{https://doi.org/10.1103/PhysRevLett.111.180601} {\bibfield  {journal}
		{\bibinfo  {journal} {Phys. Rev. Lett.}\ }\textbf {\bibinfo {volume} {111}},\
		\bibinfo {pages} {180601} (\bibinfo {year} {2013})}\BibitemShut {NoStop}%
	\bibitem [{\citenamefont {Whitney}\ \emph {et~al.}(2016)\citenamefont
		{Whitney}, \citenamefont {Sánchez}, \citenamefont {Haupt},\ and\
		\citenamefont {Splettstoesser}}]{whitney2016thermoelectricity}%
	\BibitemOpen
	\bibfield  {author} {\bibinfo {author} {\bibfnamefont {R.~S.}\ \bibnamefont
			{Whitney}}, \bibinfo {author} {\bibfnamefont {R.}~\bibnamefont {Sánchez}},
		\bibinfo {author} {\bibfnamefont {F.}~\bibnamefont {Haupt}},\ and\ \bibinfo
		{author} {\bibfnamefont {J.}~\bibnamefont {Splettstoesser}},\ }\bibfield
	{title} {\bibinfo {title} {Thermoelectricity without absorbing energy from
			the heat sources},\ }\href
	{https://doi.org/https://doi.org/10.1016/j.physe.2015.09.025} {\bibfield
		{journal} {\bibinfo  {journal} {Physica E: Low-dimensional Systems and
				Nanostructures}\ }\textbf {\bibinfo {volume} {75}},\ \bibinfo {pages} {257}
		(\bibinfo {year} {2016})}\BibitemShut {NoStop}%
	\bibitem [{\citenamefont {Werlang}\ \emph {et~al.}(2014)\citenamefont
		{Werlang}, \citenamefont {Marchiori}, \citenamefont {Cornelio},\ and\
		\citenamefont {Valente}}]{werlang2014optimal}%
	\BibitemOpen
	\bibfield  {author} {\bibinfo {author} {\bibfnamefont {T.}~\bibnamefont
			{Werlang}}, \bibinfo {author} {\bibfnamefont {M.~A.}\ \bibnamefont
			{Marchiori}}, \bibinfo {author} {\bibfnamefont {M.~F.}\ \bibnamefont
			{Cornelio}},\ and\ \bibinfo {author} {\bibfnamefont {D.}~\bibnamefont
			{Valente}},\ }\bibfield  {title} {\bibinfo {title} {Optimal rectification in
			the ultrastrong coupling regime},\ }\href
	{https://doi.org/10.1103/PhysRevE.89.062109} {\bibfield  {journal} {\bibinfo
			{journal} {Phys. Rev. E}\ }\textbf {\bibinfo {volume} {89}},\ \bibinfo
		{pages} {062109} (\bibinfo {year} {2014})}\BibitemShut {NoStop}%
	\bibitem [{\citenamefont {Zhang}\ \emph {et~al.}(2017)\citenamefont {Zhang},
		\citenamefont {Zhang}, \citenamefont {Ye}, \citenamefont {Lin},\ and\
		\citenamefont {Chen}}]{zhang2017three}%
	\BibitemOpen
	\bibfield  {author} {\bibinfo {author} {\bibfnamefont {Y.}~\bibnamefont
			{Zhang}}, \bibinfo {author} {\bibfnamefont {X.}~\bibnamefont {Zhang}},
		\bibinfo {author} {\bibfnamefont {Z.}~\bibnamefont {Ye}}, \bibinfo {author}
		{\bibfnamefont {G.}~\bibnamefont {Lin}},\ and\ \bibinfo {author}
		{\bibfnamefont {J.}~\bibnamefont {Chen}},\ }\bibfield  {title} {\bibinfo
		{title} {Three-terminal quantum-dot thermal management devices},\ }\href
	{https://doi.org/10.1063/1.4979977} {\bibfield  {journal} {\bibinfo
			{journal} {Applied Physics Letters}\ }\textbf {\bibinfo {volume} {110}},\
		\bibinfo {pages} {153501} (\bibinfo {year} {2017})}\BibitemShut {NoStop}%
	\bibitem [{\citenamefont {Aligia}\ \emph {et~al.}(2020)\citenamefont {Aligia},
		\citenamefont {Daroca}, \citenamefont {Arrachea},\ and\ \citenamefont
		{Roura-Bas}}]{aligia2020heat}%
	\BibitemOpen
	\bibfield  {author} {\bibinfo {author} {\bibfnamefont {A.~A.}\ \bibnamefont
			{Aligia}}, \bibinfo {author} {\bibfnamefont {D.~P.}\ \bibnamefont {Daroca}},
		\bibinfo {author} {\bibfnamefont {L.}~\bibnamefont {Arrachea}},\ and\
		\bibinfo {author} {\bibfnamefont {P.}~\bibnamefont {Roura-Bas}},\ }\bibfield
	{title} {\bibinfo {title} {Heat current across a capacitively coupled double
			quantum dot},\ }\href {https://doi.org/10.1103/PhysRevB.101.075417}
	{\bibfield  {journal} {\bibinfo  {journal} {Phys. Rev. B}\ }\textbf {\bibinfo
			{volume} {101}},\ \bibinfo {pages} {075417} (\bibinfo {year}
		{2020})}\BibitemShut {NoStop}%
	\bibitem [{\citenamefont {Esposito}\ \emph {et~al.}(2012)\citenamefont
		{Esposito}, \citenamefont {Kumar}, \citenamefont {Lindenberg},\ and\
		\citenamefont {Van~den Broeck}}]{esposito2012stochastically}%
	\BibitemOpen
	\bibfield  {author} {\bibinfo {author} {\bibfnamefont {M.}~\bibnamefont
			{Esposito}}, \bibinfo {author} {\bibfnamefont {N.}~\bibnamefont {Kumar}},
		\bibinfo {author} {\bibfnamefont {K.}~\bibnamefont {Lindenberg}},\ and\
		\bibinfo {author} {\bibfnamefont {C.}~\bibnamefont {Van~den Broeck}},\
	}\bibfield  {title} {\bibinfo {title} {Stochastically driven single-level
			quantum dot: A nanoscale finite-time thermodynamic machine and its various
			operational modes},\ }\href {https://doi.org/10.1103/PhysRevE.85.031117}
	{\bibfield  {journal} {\bibinfo  {journal} {Phys. Rev. E}\ }\textbf {\bibinfo
			{volume} {85}},\ \bibinfo {pages} {031117} (\bibinfo {year}
		{2012})}\BibitemShut {NoStop}%
	\bibitem [{\citenamefont {Dutta}\ \emph {et~al.}(2017)\citenamefont {Dutta},
		\citenamefont {Peltonen}, \citenamefont {Antonenko}, \citenamefont {Meschke},
		\citenamefont {Skvortsov}, \citenamefont {Kubala}, \citenamefont {K\"onig},
		\citenamefont {Winkelmann}, \citenamefont {Courtois},\ and\ \citenamefont
		{Pekola}}]{dutta2017thermal}%
	\BibitemOpen
	\bibfield  {author} {\bibinfo {author} {\bibfnamefont {B.}~\bibnamefont
			{Dutta}}, \bibinfo {author} {\bibfnamefont {J.~T.}\ \bibnamefont {Peltonen}},
		\bibinfo {author} {\bibfnamefont {D.~S.}\ \bibnamefont {Antonenko}}, \bibinfo
		{author} {\bibfnamefont {M.}~\bibnamefont {Meschke}}, \bibinfo {author}
		{\bibfnamefont {M.~A.}\ \bibnamefont {Skvortsov}}, \bibinfo {author}
		{\bibfnamefont {B.}~\bibnamefont {Kubala}}, \bibinfo {author} {\bibfnamefont
			{J.}~\bibnamefont {K\"onig}}, \bibinfo {author} {\bibfnamefont {C.~B.}\
			\bibnamefont {Winkelmann}}, \bibinfo {author} {\bibfnamefont
			{H.}~\bibnamefont {Courtois}},\ and\ \bibinfo {author} {\bibfnamefont
			{J.~P.}\ \bibnamefont {Pekola}},\ }\bibfield  {title} {\bibinfo {title}
		{Thermal conductance of a single-electron transistor},\ }\href
	{https://doi.org/10.1103/PhysRevLett.119.077701} {\bibfield  {journal}
		{\bibinfo  {journal} {Phys. Rev. Lett.}\ }\textbf {\bibinfo {volume} {119}},\
		\bibinfo {pages} {077701} (\bibinfo {year} {2017})}\BibitemShut {NoStop}%
	\bibitem [{\citenamefont {Dutta}\ \emph {et~al.}(2020)\citenamefont {Dutta},
		\citenamefont {Majidi}, \citenamefont {Talarico}, \citenamefont {Lo~Gullo},
		\citenamefont {Courtois},\ and\ \citenamefont
		{Winkelmann}}]{dutta2020single}%
	\BibitemOpen
	\bibfield  {author} {\bibinfo {author} {\bibfnamefont {B.}~\bibnamefont
			{Dutta}}, \bibinfo {author} {\bibfnamefont {D.}~\bibnamefont {Majidi}},
		\bibinfo {author} {\bibfnamefont {N.~W.}\ \bibnamefont {Talarico}}, \bibinfo
		{author} {\bibfnamefont {N.}~\bibnamefont {Lo~Gullo}}, \bibinfo {author}
		{\bibfnamefont {H.}~\bibnamefont {Courtois}},\ and\ \bibinfo {author}
		{\bibfnamefont {C.~B.}\ \bibnamefont {Winkelmann}},\ }\bibfield  {title}
	{\bibinfo {title} {Single-quantum-dot heat valve},\ }\href
	{https://doi.org/10.1103/PhysRevLett.125.237701} {\bibfield  {journal}
		{\bibinfo  {journal} {Phys. Rev. Lett.}\ }\textbf {\bibinfo {volume} {125}},\
		\bibinfo {pages} {237701} (\bibinfo {year} {2020})}\BibitemShut {NoStop}%
	\bibitem [{\citenamefont {Ruokola}\ and\ \citenamefont
		{Ojanen}(2011)}]{ruokola2011single}%
	\BibitemOpen
	\bibfield  {author} {\bibinfo {author} {\bibfnamefont {T.}~\bibnamefont
			{Ruokola}}\ and\ \bibinfo {author} {\bibfnamefont {T.}~\bibnamefont
			{Ojanen}},\ }\bibfield  {title} {\bibinfo {title} {Single-electron heat
			diode: Asymmetric heat transport between electronic reservoirs through
			coulomb islands},\ }\href {https://doi.org/10.1103/PhysRevB.83.241404}
	{\bibfield  {journal} {\bibinfo  {journal} {Phys. Rev. B}\ }\textbf {\bibinfo
			{volume} {83}},\ \bibinfo {pages} {241404} (\bibinfo {year}
		{2011})}\BibitemShut {NoStop}%
	\bibitem [{\citenamefont {Zhang}\ \emph {et~al.}(2018)\citenamefont {Zhang},
		\citenamefont {Yang}, \citenamefont {Zhang}, \citenamefont {Lin},
		\citenamefont {Lin},\ and\ \citenamefont {Chen}}]{zhang2018coulomb}%
	\BibitemOpen
	\bibfield  {author} {\bibinfo {author} {\bibfnamefont {Y.}~\bibnamefont
			{Zhang}}, \bibinfo {author} {\bibfnamefont {Z.}~\bibnamefont {Yang}},
		\bibinfo {author} {\bibfnamefont {X.}~\bibnamefont {Zhang}}, \bibinfo
		{author} {\bibfnamefont {B.}~\bibnamefont {Lin}}, \bibinfo {author}
		{\bibfnamefont {G.}~\bibnamefont {Lin}},\ and\ \bibinfo {author}
		{\bibfnamefont {J.}~\bibnamefont {Chen}},\ }\bibfield  {title} {\bibinfo
		{title} {Coulomb-coupled quantum-dot thermal transistors},\ }\href
	{https://doi.org/10.1209/0295-5075/122/17002} {\bibfield  {journal} {\bibinfo
			{journal} {{EPL} (Europhysics Letters)}\ }\textbf {\bibinfo {volume}
			{122}},\ \bibinfo {pages} {17002} (\bibinfo {year} {2018})}\BibitemShut
	{NoStop}%
	\bibitem [{\citenamefont {Benenti}\ \emph
		{et~al.}(2017{\natexlab{b}})\citenamefont {Benenti}, \citenamefont {Casati},
		\citenamefont {Saito},\ and\ \citenamefont
		{Whitney}}]{benenti2017ffundamental}%
	\BibitemOpen
	\bibfield  {author} {\bibinfo {author} {\bibfnamefont {G.}~\bibnamefont
			{Benenti}}, \bibinfo {author} {\bibfnamefont {G.}~\bibnamefont {Casati}},
		\bibinfo {author} {\bibfnamefont {K.}~\bibnamefont {Saito}},\ and\ \bibinfo
		{author} {\bibfnamefont {R.~S.}\ \bibnamefont {Whitney}},\ }\bibfield
	{title} {\bibinfo {title} {fundamental aspects of steady-state conversion of
			heat to work at the nanoscale},\ }\href@noop {} {\bibfield  {journal}
		{\bibinfo  {journal} {physics reports}\ }\textbf {\bibinfo {volume} {694}},\
		\bibinfo {pages} {1} (\bibinfo {year} {2017}{\natexlab{b}})}\BibitemShut
	{NoStop}%
	\bibitem [{\citenamefont {Perez~Daroca}\ \emph {et~al.}(2025)\citenamefont
		{Perez~Daroca}, \citenamefont {Roura-Bas},\ and\ \citenamefont
		{Aligia}}]{daroca2025role}%
	\BibitemOpen
	\bibfield  {author} {\bibinfo {author} {\bibfnamefont {D.}~\bibnamefont
			{Perez~Daroca}}, \bibinfo {author} {\bibfnamefont {P.}~\bibnamefont
			{Roura-Bas}},\ and\ \bibinfo {author} {\bibfnamefont {A.~A.}\ \bibnamefont
			{Aligia}},\ }\bibfield  {title} {\bibinfo {title} {Role of asymmetry in
			thermoelectric properties of a double quantum dot out of equilibrium},\
	}\href {https://doi.org/10.1103/PhysRevB.111.045134} {\bibfield  {journal}
		{\bibinfo  {journal} {Phys. Rev. B}\ }\textbf {\bibinfo {volume} {111}},\
		\bibinfo {pages} {045134} (\bibinfo {year} {2025})}\BibitemShut {NoStop}%
	\bibitem [{\citenamefont {Hamo}\ \emph {et~al.}(2016)\citenamefont {Hamo},
		\citenamefont {Benyamini}, \citenamefont {Shapir}, \citenamefont {Khivrich},
		\citenamefont {Waissman}, \citenamefont {Kaasbjerg}, \citenamefont {Oreg},
		\citenamefont {von Oppen},\ and\ \citenamefont {Ilani}}]{hamo2016electron}%
	\BibitemOpen
	\bibfield  {author} {\bibinfo {author} {\bibfnamefont {A.}~\bibnamefont
			{Hamo}}, \bibinfo {author} {\bibfnamefont {A.}~\bibnamefont {Benyamini}},
		\bibinfo {author} {\bibfnamefont {I.}~\bibnamefont {Shapir}}, \bibinfo
		{author} {\bibfnamefont {I.}~\bibnamefont {Khivrich}}, \bibinfo {author}
		{\bibfnamefont {J.}~\bibnamefont {Waissman}}, \bibinfo {author}
		{\bibfnamefont {K.}~\bibnamefont {Kaasbjerg}}, \bibinfo {author}
		{\bibfnamefont {Y.}~\bibnamefont {Oreg}}, \bibinfo {author} {\bibfnamefont
			{F.}~\bibnamefont {von Oppen}},\ and\ \bibinfo {author} {\bibfnamefont
			{S.}~\bibnamefont {Ilani}},\ }\bibfield  {title} {\bibinfo {title} {Electron
			attraction mediated by coulomb repulsion},\ }\href
	{https://doi.org/10.1038/nature18639} {\bibfield  {journal} {\bibinfo
			{journal} {Nature}\ }\textbf {\bibinfo {volume} {535}},\ \bibinfo {pages}
		{395} (\bibinfo {year} {2016})}\BibitemShut {NoStop}%
	\bibitem [{\citenamefont {Tabatabaei}(2018)}]{tabatabaei2018charge}%
	\BibitemOpen
	\bibfield  {author} {\bibinfo {author} {\bibfnamefont {S.~M.}\ \bibnamefont
			{Tabatabaei}},\ }\bibfield  {title} {\bibinfo {title} {Charge-kondo effect
			mediated by repulsive interactions},\ }\href
	{https://doi.org/10.1103/PhysRevB.97.235131} {\bibfield  {journal} {\bibinfo
			{journal} {Phys. Rev. B}\ }\textbf {\bibinfo {volume} {97}},\ \bibinfo
		{pages} {235131} (\bibinfo {year} {2018})}\BibitemShut {NoStop}%
	\bibitem [{\citenamefont {Little}(1964)}]{little1964possibility}%
	\BibitemOpen
	\bibfield  {author} {\bibinfo {author} {\bibfnamefont {W.~A.}\ \bibnamefont
			{Little}},\ }\bibfield  {title} {\bibinfo {title} {Possibility of
			synthesizing an organic superconductor},\ }\href
	{https://doi.org/10.1103/PhysRev.134.A1416} {\bibfield  {journal} {\bibinfo
			{journal} {Physical Review}\ }\textbf {\bibinfo {volume} {134}},\ \bibinfo
		{pages} {A1416} (\bibinfo {year} {1964})}\BibitemShut {NoStop}%
	\bibitem [{\citenamefont {Prawiroatmodjo}\ \emph {et~al.}(2017)\citenamefont
		{Prawiroatmodjo}, \citenamefont {Leijnse}, \citenamefont {Trier},
		\citenamefont {Chen}, \citenamefont {Christensen}, \citenamefont {von
			Soosten}, \citenamefont {Pryds},\ and\ \citenamefont
		{Jespersen}}]{prawiroatmodjo2017negativeU}%
	\BibitemOpen
	\bibfield  {author} {\bibinfo {author} {\bibfnamefont {G.~E. D.~K.}\
			\bibnamefont {Prawiroatmodjo}}, \bibinfo {author} {\bibfnamefont
			{M.}~\bibnamefont {Leijnse}}, \bibinfo {author} {\bibfnamefont
			{F.}~\bibnamefont {Trier}}, \bibinfo {author} {\bibfnamefont
			{Y.}~\bibnamefont {Chen}}, \bibinfo {author} {\bibfnamefont {D.~V.}\
			\bibnamefont {Christensen}}, \bibinfo {author} {\bibfnamefont
			{M.}~\bibnamefont {von Soosten}}, \bibinfo {author} {\bibfnamefont
			{N.}~\bibnamefont {Pryds}},\ and\ \bibinfo {author} {\bibfnamefont {T.~S.}\
			\bibnamefont {Jespersen}},\ }\bibfield  {title} {\bibinfo {title} {Transport
			and excitations in a negative-u quantum dot at the laalo$_3$/srtio$_3$
			interface},\ }\href {https://doi.org/10.1038/s41467-017-00495-7} {\bibfield
		{journal} {\bibinfo  {journal} {Nature Communications}\ }\textbf {\bibinfo
			{volume} {8}},\ \bibinfo {pages} {395} (\bibinfo {year} {2017})}\BibitemShut
	{NoStop}%
	\bibitem [{\citenamefont {ichi UCHIDA}(2021)}]{uchida2021transport}%
	\BibitemOpen
	\bibfield  {author} {\bibinfo {author} {\bibfnamefont {K.}~\bibnamefont {ichi
				UCHIDA}},\ }\bibfield  {title} {\bibinfo {title} {Transport phenomena in spin
			caloritronics},\ }\href {https://doi.org/10.2183/pjab.97.004} {\bibfield
		{journal} {\bibinfo  {journal} {Proceedings of the Japan Academy, Series B}\
		}\textbf {\bibinfo {volume} {97}},\ \bibinfo {pages} {69} (\bibinfo {year}
		{2021})}\BibitemShut {NoStop}%
	\bibitem [{\citenamefont {Contamin}\ \emph {et~al.}(2023)\citenamefont
		{Contamin}, \citenamefont {Cubaynes}, \citenamefont {Legrand}, \citenamefont
		{Marganska}, \citenamefont {Villiers}, \citenamefont {Desjardins},
		\citenamefont {Dartiailh}, \citenamefont {Vinel}, \citenamefont {Leghtas},
		\citenamefont {Thiaville}, \citenamefont {Rohart}, \citenamefont {Cottet},
		\citenamefont {Delbecq},\ and\ \citenamefont
		{Kontos}}]{contamin2023inhomogeneous}%
	\BibitemOpen
	\bibfield  {author} {\bibinfo {author} {\bibfnamefont {L.~C.}\ \bibnamefont
			{Contamin}}, \bibinfo {author} {\bibfnamefont {T.}~\bibnamefont {Cubaynes}},
		\bibinfo {author} {\bibfnamefont {W.}~\bibnamefont {Legrand}}, \bibinfo
		{author} {\bibfnamefont {M.}~\bibnamefont {Marganska}}, \bibinfo {author}
		{\bibfnamefont {M.}~\bibnamefont {Villiers}}, \bibinfo {author}
		{\bibfnamefont {M.~M.}\ \bibnamefont {Desjardins}}, \bibinfo {author}
		{\bibfnamefont {M.~C.}\ \bibnamefont {Dartiailh}}, \bibinfo {author}
		{\bibfnamefont {V.}~\bibnamefont {Vinel}}, \bibinfo {author} {\bibfnamefont
			{Z.}~\bibnamefont {Leghtas}}, \bibinfo {author} {\bibfnamefont
			{A.}~\bibnamefont {Thiaville}}, \bibinfo {author} {\bibfnamefont
			{S.}~\bibnamefont {Rohart}}, \bibinfo {author} {\bibfnamefont
			{A.}~\bibnamefont {Cottet}}, \bibinfo {author} {\bibfnamefont {M.~R.}\
			\bibnamefont {Delbecq}},\ and\ \bibinfo {author} {\bibfnamefont
			{T.}~\bibnamefont {Kontos}},\ }\bibfield  {title} {\bibinfo {title}
		{Inhomogeneous magnetic fields interacting with spinful states in a double
			quantum dot: Evidence for a staggered spin-orbit interaction},\ }\href
	{https://doi.org/10.1103/PhysRevB.107.085152} {\bibfield  {journal} {\bibinfo
			{journal} {Phys. Rev. B}\ }\textbf {\bibinfo {volume} {107}},\ \bibinfo
		{pages} {085152} (\bibinfo {year} {2023})}\BibitemShut {NoStop}%
	\bibitem [{\citenamefont {Le~Breton}\ \emph {et~al.}(2011)\citenamefont
		{Le~Breton}, \citenamefont {Sharma}, \citenamefont {Saito}, \citenamefont
		{Yuasa},\ and\ \citenamefont {Jansen}}]{breton2011thermal}%
	\BibitemOpen
	\bibfield  {author} {\bibinfo {author} {\bibfnamefont {J.-C.}\ \bibnamefont
			{Le~Breton}}, \bibinfo {author} {\bibfnamefont {S.}~\bibnamefont {Sharma}},
		\bibinfo {author} {\bibfnamefont {H.}~\bibnamefont {Saito}}, \bibinfo
		{author} {\bibfnamefont {S.}~\bibnamefont {Yuasa}},\ and\ \bibinfo {author}
		{\bibfnamefont {R.}~\bibnamefont {Jansen}},\ }\bibfield  {title} {\bibinfo
		{title} {Thermal spin current from a ferromagnet to silicon by seebeck spin
			tunnelling},\ }\href {https://doi.org/10.1038/nature10224} {\bibfield
		{journal} {\bibinfo  {journal} {Nature}\ }\textbf {\bibinfo {volume} {475}},\
		\bibinfo {pages} {82} (\bibinfo {year} {2011})}\BibitemShut {NoStop}%
	\bibitem [{\citenamefont {Uchida}\ \emph {et~al.}(2010)\citenamefont {Uchida},
		\citenamefont {Xiao}, \citenamefont {Adachi}, \citenamefont {Ohe},
		\citenamefont {Takahashi}, \citenamefont {Ieda}, \citenamefont {Ota},
		\citenamefont {Kajiwara}, \citenamefont {Umezawa}, \citenamefont {Kawai},
		\citenamefont {Bauer}, \citenamefont {Maekawa},\ and\ \citenamefont
		{Saitoh}}]{uchida2010spin}%
	\BibitemOpen
	\bibfield  {author} {\bibinfo {author} {\bibfnamefont {K.}~\bibnamefont
			{Uchida}}, \bibinfo {author} {\bibfnamefont {J.}~\bibnamefont {Xiao}},
		\bibinfo {author} {\bibfnamefont {H.}~\bibnamefont {Adachi}}, \bibinfo
		{author} {\bibfnamefont {J.}~\bibnamefont {Ohe}}, \bibinfo {author}
		{\bibfnamefont {S.}~\bibnamefont {Takahashi}}, \bibinfo {author}
		{\bibfnamefont {J.}~\bibnamefont {Ieda}}, \bibinfo {author} {\bibfnamefont
			{T.}~\bibnamefont {Ota}}, \bibinfo {author} {\bibfnamefont {Y.}~\bibnamefont
			{Kajiwara}}, \bibinfo {author} {\bibfnamefont {H.}~\bibnamefont {Umezawa}},
		\bibinfo {author} {\bibfnamefont {H.}~\bibnamefont {Kawai}}, \bibinfo
		{author} {\bibfnamefont {G.~E.~W.}\ \bibnamefont {Bauer}}, \bibinfo {author}
		{\bibfnamefont {S.}~\bibnamefont {Maekawa}},\ and\ \bibinfo {author}
		{\bibfnamefont {E.}~\bibnamefont {Saitoh}},\ }\bibfield  {title} {\bibinfo
		{title} {Spin seebeck insulator},\ }\href {https://doi.org/10.1038/nmat2856}
	{\bibfield  {journal} {\bibinfo  {journal} {Nature Materials}\ }\textbf
		{\bibinfo {volume} {9}},\ \bibinfo {pages} {894} (\bibinfo {year}
		{2010})}\BibitemShut {NoStop}%
	\bibitem [{\citenamefont {Benenti}\ \emph {et~al.}(2022)\citenamefont
		{Benenti}, \citenamefont {Casati}, \citenamefont {Marchesoni},\ and\
		\citenamefont {Wang}}]{benenti2022autonomous}%
	\BibitemOpen
	\bibfield  {author} {\bibinfo {author} {\bibfnamefont {G.}~\bibnamefont
			{Benenti}}, \bibinfo {author} {\bibfnamefont {G.}~\bibnamefont {Casati}},
		\bibinfo {author} {\bibfnamefont {F.}~\bibnamefont {Marchesoni}},\ and\
		\bibinfo {author} {\bibfnamefont {J.}~\bibnamefont {Wang}},\ }\bibfield
	{title} {\bibinfo {title} {Autonomous circular heat engine},\ }\href
	{https://doi.org/10.1103/PhysRevE.106.044104} {\bibfield  {journal} {\bibinfo
			{journal} {Phys. Rev. E}\ }\textbf {\bibinfo {volume} {106}},\ \bibinfo
		{pages} {044104} (\bibinfo {year} {2022})}\BibitemShut {NoStop}%
	\bibitem [{\citenamefont {Lieb}\ and\ \citenamefont
		{Liniger}(1963)}]{lieb-liniger1963exact}%
	\BibitemOpen
	\bibfield  {author} {\bibinfo {author} {\bibfnamefont {E.~H.}\ \bibnamefont
			{Lieb}}\ and\ \bibinfo {author} {\bibfnamefont {W.}~\bibnamefont {Liniger}},\
	}\bibfield  {title} {\bibinfo {title} {Exact analysis of an interacting bose
			gas. i. the general solution and the ground state},\ }\href
	{https://doi.org/10.1103/PhysRev.130.1605} {\bibfield  {journal} {\bibinfo
			{journal} {Phys. Rev.}\ }\textbf {\bibinfo {volume} {130}},\ \bibinfo {pages}
		{1605} (\bibinfo {year} {1963})}\BibitemShut {NoStop}%
	\bibitem [{\citenamefont {Lieb}(1963)}]{lieb1963exact}%
	\BibitemOpen
	\bibfield  {author} {\bibinfo {author} {\bibfnamefont {E.~H.}\ \bibnamefont
			{Lieb}},\ }\bibfield  {title} {\bibinfo {title} {Exact analysis of an
			interacting bose gas. ii. the excitation spectrum},\ }\href
	{https://doi.org/10.1103/PhysRev.130.1616} {\bibfield  {journal} {\bibinfo
			{journal} {Phys. Rev.}\ }\textbf {\bibinfo {volume} {130}},\ \bibinfo {pages}
		{1616} (\bibinfo {year} {1963})}\BibitemShut {NoStop}%
	\bibitem [{\citenamefont {Alicki}\ and\ \citenamefont
		{Fannes}(2013)}]{alicki2013entanglement}%
	\BibitemOpen
	\bibfield  {author} {\bibinfo {author} {\bibfnamefont {R.}~\bibnamefont
			{Alicki}}\ and\ \bibinfo {author} {\bibfnamefont {M.}~\bibnamefont
			{Fannes}},\ }\bibfield  {title} {\bibinfo {title} {Entanglement boost for
			extractable work from ensembles of quantum batteries},\ }\href
	{https://doi.org/10.1103/PhysRevE.87.042123} {\bibfield  {journal} {\bibinfo
			{journal} {Phys. Rev. E}\ }\textbf {\bibinfo {volume} {87}},\ \bibinfo
		{pages} {042123} (\bibinfo {year} {2013})}\BibitemShut {NoStop}%
	\bibitem [{\citenamefont {Lu}\ \emph {et~al.}(2025)\citenamefont {Lu},
		\citenamefont {Tian}, \citenamefont {Lü},\ and\ \citenamefont
		{Shang}}]{lu2025topological}%
	\BibitemOpen
	\bibfield  {author} {\bibinfo {author} {\bibfnamefont {Z.-G.}\ \bibnamefont
			{Lu}}, \bibinfo {author} {\bibfnamefont {G.}~\bibnamefont {Tian}}, \bibinfo
		{author} {\bibfnamefont {X.-Y.}\ \bibnamefont {Lü}},\ and\ \bibinfo {author}
		{\bibfnamefont {C.}~\bibnamefont {Shang}},\ }\bibfield  {title} {\bibinfo
		{title} {Topological quantum batteries},\ }\bibfield  {journal} {\bibinfo
		{journal} {Physical Review Letters}\ }\textbf {\bibinfo {volume} {134}},\
	\href {https://doi.org/10.1103/physrevlett.134.180401}
	{10.1103/physrevlett.134.180401} (\bibinfo {year} {2025})\BibitemShut
	{NoStop}%
	\bibitem [{\citenamefont {Li}\ \emph {et~al.}(2026)\citenamefont {Li},
		\citenamefont {Shang}, \citenamefont {Kuwahara},\ and\ \citenamefont
		{Vu}}]{li2026macroscopic}%
	\BibitemOpen
	\bibfield  {author} {\bibinfo {author} {\bibfnamefont {H.}~\bibnamefont
			{Li}}, \bibinfo {author} {\bibfnamefont {C.}~\bibnamefont {Shang}}, \bibinfo
		{author} {\bibfnamefont {T.}~\bibnamefont {Kuwahara}},\ and\ \bibinfo
		{author} {\bibfnamefont {T.~V.}\ \bibnamefont {Vu}},\ }\bibfield  {title}
	{\bibinfo {title} {Macroscopic particle transport in dissipative long-range
			bosonic systems},\ }\bibfield  {journal} {\bibinfo  {journal} {Nature
			Communications}\ }\textbf {\bibinfo {volume} {17}},\ \href
	{https://doi.org/10.1038/s41467-026-70881-7} {10.1038/s41467-026-70881-7}
	(\bibinfo {year} {2026})\BibitemShut {NoStop}%
	\bibitem [{\citenamefont {Ghosh}\ \emph
		{et~al.}(2026{\natexlab{b}})\citenamefont {Ghosh}, \citenamefont {Ghosh},
		\citenamefont {Dutta},\ and\ \citenamefont {Maity}}]{ghosh2026quantum}%
	\BibitemOpen
	\bibfield  {author} {\bibinfo {author} {\bibfnamefont {S.}~\bibnamefont
			{Ghosh}}, \bibinfo {author} {\bibfnamefont {A.}~\bibnamefont {Ghosh}},
		\bibinfo {author} {\bibfnamefont {B.}~\bibnamefont {Dutta}},\ and\ \bibinfo
		{author} {\bibfnamefont {P.}~\bibnamefont {Maity}},\ }\href
	{https://doi.org/10.48550/ARXIV.2606.06432} {\bibinfo {title} {Quantum
			thermal logic gates}} (\bibinfo {year} {2026}{\natexlab{b}})\BibitemShut
	{NoStop}%
	\bibitem [{\citenamefont {Ghosh}\ \emph
		{et~al.}(2026{\natexlab{c}})\citenamefont {Ghosh}, \citenamefont {Mandal},\
		and\ \citenamefont {Ghosh}}]{ghosh2026information}%
	\BibitemOpen
	\bibfield  {author} {\bibinfo {author} {\bibfnamefont {S.}~\bibnamefont
			{Ghosh}}, \bibinfo {author} {\bibfnamefont {S.}~\bibnamefont {Mandal}},\ and\
		\bibinfo {author} {\bibfnamefont {A.}~\bibnamefont {Ghosh}},\ }\href@noop {}
	{\bibinfo {title} {A feedback-controlled maxwell demon refrigerator in
			coupled quantum dots}} (\bibinfo {year} {2026}{\natexlab{c}}),\ \Eprint
	{https://arxiv.org/abs/(Manuscript under preparation)} {(Manuscript under
		preparation)} \BibitemShut {NoStop}%
	\bibitem [{\citenamefont {Satrya}\ \emph {et~al.}(2025)\citenamefont {Satrya},
		\citenamefont {Strelnikov}, \citenamefont {Magazzù}, \citenamefont {Chang},
		\citenamefont {Upadhyay}, \citenamefont {Peltonen}, \citenamefont {Karimi},\
		and\ \citenamefont {Pekola}}]{satrya2025heat}%
	\BibitemOpen
	\bibfield  {author} {\bibinfo {author} {\bibfnamefont {C.~D.}\ \bibnamefont
			{Satrya}}, \bibinfo {author} {\bibfnamefont {A.~S.}\ \bibnamefont
			{Strelnikov}}, \bibinfo {author} {\bibfnamefont {L.}~\bibnamefont
			{Magazzù}}, \bibinfo {author} {\bibfnamefont {Y.-C.}\ \bibnamefont {Chang}},
		\bibinfo {author} {\bibfnamefont {R.}~\bibnamefont {Upadhyay}}, \bibinfo
		{author} {\bibfnamefont {J.~T.}\ \bibnamefont {Peltonen}}, \bibinfo {author}
		{\bibfnamefont {B.}~\bibnamefont {Karimi}},\ and\ \bibinfo {author}
		{\bibfnamefont {J.~P.}\ \bibnamefont {Pekola}},\ }\href
	{https://arxiv.org/abs/2510.23092} {\bibinfo {title} {Heat measurement of
			quantum interference}} (\bibinfo {year} {2025}),\ \Eprint
	{https://arxiv.org/abs/2510.23092} {arXiv:2510.23092 [cond-mat.mes-hall]}
	\BibitemShut {NoStop}%
\end{thebibliography}
\end{document}